\documentclass[aps,twocolumn,a4paper,showkeys,nofootinbib,longbibliography,10pt]{revtex4-2}
\usepackage[utf8]{inputenc}
\usepackage{filecontents}
\usepackage{natbib}
\usepackage{amsmath,amssymb,bm,amsthm}
\usepackage{subfigure}
\usepackage{graphicx}% Include figure files
\usepackage{dcolumn}% Align table columns on decimal point
\usepackage{bm}% bold math
\usepackage[mathlines]{lineno}% Enable numbering of text and display math
\usepackage{booktabs}
\usepackage{color}
\newcounter{one}
\usepackage{url}

\usepackage{textcase}

\usepackage{colortbl}
\usepackage{tabularx}
\usepackage{verbatim}
\usepackage{multirow}

\usepackage[T1]{fontenc} 
\usepackage{lmodern}
\usepackage{bbm}
\usepackage[utf8]{inputenc}
\usepackage{amsfonts}
\usepackage{array}

\usepackage{bbm}
\usepackage{enumitem}
\usepackage{umoline}
\usepackage[usenames,svgnames]{xcolor}
\usepackage{natbib}
\usepackage[hyperindex,breaklinks]{hyperref}
\hypersetup{
     colorlinks=true,       		% false: boxed links; true: colored links
     linkcolor=Navy,          	% color of internal links
     citecolor=Navy,            % color of links to bibliography
     filecolor=Navy,      		% color of file links
     urlcolor=Navy,           	% color of external links
    runcolor=cyan,
 }
\usepackage{graphicx}
\usepackage{amsfonts}
\usepackage{amssymb}
\usepackage{amsmath}
\usepackage{bbm}
\usepackage{enumitem}

\newcommand{\bra}[1]{\langle #1 |}
\newcommand{\ket}[1]{| #1 \rangle}

\newcommand{\tr}[0]{ {\rm tr}}

\newcommand{\half}[1]{{ \rm h}}
\newcommand{\Oorderof}{\mathcal{O}}
\newcommand{\orderof}[1]{\Oorderof(#1)} 

\newcommand{\for}[0]{\quad \textrm{for} \quad}

\newcommand{\dist}{d}

\newcommand{\co}{{\rm c}}

\newcommand{\diam}{{\rm diam}}

\newcommand{\ad}{{\rm ad}}

\def\beq{\begin{equation}}
\def\eeq{\end{equation}}
\def\nbeq{\begin{equation*}}
\def\neeq{\end{equation*}}
\def\<{\langle}
\def\>{\rangle}

\def\tr{{\rm tr}}

\newtheorem{theorem}{Theorem}
\newtheorem{subtheorem}{Subtheorem}
\newtheorem{lemma}{Lemma}
\newtheorem{corol}[lemma]{Corollary}
\newtheorem{assump}[lemma]{Assumption} 

\newtheorem{prop}[lemma]{Proposition} 
\newtheorem{claim}[lemma]{Claim} 

\newcommand{\bal}[2]{#1[#2]}

\newcommand{\nb}{\hat{n}}

\newcommand{\sectionprl}[1]{{\par\it #1.---}}

\newcommand{\br}[1]{\left( #1 \right)}
 \newcommand{\norm}[1]{\left \|  #1 \right \|}

\usepackage{mathtools}
\def\multiset#1#2{\ensuremath{\left(\kern-.3em\left(\genfrac{}{}{0pt}{}{#1}{#2}\right)\kern-.3em\right)}}

\usepackage{minitoc}
\usepackage[toc,page,titletoc]{appendix}
\usepackage[capitalise,nameinlink]{cleveref}

\crefname{supp}{Supplement}{Supplements}

\begin{document}

%\title{Polynomial growth of out-of-time-order correlator in arbitrary realistic long-range interacting systems}
\title{Lieb-Robinson bound and almost-linear light-cone in interacting boson systems}

\author{Tomotaka Kuwahara$^{1}$ and Keiji Saito$^{2}$}
%\email{tomotaka.kuwahara@riken.jp}
%\altaffiliation{Present address: Mathematical Science Team, RIKEN Center for Advanced Intelligence Project (AIP),1-4-1 Nihonbashi, Chuo-ku, Tokyo 103-0027, Japan}
\affiliation{$^{1}$
Mathematical Science Team, RIKEN Center for Advanced Intelligence Project (AIP),1-4-1 Nihonbashi, Chuo-ku, Tokyo 103-0027, Japan
}
%\affiliation{$^{2}$Department of Mathematics, Faculty of Science and Technology, Keio University, 3-14-1 Hiyoshi, Kouhoku-ku, Yokohama 223-8522, Japan}
%\affiliation{$^{2}$Interdisciplinary Theoretical \& Mathematical Sciences Program (iTHEMS) RIKEN 2-1, Hirosawa, Wako, Saitama 351-0198, Japan}

%\affiliation{
%Department of Physics, Graduate School of Science,
%University of Tokyo, Kashiwa 277-8574, Japan
%}
%\author{Keiji Saito}
%%\email{saitoh@rk.phys.keio.ac.jp}
\affiliation{$^{2}$Department of Physics, Keio University, Yokohama 223-8522, Japan}

\begin{abstract}

In this work, we investigate how quickly local perturbations propagate in interacting boson systems with Bose-Hubbard-type Hamiltonians. In general, these systems have unbounded local energies, and arbitrarily fast information propagation may occur.  
We focus on a specific but experimentally natural situation in which
the number of bosons at any one site in the unperturbed initial state is approximately limited. We rigorously prove the existence of an almost-linear information-propagation light-cone, thus establishing a Lieb--Robinson bound: the wave-front grows at most as $t\log^2 (t)$. We prove the clustering theorem for gapped ground states and study the time complexity of classically simulating one-dimensional quench dynamics, a topic of great practical interest.

\end{abstract}

\maketitle

%1st graph:

%\section{Introduction}
\sectionprl{Introduction}
In non-relativistic quantum many-body systems, the speed limit of information propagation is characterized by the Lieb--Robinson bound~\cite{ref:LR-bound72,PhysRevLett.97.050401,doi:10.1063/1.5095769},
an effective light-cone outside which the amount of transferred information rapidly decays with distance. 
In standard spin models such as the transverse Ising model, 
the light-cone is linear over time and characterized by the Lieb--Robinson velocity, which depends only on the system details.
As a fundamental restriction applied to generic many-body quantum systems, 
the Lieb--Robinson bound has been utilized to establish the clustering theorem on bi-partite correlations in ground states~\cite{ref:Hastings2004-Markov,ref:Hastings2006-ExpDec,ref:Nachtergaele2006-LR} and an
efficient classical/quantum algorithm to simulate quantum many-body dynamics~\cite{PhysRevLett.97.157202,8555119,PhysRevX.9.031006}. 
It has featured in many fields of quantum many-body physics including condensed matter theory~\cite{PhysRevB.69.104431,Nachtergaele2007,PhysRevB.72.045141,ref:Hastings-AL07,bravyi2010topological,PhysRevLett.111.080401,hastings2015quantization}, 
statistical mechanics~\cite{ref:Mueller2013-Thermal,PhysRevLett.113.127202,PhysRevLett.119.100601,KUWAHARA201696,PhysRevB.95.014112,PhysRevLett.119.060201,kuwahara2020improved}, 
high-energy physics~\cite{Maldacena2016,PhysRevLett.117.091602,PhysRevLett.124.240505,PhysRevLett.124.180601,chen2019finite,PhysRevLett.126.030604,chen2021concentration}, and quantum information~\cite{PhysRevB.76.201102,PhysRevLett.97.150404,PhysRevLett.116.080503,PhysRevLett.121.030501,PhysRevX.11.011020}.

The Lieb--Robinson bound and the existence of a linear light-cone are well-understood under the following two conditions ~\cite{Nachtergaele2006,ref:Hastings2006-ExpDec,ref:Nachtergaele2006-LR,nachtergaele2010lieb,doi:10.1063/1.5095769}: i) the interaction is short-range, and ii) the Hamiltonian is locally bounded.
% \red{(explanatiaon) , namely (unnecessary?)}. 
If either of these conditions is broken, as often happens in real-world quantum systems, the shape of the linear light-cone becomes quite complicated. 
When there are long-range interactions, breaking the first condition,  a comprehensive characterization of the shape of the light-cone 
has been achieved~\cite{PhysRevLett.111.260401,PhysRevLett.114.157201,chen2019finite,PhysRevX.10.031009,PhysRevX.10.031010,tran2020optimal,PhysRevLett.126.030604}. 
However, it remains challenging to clarify the Lieb--Robinson bound when the second condition breaks down. 

Quantum boson systems are representative examples of the breakdown of this second condition with locally unbounded Hamiltonians.
%; other examples include semi-classical spin systems~\cite{PRXQuantum.1.010303,yin2020quantum} and continuous fermion systems. 
The difficulty lies in the fact that the standard approach for the Lieb--Robinson bound necessarily results in a Lieb--Robinson velocity proportional to the norm of the local energy. 
When $N$ bosons clump at a single location, the on-site energy can be as large as ${\rm poly}(N)$, leading to an infinite Lieb--Robinson velocity as $N\to \infty$.
Even though it is quite unlikely that many bosons will clump together in realistic experiments, the theoretical possibility of such situations must be taken into account.
If harmonic and anharmonic systems~\cite{cramer2008locality,Nachtergaele2009,Raz2009,doi:10.1142/S0129055X1000393X,nachtergaele2014dynamics} 
and spin boson models~\cite{PhysRevLett.111.230404,PhysRevLett.115.130401,doi:10.1063/1.4940436} are considered, 
the Lieb--Robinson bound with the linear light-cone has been established.
However, we have no hope of unconditionally proving the existence of a Lieb--Robinson bound  without restricting the form of Hamiltonians or initial states. (In Ref.~\cite{PhysRevLett.102.240501}, Eisert and Gross provided 1D quantum boson systems with nearest-neighbor interactions, 
inducing an exponential speed of information propagation.) % for initial states in low-energy sectors. 

Recent experiments have focused on interacting bosonic systems of the Bose-Hubbard type~\cite{RevModPhys.80.885,Sherson2010,Bakr547,cheneau2012light,langen2013local,Braun3641,Islam2015,Choi1547,PhysRevLett.116.205301,Baier201,PhysRevLett.123.050502,Yan753,PhysRevX.10.021044,Yang2020,Takasueaba9255}, which typically appear in cold atom setups.
Since the earliest experiments on the Lieb--Robinson bound~\cite{cheneau2012light,langen2013local}, 
there have been many attempts to clarify information propagation 
in these models rigorously. 
However, with a few exceptions~\cite{PhysRevA.84.032309,PRXQuantum.1.010303}, establishing the Lieb--Robinson bound in Bose-Hubbard-type models remains an open problem.
A previous rigorous study~\cite{PhysRevA.84.032309} showed that initially concentrated bosons in the vacuum spread at a finite speed. 
In Ref.~\cite{PRXQuantum.1.010303}, the Lieb--Robinson velocity was qualitatively improved from $\orderof{N}$ to $\orderof{\sqrt{N}}$ (still infinitely large in the limit of $N\to \infty$), where $N$ is the total number of bosons. 
On the other hand, numerical calculations and theoretical case studies indicate 
that a linear light-cone should be observed in practical settings such as quench dynamics ~\cite{L_uchli_2008,PhysRevA.85.053625,PhysRevA.87.063616,PhysRevA.89.031602,PhysRevA.92.063619,PhysRevLett.120.020401,PhysRevA.98.053618,PhysRevA.103.023334}. 
The most natural condition is to require a finite number of bosons at any one site in the initial state, for example, a
Mott state.
However, this condition can break down over time, and a large bias in the boson distribution may cause an unexpected acceleration of information propagation~\cite{PhysRevA.84.032309}.   
Until now, no theoretical tools have been developed to overcome this obstacle.

In this work, we establish the Lieb--Robinson bound with an almost linear light-cone when a local perturbation is added to quantum states that are initially time-independent and have low boson density [see the condition~\eqref{main_condition_for_moment_generating}]. 
Our Lieb--Robinson bound characterizes a wave-front that propagates as $t \log^2(t)$ with time. 
As a practical application, we derive the clustering theorem for non-critical ground states by extending the technique in \cite{ref:Hastings2004-Markov,ref:Hastings2006-ExpDec,ref:Nachtergaele2006-LR}.
In addition, 
we extend our theory to analyze the time complexity of computing quantum dynamics by quenching the Hamiltonian parameter, 
a topic of major research interest~\cite{PhysRevLett.98.180601,L_uchli_2008,PhysRevLett.100.030602,PhysRevLett.101.063001,PhysRevA.87.063616,PhysRevA.89.031602,PhysRevA.92.063619,PhysRevA.98.053618,PhysRevA.79.021608,PhysRevA.82.063603,Enss_2012,PhysRevA.85.053625,PhysRevA.85.033641,PhysRevA.90.033606,PhysRevLett.112.065301,PhysRevA.89.031602,Geiger_2014,Krutitsky2014,PhysRevB.91.235132,PhysRevB.96.054503,PhysRevLett.121.250404,PhysRevLett.120.020401,PhysRevA.98.053618,PhysRevA.99.023620,Despres2019,PhysRevA.102.033337,PhysRevA.103.023334}.
We rigorously establish the time complexity of $e^{t \log^3(t)}$ to simulate local quench dynamics for one-dimensional Bose-Hubbard-type Hamiltonians.
%\red{It extends the results for quantum spin systems~\cite{PhysRevLett.97.157202}.}
%

%\section{Setup and main result}
\sectionprl{Setup and main result}
We consider a quantum system on a finite-dimensional lattice (graph), where bosons interact with each other. 
An unbounded number of bosons can sit on each of the sites, and the local Hilbert dimension is thus infinitely large.
We denote by $\Lambda$ the set of all sites on the lattice. For an arbitrary partial set $X\subseteq \Lambda$, we denote the cardinality (the number of sites contained in $X$) by $|X|$.
For arbitrary subsets $X, Y \subseteq \Lambda$, we define $\dist_{X,Y}$ as the shortest path-length on the graph that connects $X$ and $Y$.
For a subset $X\subseteq \Lambda$, we define the extended subset $\bal{X}{r}$ by length $r$ as
\begin{align}
\bal{X}{r}:= \{i\in \Lambda| \dist_{X,i} \le r \} , \label{main_def:bal_X_r}
\end{align}
where $\bal{X}{0}=X$ and $r$ is an arbitrary positive number (i.e., $r\in \mathbb{R}^+$).

We define $b_i$ and $b_i^\dagger$ as the annihilation and creation operators of the boson, respectively.
We also define $\nb_i:=b_i^\dagger b_i$ as the number operator of bosons on site $i$.
We consider a Hamiltonian of the form 
\begin{align}
&H:= \sum_{\langle i, j \rangle} J_{i,j} (b_i b_j^\dagger +{\rm h.c.} )+  \sum_{Z \subset \Lambda: |Z|\le k} v_Z  , \label{main_def:Ham} 
\end{align}
where $|J_{i,j}| \le \bar{J}$ and $\sum_{\langle i, j \rangle}$ denotes summation over all pairs of adjacent sites $\{i,j\}$ on the lattice. 
Here, $v_Z$ consists of finite-range boson-boson interactions on subset $Z$.
We now assume that $v_Z$ is given as a function of the number operators $\{\nb_i\}_{i\in Z}$.
The simplest example is the Bose-Hubbard model: 
\begin{align}
H= \sum_{\langle i, j \rangle}J (b_i b_j^\dagger +{\rm h.c.} ) + \frac{U}{2} \sum_{i\in \Lambda} \nb_i(\nb_i-1) 
-\mu\sum_{i\in \Lambda} \nb_i  , \notag 
\end{align}
where $U$ and $\mu$ are $\orderof{1}$ constants.
For an arbitrary operator $O$, the time-evolution due to another operator $A$ is
\begin{align}
\label{notation_time_evolve_A}
O(A,t) := e^{iAt} O e^{-iA t}.
\end{align}
(We abbreviate $O(H,t)$ as $O(t)$ for simplicity.)

Let $\rho_0$ be a time-independent quantum state, i.e., $[\rho_0,H]=0$.
We consider propagation of a local perturbation to $\rho_0$ such as
$
\rho \to O_{i_0} \rho_0 O_{i_0}^\dagger ,
$
where $i_0\in \Lambda$ and $O_{i_0}$ can take the form of a projection onto site $i_0$. 
We are interested in how fast this perturbation propagates. 
Mathematically, after the time evolution, $\rho(t)$ is given by $O_{i_0}(t) \rho_0 O_{i_0}(t)^\dagger$. 
Thus, we must estimate the approximation error of 
\begin{align}
O_{i_0}(t) \rho_0 \approx O^{(t)}_{i_0[R]} \rho_0  ,
\end{align}
where $O^{(t)}_{i_0[R]}$ is an appropriate operator supported on subset $i_0[R]$ [see the notation~\eqref{main_def:bal_X_r}]. 
Our main result concerns the approximation error for finite $R$ (see  Sec.~S.II. in Supplementary materials~\cite{Supplement_boson} for the formal expression).

Following Ref.~\cite{PhysRevX.10.031010}, we define the shape of the light cone in the following sense.
We say that the Hamiltonian dynamics $e^{-iHt}$ have an effective light cone with velocity $ v_{t,\delta}$ if the following inequality holds for an arbitrary error $\delta \in \mathbb{R}$ and $t$:
 \begin{align}
&\| [O_{i_0}(t)-O^{(t)}_{i_0[R]} ]\|\le \delta \norm{O_{i_0}} \for R\ge v_{t,\delta} |t|. \label{def:linear_light_cone}
\end{align}
When $v_{t,\delta}$ converges to a finite value for $t\to \infty$ (i.e., $v_{\infty,\delta} = {\rm const.}$), we say that the effective light cone is linear.
 From the definition, the amount of information propagation is smaller than $\delta$ outside the region separated by the distance $v_{t,\delta}|t|$.

{\bf Main Theorem.}
\textit{Let us assume that the number of boson creations by $O_{i_0}$ is finitely bounded.
Then, for an arbitrary time-independent quantum state $\rho_0$ satisfying the low-boson-density condition 
\begin{align}
\label{main_condition_for_moment_generating}
\max_{i\in \Lambda} \tr (e^{c_{0} (\nb_i -\bar{q}) } \rho_0) \le  1 \quad c_{0} \le 1,
\end{align}
we can approximate $O_{i_0}(t) \rho_0$ by another operator $O^{(t)}_{i_0[R]}$ supported on $i_0[R]$ with the following approximation error:
\begin{align}
\label{main_main_theorem_short_time_LR_main_ineq}
&\left \| \br{ O_{i_0}(t)-O^{(t)}_{i_0[R]} }\rho_0 \right\|  \notag \\
&\le \| O_{i_0}\| \exp\br{c_{0}\bar{q}- C_1 \frac{R}{t\log (R)} + C_2 \log(R)}  ,
\end{align}
where $t\ge 1$, and $C_1$ and $C_2$ are constants of $\orderof{1}$ that are independent of $\bar{q}$ and only depend on the details of the system. 
For a general operator $O_{X_0}$, we can obtain a similar inequality by slightly changing~\eqref{main_main_theorem_short_time_LR_main_ineq}.
}

Condition ~\eqref{main_condition_for_moment_generating} ensures that the probability for many bosons to be concentrated on one site is exponentially small in the initial state $\rho_0$. 
We notice that the condition can break down as time increases. 
By applying the inequality~\eqref{main_main_theorem_short_time_LR_main_ineq} to \eqref{def:linear_light_cone}, we obtain $v_{t,\delta} \propto \log^2(t) [\log(1/\delta) +c_0\bar{q}]$. Hence, information propagation is restricted in the region that is separated from $i_0$ by at most $\orderof{\bar{q}} t\log^2(t)$.  
Therefore, we can ensure that the acceleration of information propagation observed in Ref.~\cite{PhysRevLett.102.240501} cannot occur in our model, 
because the speed of information becomes at most polylogarithmically large with time, i.e., $\le \log^2(t)$.
%\red{As shown below, the key point is that the boson density }

\sectionprl{Clustering Theorem}  
As an immediate application of the main theorem, 
we consider the exponential decay of bi-partite correlations in gapped ground states, i.e., the clustering theorem.
Here, we denote the non-degenerate ground state by $\ket{E_0}$ and the spectral gap by $\Delta E$. 
We prove an upper-bound on the correlation function ${\rm Cor}(O_X,O_Y):= \bra{E_0}O_XO_Y\ket{E_0} - \bra{E_0}O_X\ket{E_0} \bra{E_0}O_Y\ket{E_0}$, where $O_X$ and $O_Y$ are operators supported on $X$ and $Y$.
For simplicity, we let $\bar{q}=\orderof{1}$. 
Then, the following inequality holds if $\ket{E_0}$ satisfies condition~\eqref{main_condition_for_moment_generating} (see Sec.~S.III. in Supplementary materials~\cite{Supplement_boson}):
\begin{align}
{\rm Cor}(O_X,O_Y)\le 
C_3 \|O_X\| \cdot \|O_Y\| \exp\br{-\sqrt{\frac{C_3' \Delta E}{ \log(R)}R}},  
\label{clustering_gs_boson}
\end{align}
where $C_3$, $C_3'$ and $C_3''$ are $\orderof{1}$ constants. 
From the inequality, the bi-partite correlations decay beyond $R\approx \tilde{\mathcal{O}}(1/\Delta E)$. 
This sub-exponential decay, which is weaker than the exponential decay described in Ref.~\cite{ref:Hastings2004-Markov,ref:Hastings2006-ExpDec,ref:Nachtergaele2006-LR}, is a consequence of the asymptotic form of $e^{-\orderof{R/(t\log R)}}$ in our Lieb--Robinson bound~\eqref{main_main_theorem_short_time_LR_main_ineq}.

%\section{Quench } 

\sectionprl{Application to quench dynamics}  
We next consider the application of our results to quench dynamics, 
the most popular setup in the study of non-equilibrium quantum systems. 
Here, a system is initially prepared in a steady state $\rho_0$ (e.g., the ground state), and then evolves unitarily in time under
the sudden change of the Hamiltonian $H \to H'$.  
We consider the case where the Hamiltonian $H'$ is given by $H'= H + h_{X_0}$, where we assume $H'$ still has the form of Eq.~\eqref{main_def:Ham}.
In addition, the interaction $h_{X_0}$ includes only polynomials of finite degree in $\{\nb_i\}_{i\in X_0}$, such as $\nb_i^2$ and $\nb_i^2 \nb_j^3$, etc.

Our purpose is to find an appropriate unitary operator $U_{i_0[R]}$ supported on $i_0[R]$ that gives $\rho_0(H',t)  \approx U_{i_0[R]} \rho_0   U^\dagger_{i_0[R]}$.
We can prove the following theorem (see Sec.~S.IX. in the Supplementary materials~\cite{Supplement_boson} for details):

{\bf Quench theorem.}
\textit{For initial state $\rho_0$ with the conditions $[\rho_0,H]=0$ and \eqref{main_condition_for_moment_generating}, 
we have 
\begin{align}
\label{main_theorem_short_time_LR_main_ineqquench}
&\norm{\rho_0(H',t)  - U_{i_0[R]} \rho_0   U^\dagger_{i_0[R]} }_1 \notag \\
& \le \exp\br{c_0\bar{q}- C'_1 \frac{(R-r_0)}{t\log (R)} + C'_2 \log(R)}  ,
\end{align}
where we define $r_0$ such that $X_0 \in i_0[r_0]$ for an appropriate $i_0\in \Lambda$, and $C'_1$ and $C'_2$ are constants of $\orderof{1}$ that are independent of $\bar{q}$ and only depend on the details of the system. 
Moreover, the computational cost of constructing the unitary operator $U_{i_0[R]}$ is at most 
$
 \exp\left [ \mathcal{O}\br {R^D  \log(R)} \right]  .
$}

This theorem immediately gives the following corollary on the time complexity of preparing $U_{i_0[R]}$:\\
{\bf Corollary.} \textit{
The computational cost of calculating the quench dynamics on 1D chains up to an error $\epsilon$ is at most
 \begin{align}
 \label{time_complexity_quench}
\exp \left[ t \log^3(t) +t \log(1/\epsilon) \log\log^2(1/\epsilon) \right] ,
\end{align}
where we assume $r_0=\orderof{1}$ and $\bar{q}=\orderof{1}$.}
When the error $\epsilon$ is fixed, we have a time complexity of $e^{t \log^3(t)}$. 
This is the first rigorous result on the efficiency of the classical simulation of interacting boson systems.

\sectionprl{Proof of the main theorem}
%\sectionprl{Approach to proving main theorem}
For the proof, we connect the Lieb--Robinson bounds for small time evolutions step by step, based on previous analyses of the Lieb--Robinson bound in long-range interacting systems~\cite{Kuwahara_2016_njp,PhysRevLett.126.030604}.  
The great merit of this approach is that we have to derive the Lieb--Robinson bound \textit{only for short-time evolution.} 
We decompose the total time $t$ into $m_t$ pieces and define $\Delta t:=t/m_t$ with $m_t=\orderof{t}$. 
Note that we can make $\Delta t$ arbitrarily small by making $m_t$ sufficiently large. 
For a fixed $R$, we define the subset $X_m$ as follows: 
\begin{align}
X_m:= i_0[m\Delta r], \quad \Delta r = \left \lfloor R/m_t \right \rfloor , \notag 
\end{align}
where $X_m = X_0[m\Delta r]$ and $X_{m_t} \subseteq i_0[R]$. 

We connect the step-by-step approximations of the short-time evolution to reach the final approximation.
Under the assumption of the time invariance of $\rho_0$ (i.e., $\rho_0(t)=\rho$), we can derive the following inequality~\cite{PhysRevLett.126.030604}: 
 \begin{align}
 \label{main_unitary_connect_upper_bound}
&\norm{\left [O_{i_0} (m_t\Delta t) - O_{X_{m_t}}^{(m_t)}\right]\rho_0 }_1 \notag \\
& \le \sum_{m=1}^{m_t} \norm{\left [ O_{X_{m-1}}^{(m-1)} (\Delta t) - O_{X_{m}}^{(m)} \right]\rho_0 }_1,
\end{align}
where $O_{X_0}^{(0)} = O_{X_0}$, and $O_{X_{m}}^{(m)}$ is recursively defined by approximating $O_{X_m}^{(m)} (\Delta t)$. 
When $\rho_0$ depends on the time, a severe modification is required in the inequality~\eqref{main_unitary_connect_upper_bound} 
(see Sec. IV. B in Supplementary materials~\cite{Supplement_boson}). 
In order to reduce~\eqref{main_unitary_connect_upper_bound} to the main inequality~\eqref{main_main_theorem_short_time_LR_main_ineq}, 
we need to obtain 
 \begin{align}
 \label{short_time_evolution_approx}
O_{X_{m-1}}^{(m-1)} (\Delta t)  \rho_0 \approx U_{X_{m}}^{(m)\dagger}O_{X_{m-1}}^{(m-1)} U_{X_{m}}^{(m)} \rho_0 = O_{X_{m}}^{(m)} \rho_0 , 
\end{align}
by using an appropriate unitary operator supported on $X_m$. 

Therefore, our primary task is to estimate the approximation error of~\eqref{short_time_evolution_approx}, 
which gives the Lieb--Robinson bound for the short time $\Delta t$. 
We can prove that, for a general operator $O_X$ with $X\subseteq i[r]$ ($i\in \Lambda$), there exists a unitary operator $ U_{X[\ell]}$ supported on $X[\ell]$ such that 
\begin{align}
\label{main_ineq:main_theorem_short_time_LR_time_dependent}
&\left \| \left( O_X (t) - U_{X[\ell]}^\dagger O_X U_{X[\ell]} \right) \rho_0 \right\|_1 \notag \\
&\le \| O_X\| e^{c_0\bar{q}-\ell/\log (r)+C_0\log(r)} ,
\end{align}
for $t\le \Delta t_0$ (see Subtheorem~1 in Supplementary materials~\cite{Supplement_boson}),
where $C_0$ and $\Delta t_0$ are $\orderof{1}$ constants. 
We here choose the time width $\Delta t$ such that $\Delta t\le \Delta t_0$. 
By using the inequality~\eqref{main_ineq:main_theorem_short_time_LR_time_dependent} with $\ell=\Delta r$ and $t=\Delta t$, 
we can reduce the inequality~\eqref{main_unitary_connect_upper_bound} to the desired form~\eqref{main_main_theorem_short_time_LR_main_ineq} by choosing $C_1$ and $C_2$ appropriately. 
This completes the proof of the main theorem.  $\square$

%\section{Short-time Lieb--Robinson bound}

 \begin{figure}[tt]
\centering
\includegraphics[clip, scale=0.28]{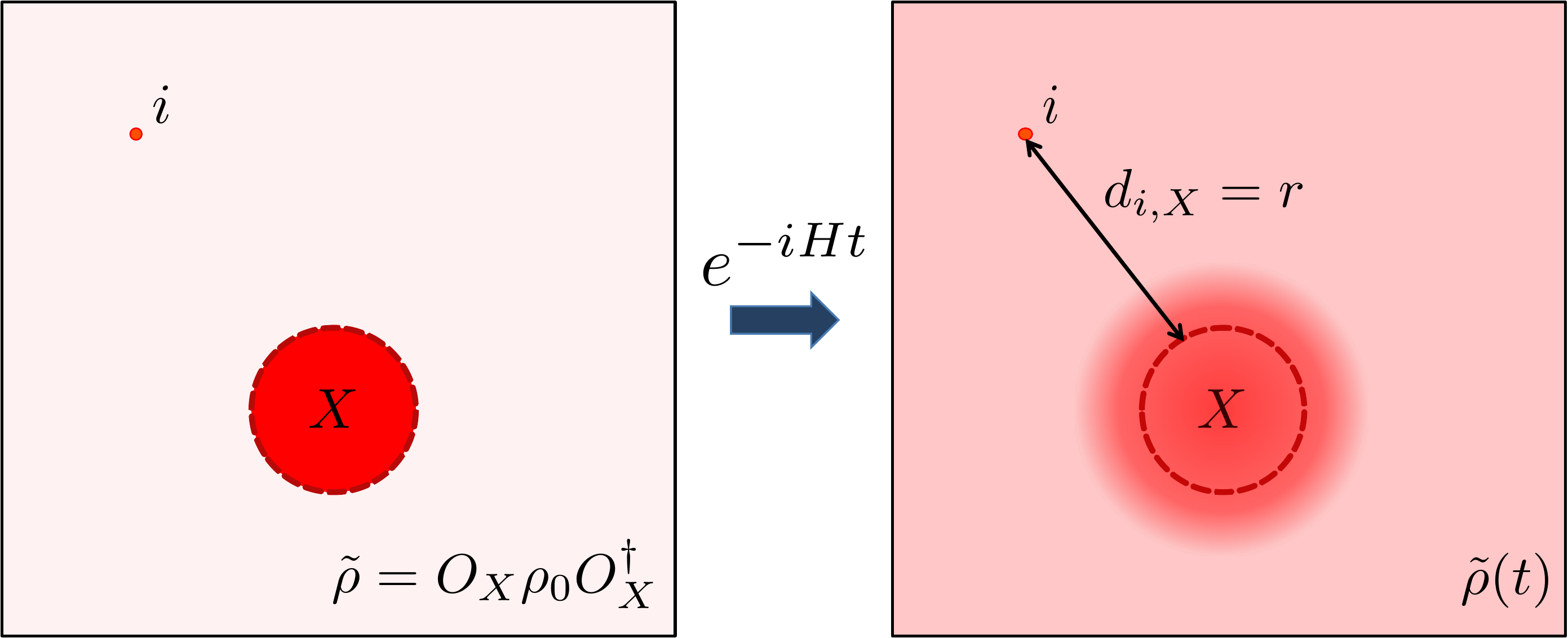}
\caption{Boson density after time evolution. 
In Ref.~\cite{PhysRevA.84.032309}, all bosons were initially concentrated in a region $X$ on the vacuum and were shown to spread beyond it with a finite velocity.
However, if there is initially a finite number of bosons outside $X$, the upper bound of the boson number increases exponentially with $t$. 
This spoils the approach of Ref.~\cite{PhysRevA.84.032309} in general setups for long-time evolution.  
Our approach only considers the short time $t=\orderof{1}$, when the exponential increase $e^{\orderof{t}}$ is still $\orderof{1}$. 
We then ensure that the boson number distribution for $\nb_i$ exponentially decays if the site $i$ of interest is sufficiently separated from the region $X$.
}
\label{fig_density_boson}
\end{figure}

\sectionprl{Short-time Lieb--Robinson bound}
We have seen that the bosonic Lieb--Robinson bound can be immediately derived if we can prove the inequality~\eqref{main_ineq:main_theorem_short_time_LR_time_dependent}, 
which includes all the difficulties in our proof.  
We will now provide a sketch of the proof; a fuller and more formal presentation can be found in the Supplementary materials~\cite{Supplement_boson} (Secs.~S.V.,~S.VI.,~S.VII. and S.VIII.). 

We first consider the boson density after short-time evolution (see Sec.~S.VI. in Supplementary materials~\cite{Supplement_boson}). 
For this purpose, we need to estimate 
\begin{align}
\label{def_tilde_rho_t}
\tr \left [ \nb_i^s \tilde{\rho}(t) \right] ,\quad 
\tilde{\rho}(t)= e^{-iHt} O_X \rho_0 O_X^\dagger e^{iHt} ,
\end{align}
with $s\in \mathbb{N}$. 
This quantity characterizes the influence of the perturbation $O_X$ on the boson density after time evolution. 
In the state $\tilde{\rho}(0)$, the boson number $\nb_i$ ($i\notin X$) is exponentially suppressed because of condition~\eqref{main_condition_for_moment_generating}, 
while the bosons may be highly concentrated in the region $X$.  
Time evolution will cause these concentrated bosons to spread outside $X$ (see Fig.~\ref{fig_density_boson}).

In order to characterize the dynamics of the bosons,  
we utilize the method in Ref.~\cite{PhysRevA.84.032309}.
We can prove that 
\begin{align}
\label{ineq:boson_density_main}
\frac{\tr \left [ \nb_i^s \tilde{\rho}(t) \right]}{\|O_X\|^2} \le c'_{1} e^{c_0\bar{q}}  |X|^3 (c_{1}s |X| )^{s}   e^{-\dist_{i,X}}  +c''_{1}e^{c_0\bar{q}}  (c_{1}s)^{s},
\end{align}
where $c_1$, $c_1'$, and $c_1''$ depend on the time as $e^{\orderof{t}}$. 
The above upper bound induces an exponential increase of the boson density with time; hence we cannot use it for arbitrarily large $t$.  
However, the key point of our proof-method is that we only need to treat the short-time evolution, where the coefficients $c_1$, $c_1'$, and $c_1''$ are $\orderof{1}$ constants.
By using Markov's inequality, we can ensure that the probability distribution of the boson number $\nb_i$ obeys 
\begin{align}
\label{upper_bound_prob_site_i_Z_0}
P_{i,\ge z_0}^{(t)}  \le 2c''_{1} e^{c_0\bar{q}} \|O_X\|^2 \br{\frac{\tilde{c}_1 \dist_{i,X} }{z_0  }}^{\tilde{c}'_1 \dist_{i,X}/\log (r)} 
\end{align}
under the condition $\dist_{i,X} \gtrsim \log(r)$,
where $P_{i,\ge z_0}^{(t)}$ is the probability that $z_0$ or more bosons are observed at the site $i$.  
(Recall that by definition $X\subseteq i[r]$.)
Finally, we remark that it is essential to the proof that the Hamiltonian be the form~\eqref{main_def:Ham}; if the Hamiltonian includes interactions such as $\nb_i \nb_j b_i b_j^\dagger$, 
the inequality~\eqref{ineq:boson_density_main} may break down even for small $t$.

 \begin{figure}[tt]
\centering
\includegraphics[clip, scale=0.3]{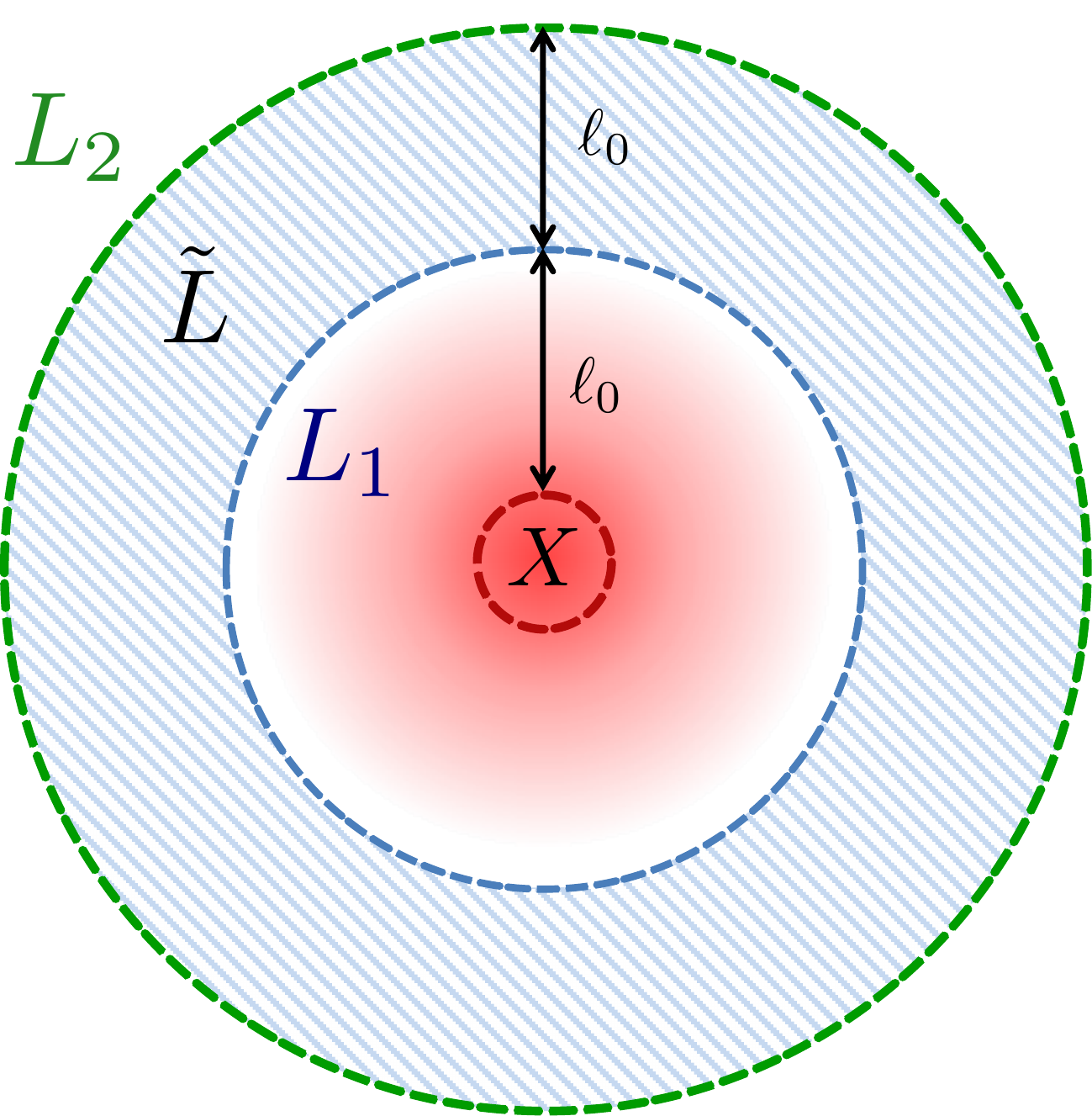}
\caption{
Schematic picture of the region where the boson number is truncated. 
In the region $L_1$ ($=X[\ell_0]$), we cannot restrict the boson number distribution.
On the other hand, as long as $\ell_0$ is sufficiently large, 
the boson number outside $L_1$ can be truncated up to a finite value $q$.
We perform the boson number truncation in the shaded region $\tilde{L} = L_2 \setminus L_1$ with $L_2=X[2\ell_0]$. 
By using an effective Hamiltonian $\tilde{H}[\tilde{L},q]$ as in Eq.~\eqref{main_def:tilde_Ham_effective}, we can approximate the 
exact dynamics by choosing the appropriate $q$ [i.e., $\eta \ell_0$; see the inequality~\eqref{main_ineq:prop:error_time_evolution_effective_Ham}].
}
\label{fig_Boson_truncation}
\end{figure}

In the second technique, we construct an effective Hamiltonian 
that has bounded local energy in a specific region and approximates the exact dynamics (see Sec.~S.VII. in Supplementary materials~\cite{Supplement_boson}). 
The inequality~\eqref{upper_bound_prob_site_i_Z_0} 
implies that the boson number $\nb_i$ is strongly suppressed when the site $i$ is sufficiently separated from the region $X$. 
Hence, we expect that, in the original Hamiltonian $H$, the maximum boson number at one site can be truncated during short-time evolution. 
We first define two regions $L_1:=X[\ell_0]$ and $L_2:=X[2\ell_0]$, where the length $\ell_0$ is appropriately chosen. 
We then consider the boson truncation in the region $\tilde{L}$ which is defined as (see Fig.~\ref{fig_Boson_truncation})
\begin{align}
\label{main_def:tilde_L}
\tilde{L} : = L_2 \setminus L_1. 
\end{align}

We now define $\bar{\Pi}_{\tilde{L},q}$ as the projection onto the eigenspace such that the boson number $\nb_i$ ($\forall i \in \tilde{L}$) is truncated up to $q$, 
i.e., $\| \nb_i \bar{\Pi}_{\tilde{L},q} \| \le q$. 
We then approximate the time-evolution operator $e^{-iHt}$ by using an effective Hamiltonian $\tilde{H}[\tilde{L},q]$, defined by 
 \begin{align}
 \label{main_def:tilde_Ham_effective}
&\tilde{H}[\tilde{L},q] :=\bar{\Pi}_{\tilde{L},q} H  \bar{\Pi}_{\tilde{L},q}  ,
\end{align}
with a bounded local energy in the region $\tilde{L}$. 
In general, the time evolution $O_X(t)$ cannot be approximated by $O_X(\tilde{H}[\tilde{L},q],t)$ at all, where we have used the notation~\eqref{notation_time_evolve_A}.  
However, we are only interested in the norm difference between $O_X(t)  \rho_0$ and $O_X(\tilde{H}[\tilde{L},q],t) \rho_0$. 
We can prove 
 \begin{align}
 \label{main_ineq:prop:error_time_evolution_effective_Ham}
&\left \| \left[ O_X(t) -O_X(\tilde{H}[\tilde{L},\eta\ell_0],t)\right] \rho_0 \right\|_1  \notag \\
&\le \frac{\|O_X\|}{2}  e^{c_0\bar{q}} e^{-2\ell_0/\log (r) }
\end{align}
for $q=\eta \ell_0$ and $\ell_0\ge C_0 \log^2 (r)$, 
where $\eta$ and $C_0$ are $\orderof{1}$ constants which are independent of $\bar{q}$, and $r$ has been defined by $X\subseteq i[r]$.  
From this upper bound, we can see that the error exponentially decreases with the number of the boson truncation. 
Thus, by using the Hamiltonian $\tilde{H}[\tilde{L},\eta\ell_0]$, the biggest obstacle, namely the unboundedness of the interaction norms, has been removed, at least in the region $\tilde{L}$.
However, outside this region, the norm is still unbounded. 
We thus need to consider how to derive the Lieb--Robinson bound for $e^{-i \tilde{H}[\tilde{L},\eta\ell_0] t}$ only from the finiteness of the Hamiltonian norm in the region $\tilde{L}$.

Our final task is to approximate the time evolution $O_X(\tilde{H}[\tilde{L},\eta\ell_0],t)$ by $U_{L_2}^\dagger O_X U_{L_2}$, 
where $U_{L_2}$ is an appropriate unitary operator supported on the subset $L_2$ (see Sec.~S.VIII. in Supplementary materials~\cite{Supplement_boson}).
By a careful calculation based on the standard approach to deriving the Lieb--Robinson bound, 
we can show that the approximation error obeys 
 \begin{align}
  \label{main_ineq:prop:short_time_Lieb--Robinson}
&\left \| O_X(\tilde{H}[\tilde{L},\eta\ell_0],t) - U_{L_2}^\dagger O_X U_{L_2}    \right\| 
 \le \frac{\|O_X\|}{2} e^{-2\ell_0/\log (r) } ,
\end{align} 
assuming $t\le \Delta t_0$ with $\Delta t_0$ an $\orderof{1}$ constant. 
Therefore, under the conditions $\ell_0\ge C_0\log^2 (r)$ and $t\le \Delta t_0$, we have the inequalities~\eqref{main_ineq:prop:error_time_evolution_effective_Ham} and \eqref{main_ineq:prop:short_time_Lieb--Robinson}, which together yield the desired inequality~\eqref{main_ineq:main_theorem_short_time_LR_time_dependent} since $e^{c_0\bar{q}}\ge 1$.

\sectionprl{Conclusion}
In this work, we have established the Lieb--Robinson bound~\eqref{main_main_theorem_short_time_LR_main_ineq}
with an almost-linear light cone $R\propto t\log^2(t)$ 
for arbitrary initial steady states under the condition~\eqref{main_condition_for_moment_generating}. 
Our bound leads to the clustering theorem~\eqref{clustering_gs_boson} for gapped ground states and the efficient simulation of the quench dynamics as in \eqref{time_complexity_quench}. 
Our result gives the first rigorous characterization of the light cone of interacting boson systems under experimentally realistic conditions.

%As a remark, the condition~\eqref{main_condition_for_moment_generating} is expected to hold in real experimental setups although it would be difficult to rigorously prove in general. 
%One of the candidates that satisfy the condition is the quantum Gibbs states; 
%as a trivial example, our theory can be applied to the infinite temperature state.  
%If we restrict a specific setup (e.g., there exists a force of repulsion between bosons), we would be able to prove the condition~\eqref{main_condition_for_moment_generating} for low-energy states by employing the techniques in~\cite{Arad_connecting_global_local_dist}.   

Nevertheless, this Lieb--Robinson bound might be further improved.
First, the asymptotic form $e^{-R/(t\log R)}$ in~\eqref{main_main_theorem_short_time_LR_main_ineq} could be changed to $e^{-R+vt}$, 
which would induce a strictly linear light cone for information propagation. 
Second, there remains the challenge to clarify the class of quantum states that rigorously satisfy the assumption~\eqref{main_condition_for_moment_generating}.
Third, regarding the time independence of $\rho_0$,  
we conjecture that an information wave-front of at least polynomial form (i.e., $R\propto t^{\zeta}$, $\zeta\ge1$) can be derived when $\rho_0$ is time-dependent. 
Although our current techniques cannot immediately accommodate these improvements,
we hope to develop a better Lieb--Robinson bound for interacting bosons in the future. 

{~}\\

\textit{Note added.}
For the readers information, we would like to refer to a subsequent study by Yin and Lucas~\cite{yin2021finite}, which proves the linear light cone for interacting boson systems in another specific setup.

\begin{acknowledgments}
The work of T. K. was supported by the RIKEN Center for AIP and JSPS KAKENHI (Grant No. 18K13475). 
TK gives thanks to God for his wisdom.
K.S. was supported by JSPS Grants-in-Aid for Scientific Research (JP16H02211 and JP19H05603).

\end{acknowledgments}

\bibliography{LR_boson}

\providecommand{\noopsort}[1]{}\providecommand{\singleletter}[1]{#1}%
\begin{thebibliography}{98}%
\makeatletter
\providecommand \@ifxundefined [1]{%
 \@ifx{#1\undefined}
}%
\providecommand \@ifnum [1]{%
 \ifnum #1\expandafter \@firstoftwo
 \else \expandafter \@secondoftwo
 \fi
}%
\providecommand \@ifx [1]{%
 \ifx #1\expandafter \@firstoftwo
 \else \expandafter \@secondoftwo
 \fi
}%
\providecommand \natexlab [1]{#1}%
\providecommand \enquote  [1]{``#1''}%
\providecommand \bibnamefont  [1]{#1}%
\providecommand \bibfnamefont [1]{#1}%
\providecommand \citenamefont [1]{#1}%
\providecommand \href@noop [0]{\@secondoftwo}%
\providecommand \href [0]{\begingroup \@sanitize@url \@href}%
\providecommand \@href[1]{\@@startlink{#1}\@@href}%
\providecommand \@@href[1]{\endgroup#1\@@endlink}%
\providecommand \@sanitize@url [0]{\catcode `\\12\catcode `\$12\catcode
  `\&12\catcode `\#12\catcode `\^12\catcode `\_12\catcode `\%12\relax}%
\providecommand \@@startlink[1]{}%
\providecommand \@@endlink[0]{}%
\providecommand \url  [0]{\begingroup\@sanitize@url \@url }%
\providecommand \@url [1]{\endgroup\@href {#1}{\urlprefix }}%
\providecommand \urlprefix  [0]{URL }%
\providecommand \Eprint [0]{\href }%
\providecommand \doibase [0]{https://doi.org/}%
\providecommand \selectlanguage [0]{\@gobble}%
\providecommand \bibinfo  [0]{\@secondoftwo}%
\providecommand \bibfield  [0]{\@secondoftwo}%
\providecommand \translation [1]{[#1]}%
\providecommand \BibitemOpen [0]{}%
\providecommand \bibitemStop [0]{}%
\providecommand \bibitemNoStop [0]{.\EOS\space}%
\providecommand \EOS [0]{\spacefactor3000\relax}%
\providecommand \BibitemShut  [1]{\csname bibitem#1\endcsname}%
\let\auto@bib@innerbib\@empty
%</preamble>
\bibitem [{\citenamefont {Lieb}\ and\ \citenamefont
  {Robinson}(1972)}]{ref:LR-bound72}%
  \BibitemOpen
  \bibfield  {author} {\bibinfo {author} {\bibfnamefont {E.~H.}\ \bibnamefont
  {Lieb}}\ and\ \bibinfo {author} {\bibfnamefont {D.~W.}\ \bibnamefont
  {Robinson}},\ }\bibfield  {title} {\bibinfo {title} {{\it The finite group
  velocity of quantum spin systems}},\ }\href
  {https://doi.org/10.1007/BF01645779} {\bibfield  {journal} {\bibinfo
  {journal} {Communications in Mathematical Physics}\ }\textbf {\bibinfo
  {volume} {28}},\ \bibinfo {pages} {251} (\bibinfo {year} {1972})}\BibitemShut
  {NoStop}%
\bibitem [{\citenamefont {Bravyi}\ \emph {et~al.}(2006)\citenamefont {Bravyi},
  \citenamefont {Hastings},\ and\ \citenamefont
  {Verstraete}}]{PhysRevLett.97.050401}%
  \BibitemOpen
  \bibfield  {author} {\bibinfo {author} {\bibfnamefont {S.}~\bibnamefont
  {Bravyi}}, \bibinfo {author} {\bibfnamefont {M.~B.}\ \bibnamefont
  {Hastings}},\ and\ \bibinfo {author} {\bibfnamefont {F.}~\bibnamefont
  {Verstraete}},\ }\bibfield  {title} {\bibinfo {title} {{\it Lieb-Robinson
  Bounds and the Generation of Correlations and Topological Quantum Order}},\
  }\href {https://doi.org/10.1103/PhysRevLett.97.050401} {\bibfield  {journal}
  {\bibinfo  {journal} {Phys. Rev. Lett.}\ }\textbf {\bibinfo {volume} {97}},\
  \bibinfo {pages} {050401} (\bibinfo {year} {2006})}\BibitemShut {NoStop}%
\bibitem [{\citenamefont {Nachtergaele}\ \emph {et~al.}(2019)\citenamefont
  {Nachtergaele}, \citenamefont {Sims},\ and\ \citenamefont
  {Young}}]{doi:10.1063/1.5095769}%
  \BibitemOpen
  \bibfield  {author} {\bibinfo {author} {\bibfnamefont {B.}~\bibnamefont
  {Nachtergaele}}, \bibinfo {author} {\bibfnamefont {R.}~\bibnamefont {Sims}},\
  and\ \bibinfo {author} {\bibfnamefont {A.}~\bibnamefont {Young}},\ }\bibfield
   {title} {\bibinfo {title} {{\it Quasi-locality bounds for quantum lattice
  systems. I. Lieb-Robinson bounds, quasi-local maps, and spectral flow
  automorphisms}},\ }\href {https://doi.org/10.1063/1.5095769} {\bibfield
  {journal} {\bibinfo  {journal} {Journal of Mathematical Physics}\ }\textbf
  {\bibinfo {volume} {60}},\ \bibinfo {pages} {061101} (\bibinfo {year}
  {2019})}\BibitemShut {NoStop}%
\bibitem [{\citenamefont
  {Hastings}(2004{\natexlab{a}})}]{ref:Hastings2004-Markov}%
  \BibitemOpen
  \bibfield  {author} {\bibinfo {author} {\bibfnamefont {M.~B.}\ \bibnamefont
  {Hastings}},\ }\bibfield  {title} {\bibinfo {title} {{\it Locality in Quantum
  and Markov Dynamics on Lattices and Networks}},\ }\href
  {https://doi.org/10.1103/PhysRevLett.93.140402} {\bibfield  {journal}
  {\bibinfo  {journal} {Phys. Rev. Lett.}\ }\textbf {\bibinfo {volume} {93}},\
  \bibinfo {pages} {140402} (\bibinfo {year} {2004}{\natexlab{a}})}\BibitemShut
  {NoStop}%
\bibitem [{\citenamefont {Hastings}\ and\ \citenamefont
  {Koma}(2006)}]{ref:Hastings2006-ExpDec}%
  \BibitemOpen
  \bibfield  {author} {\bibinfo {author} {\bibfnamefont {M.~B.}\ \bibnamefont
  {Hastings}}\ and\ \bibinfo {author} {\bibfnamefont {T.}~\bibnamefont
  {Koma}},\ }\bibfield  {title} {\bibinfo {title} {{\it Spectral Gap and
  Exponential Decay of Correlations}},\ }\href
  {https://doi.org/10.1007/s00220-006-0030-4} {\bibfield  {journal} {\bibinfo
  {journal} {Communications in Mathematical Physics}\ }\textbf {\bibinfo
  {volume} {265}},\ \bibinfo {pages} {781} (\bibinfo {year}
  {2006})}\BibitemShut {NoStop}%
\bibitem [{\citenamefont {Nachtergaele}\ and\ \citenamefont
  {Sims}(2006)}]{ref:Nachtergaele2006-LR}%
  \BibitemOpen
  \bibfield  {author} {\bibinfo {author} {\bibfnamefont {B.}~\bibnamefont
  {Nachtergaele}}\ and\ \bibinfo {author} {\bibfnamefont {R.}~\bibnamefont
  {Sims}},\ }\bibfield  {title} {\bibinfo {title} {{\it Lieb-Robinson Bounds
  and the Exponential Clustering Theorem}},\ }\href
  {https://doi.org/10.1007/s00220-006-1556-1} {\bibfield  {journal} {\bibinfo
  {journal} {Communications in Mathematical Physics}\ }\textbf {\bibinfo
  {volume} {265}},\ \bibinfo {pages} {119} (\bibinfo {year}
  {2006})}\BibitemShut {NoStop}%
\bibitem [{\citenamefont {Osborne}(2006)}]{PhysRevLett.97.157202}%
  \BibitemOpen
  \bibfield  {author} {\bibinfo {author} {\bibfnamefont {T.~J.}\ \bibnamefont
  {Osborne}},\ }\bibfield  {title} {\bibinfo {title} {{\it Efficient
  Approximation of the Dynamics of One-Dimensional Quantum Spin Systems}},\
  }\href {https://doi.org/10.1103/PhysRevLett.97.157202} {\bibfield  {journal}
  {\bibinfo  {journal} {Phys. Rev. Lett.}\ }\textbf {\bibinfo {volume} {97}},\
  \bibinfo {pages} {157202} (\bibinfo {year} {2006})}\BibitemShut {NoStop}%
\bibitem [{\citenamefont {Haah}\ \emph {et~al.}(2018)\citenamefont {Haah},
  \citenamefont {Hastings}, \citenamefont {Kothari},\ and\ \citenamefont
  {Low}}]{8555119}%
  \BibitemOpen
  \bibfield  {author} {\bibinfo {author} {\bibfnamefont {J.}~\bibnamefont
  {Haah}}, \bibinfo {author} {\bibfnamefont {M.}~\bibnamefont {Hastings}},
  \bibinfo {author} {\bibfnamefont {R.}~\bibnamefont {Kothari}},\ and\ \bibinfo
  {author} {\bibfnamefont {G.~H.}\ \bibnamefont {Low}},\ }\bibfield  {title}
  {\bibinfo {title} {{\it Quantum Algorithm for Simulating Real Time Evolution
  of Lattice Hamiltonians}},\ }in\ \href
  {https://doi.org/10.1109/FOCS.2018.00041} {\emph {\bibinfo {booktitle} {2018
  IEEE 59th Annual Symposium on Foundations of Computer Science (FOCS)}}}\
  (\bibinfo {year} {2018})\ pp.\ \bibinfo {pages} {350--360}\BibitemShut
  {NoStop}%
\bibitem [{\citenamefont {Tran}\ \emph {et~al.}(2019)\citenamefont {Tran},
  \citenamefont {Guo}, \citenamefont {Su}, \citenamefont {Garrison},
  \citenamefont {Eldredge}, \citenamefont {Foss-Feig}, \citenamefont {Childs},\
  and\ \citenamefont {Gorshkov}}]{PhysRevX.9.031006}%
  \BibitemOpen
  \bibfield  {author} {\bibinfo {author} {\bibfnamefont {M.~C.}\ \bibnamefont
  {Tran}}, \bibinfo {author} {\bibfnamefont {A.~Y.}\ \bibnamefont {Guo}},
  \bibinfo {author} {\bibfnamefont {Y.}~\bibnamefont {Su}}, \bibinfo {author}
  {\bibfnamefont {J.~R.}\ \bibnamefont {Garrison}}, \bibinfo {author}
  {\bibfnamefont {Z.}~\bibnamefont {Eldredge}}, \bibinfo {author}
  {\bibfnamefont {M.}~\bibnamefont {Foss-Feig}}, \bibinfo {author}
  {\bibfnamefont {A.~M.}\ \bibnamefont {Childs}},\ and\ \bibinfo {author}
  {\bibfnamefont {A.~V.}\ \bibnamefont {Gorshkov}},\ }\bibfield  {title}
  {\bibinfo {title} {{\it Locality and Digital Quantum Simulation of Power-Law
  Interactions}},\ }\href {https://doi.org/10.1103/PhysRevX.9.031006}
  {\bibfield  {journal} {\bibinfo  {journal} {Phys. Rev. X}\ }\textbf {\bibinfo
  {volume} {9}},\ \bibinfo {pages} {031006} (\bibinfo {year}
  {2019})}\BibitemShut {NoStop}%
\bibitem [{\citenamefont {Hastings}(2004{\natexlab{b}})}]{PhysRevB.69.104431}%
  \BibitemOpen
  \bibfield  {author} {\bibinfo {author} {\bibfnamefont {M.~B.}\ \bibnamefont
  {Hastings}},\ }\bibfield  {title} {\bibinfo {title} {{\it Lieb-Schultz-Mattis
  in higher dimensions}},\ }\href {https://doi.org/10.1103/PhysRevB.69.104431}
  {\bibfield  {journal} {\bibinfo  {journal} {Phys. Rev. B}\ }\textbf {\bibinfo
  {volume} {69}},\ \bibinfo {pages} {104431} (\bibinfo {year}
  {2004}{\natexlab{b}})}\BibitemShut {NoStop}%
\bibitem [{\citenamefont {Nachtergaele}\ and\ \citenamefont
  {Sims}(2007)}]{Nachtergaele2007}%
  \BibitemOpen
  \bibfield  {author} {\bibinfo {author} {\bibfnamefont {B.}~\bibnamefont
  {Nachtergaele}}\ and\ \bibinfo {author} {\bibfnamefont {R.}~\bibnamefont
  {Sims}},\ }\bibfield  {title} {\bibinfo {title} {{\it A Multi-Dimensional
  Lieb-Schultz-Mattis Theorem}},\ }\href
  {https://doi.org/10.1007/s00220-007-0342-z} {\bibfield  {journal} {\bibinfo
  {journal} {Communications in Mathematical Physics}\ }\textbf {\bibinfo
  {volume} {276}},\ \bibinfo {pages} {437} (\bibinfo {year}
  {2007})}\BibitemShut {NoStop}%
\bibitem [{\citenamefont {Hastings}\ and\ \citenamefont
  {Wen}(2005)}]{PhysRevB.72.045141}%
  \BibitemOpen
  \bibfield  {author} {\bibinfo {author} {\bibfnamefont {M.~B.}\ \bibnamefont
  {Hastings}}\ and\ \bibinfo {author} {\bibfnamefont {X.-G.}\ \bibnamefont
  {Wen}},\ }\bibfield  {title} {\bibinfo {title} {{\it Quasiadiabatic
  continuation of quantum states: The stability of topological ground-state
  degeneracy and emergent gauge invariance}},\ }\href
  {https://doi.org/10.1103/PhysRevB.72.045141} {\bibfield  {journal} {\bibinfo
  {journal} {Phys. Rev. B}\ }\textbf {\bibinfo {volume} {72}},\ \bibinfo
  {pages} {045141} (\bibinfo {year} {2005})}\BibitemShut {NoStop}%
\bibitem [{\citenamefont {Hastings}(2007{\natexlab{a}})}]{ref:Hastings-AL07}%
  \BibitemOpen
  \bibfield  {author} {\bibinfo {author} {\bibfnamefont {M.~B.}\ \bibnamefont
  {Hastings}},\ }\bibfield  {title} {\bibinfo {title} {{\it An area law for
  one-dimensional quantum systems}},\ }\href
  {http://stacks.iop.org/1742-5468/2007/i=08/a=P08024} {\bibfield  {journal}
  {\bibinfo  {journal} {Journal of Statistical Mechanics: Theory and
  Experiment}\ }\textbf {\bibinfo {volume} {2007}},\ \bibinfo {pages} {P08024}
  (\bibinfo {year} {2007}{\natexlab{a}})},\ \Eprint
  {https://arxiv.org/abs/arXiv:0705.2024} {arXiv:0705.2024} \BibitemShut
  {NoStop}%
\bibitem [{\citenamefont {Bravyi}\ \emph {et~al.}(2010)\citenamefont {Bravyi},
  \citenamefont {Hastings},\ and\ \citenamefont
  {Michalakis}}]{bravyi2010topological}%
  \BibitemOpen
  \bibfield  {author} {\bibinfo {author} {\bibfnamefont {S.}~\bibnamefont
  {Bravyi}}, \bibinfo {author} {\bibfnamefont {M.~B.}\ \bibnamefont
  {Hastings}},\ and\ \bibinfo {author} {\bibfnamefont {S.}~\bibnamefont
  {Michalakis}},\ }\bibfield  {title} {\bibinfo {title} {{\it Topological
  quantum order: Stability under local perturbations}},\ }\href
  {https://doi.org/http://dx.doi.org/10.1063/1.3490195} {\bibfield  {journal}
  {\bibinfo  {journal} {Journal of Mathematical Physics}\ }\textbf {\bibinfo
  {volume} {51}},\ \bibinfo {eid} {093512} (\bibinfo {year}
  {2010})}\BibitemShut {NoStop}%
\bibitem [{\citenamefont {Haegeman}\ \emph {et~al.}(2013)\citenamefont
  {Haegeman}, \citenamefont {Michalakis}, \citenamefont {Nachtergaele},
  \citenamefont {Osborne}, \citenamefont {Schuch},\ and\ \citenamefont
  {Verstraete}}]{PhysRevLett.111.080401}%
  \BibitemOpen
  \bibfield  {author} {\bibinfo {author} {\bibfnamefont {J.}~\bibnamefont
  {Haegeman}}, \bibinfo {author} {\bibfnamefont {S.}~\bibnamefont
  {Michalakis}}, \bibinfo {author} {\bibfnamefont {B.}~\bibnamefont
  {Nachtergaele}}, \bibinfo {author} {\bibfnamefont {T.~J.}\ \bibnamefont
  {Osborne}}, \bibinfo {author} {\bibfnamefont {N.}~\bibnamefont {Schuch}},\
  and\ \bibinfo {author} {\bibfnamefont {F.}~\bibnamefont {Verstraete}},\
  }\bibfield  {title} {\bibinfo {title} {{\it Elementary Excitations in Gapped
  Quantum Spin Systems}},\ }\href
  {https://doi.org/10.1103/PhysRevLett.111.080401} {\bibfield  {journal}
  {\bibinfo  {journal} {Phys. Rev. Lett.}\ }\textbf {\bibinfo {volume} {111}},\
  \bibinfo {pages} {080401} (\bibinfo {year} {2013})}\BibitemShut {NoStop}%
\bibitem [{\citenamefont {Hastings}\ and\ \citenamefont
  {Michalakis}(2014)}]{hastings2015quantization}%
  \BibitemOpen
  \bibfield  {author} {\bibinfo {author} {\bibfnamefont {M.~B.}\ \bibnamefont
  {Hastings}}\ and\ \bibinfo {author} {\bibfnamefont {S.}~\bibnamefont
  {Michalakis}},\ }\bibfield  {title} {\bibinfo {title} {{\it Quantization of
  Hall Conductance for Interacting Electrons on a Torus}},\ }\href
  {https://doi.org/10.1007/s00220-014-2167-x} {\bibfield  {journal} {\bibinfo
  {journal} {Communications in Mathematical Physics}\ }\textbf {\bibinfo
  {volume} {334}},\ \bibinfo {pages} {433} (\bibinfo {year}
  {2014})}\BibitemShut {NoStop}%
\bibitem [{\citenamefont {M{\"u}ller}\ \emph {et~al.}(2015)\citenamefont
  {M{\"u}ller}, \citenamefont {Adlam}, \citenamefont {Masanes},\ and\
  \citenamefont {Wiebe}}]{ref:Mueller2013-Thermal}%
  \BibitemOpen
  \bibfield  {author} {\bibinfo {author} {\bibfnamefont {M.~P.}\ \bibnamefont
  {M{\"u}ller}}, \bibinfo {author} {\bibfnamefont {E.}~\bibnamefont {Adlam}},
  \bibinfo {author} {\bibfnamefont {L.}~\bibnamefont {Masanes}},\ and\ \bibinfo
  {author} {\bibfnamefont {N.}~\bibnamefont {Wiebe}},\ }\bibfield  {title}
  {\bibinfo {title} {{\it Thermalization and Canonical Typicality in
  Translation-Invariant Quantum Lattice Systems}},\ }\href
  {https://doi.org/10.1007/s00220-015-2473-y} {\bibfield  {journal} {\bibinfo
  {journal} {Communications in Mathematical Physics}\ }\textbf {\bibinfo
  {volume} {340}},\ \bibinfo {pages} {499} (\bibinfo {year}
  {2015})}\BibitemShut {NoStop}%
\bibitem [{\citenamefont {Damanik}\ \emph {et~al.}(2014)\citenamefont
  {Damanik}, \citenamefont {Lemm}, \citenamefont {Lukic},\ and\ \citenamefont
  {Yessen}}]{PhysRevLett.113.127202}%
  \BibitemOpen
  \bibfield  {author} {\bibinfo {author} {\bibfnamefont {D.}~\bibnamefont
  {Damanik}}, \bibinfo {author} {\bibfnamefont {M.}~\bibnamefont {Lemm}},
  \bibinfo {author} {\bibfnamefont {M.}~\bibnamefont {Lukic}},\ and\ \bibinfo
  {author} {\bibfnamefont {W.}~\bibnamefont {Yessen}},\ }\bibfield  {title}
  {\bibinfo {title} {New anomalous lieb-robinson bounds in quasiperiodic $xy$
  chains},\ }\href {https://doi.org/10.1103/PhysRevLett.113.127202} {\bibfield
  {journal} {\bibinfo  {journal} {Phys. Rev. Lett.}\ }\textbf {\bibinfo
  {volume} {113}},\ \bibinfo {pages} {127202} (\bibinfo {year}
  {2014})}\BibitemShut {NoStop}%
\bibitem [{\citenamefont {Iyoda}\ \emph {et~al.}(2017)\citenamefont {Iyoda},
  \citenamefont {Kaneko},\ and\ \citenamefont
  {Sagawa}}]{PhysRevLett.119.100601}%
  \BibitemOpen
  \bibfield  {author} {\bibinfo {author} {\bibfnamefont {E.}~\bibnamefont
  {Iyoda}}, \bibinfo {author} {\bibfnamefont {K.}~\bibnamefont {Kaneko}},\ and\
  \bibinfo {author} {\bibfnamefont {T.}~\bibnamefont {Sagawa}},\ }\bibfield
  {title} {\bibinfo {title} {{\it Fluctuation Theorem for Many-Body Pure
  Quantum States}},\ }\href {https://doi.org/10.1103/PhysRevLett.119.100601}
  {\bibfield  {journal} {\bibinfo  {journal} {Phys. Rev. Lett.}\ }\textbf
  {\bibinfo {volume} {119}},\ \bibinfo {pages} {100601} (\bibinfo {year}
  {2017})}\BibitemShut {NoStop}%
\bibitem [{\citenamefont {Kuwahara}\ \emph {et~al.}(2016)\citenamefont
  {Kuwahara}, \citenamefont {Mori},\ and\ \citenamefont
  {Saito}}]{KUWAHARA201696}%
  \BibitemOpen
  \bibfield  {author} {\bibinfo {author} {\bibfnamefont {T.}~\bibnamefont
  {Kuwahara}}, \bibinfo {author} {\bibfnamefont {T.}~\bibnamefont {Mori}},\
  and\ \bibinfo {author} {\bibfnamefont {K.}~\bibnamefont {Saito}},\ }\bibfield
   {title} {\bibinfo {title} {{\it Floquet-Magnus theory and generic transient
  dynamics in periodically driven many-body quantum systems}},\ }\href
  {https://doi.org/10.1016/j.aop.2016.01.012} {\bibfield  {journal} {\bibinfo
  {journal} {Annals of Physics}\ }\textbf {\bibinfo {volume} {367}},\ \bibinfo
  {pages} {96 } (\bibinfo {year} {2016})}\BibitemShut {NoStop}%
\bibitem [{\citenamefont {Abanin}\ \emph {et~al.}(2017)\citenamefont {Abanin},
  \citenamefont {De~Roeck}, \citenamefont {Ho},\ and\ \citenamefont
  {Huveneers}}]{PhysRevB.95.014112}%
  \BibitemOpen
  \bibfield  {author} {\bibinfo {author} {\bibfnamefont {D.~A.}\ \bibnamefont
  {Abanin}}, \bibinfo {author} {\bibfnamefont {W.}~\bibnamefont {De~Roeck}},
  \bibinfo {author} {\bibfnamefont {W.~W.}\ \bibnamefont {Ho}},\ and\ \bibinfo
  {author} {\bibfnamefont {F.~m.~c.}\ \bibnamefont {Huveneers}},\ }\bibfield
  {title} {\bibinfo {title} {{\it Effective Hamiltonians, prethermalization,
  and slow energy absorption in periodically driven many-body systems}},\
  }\href {https://doi.org/10.1103/PhysRevB.95.014112} {\bibfield  {journal}
  {\bibinfo  {journal} {Phys. Rev. B}\ }\textbf {\bibinfo {volume} {95}},\
  \bibinfo {pages} {014112} (\bibinfo {year} {2017})}\BibitemShut {NoStop}%
\bibitem [{\citenamefont {Bachmann}\ \emph {et~al.}(2017)\citenamefont
  {Bachmann}, \citenamefont {De~Roeck},\ and\ \citenamefont
  {Fraas}}]{PhysRevLett.119.060201}%
  \BibitemOpen
  \bibfield  {author} {\bibinfo {author} {\bibfnamefont {S.}~\bibnamefont
  {Bachmann}}, \bibinfo {author} {\bibfnamefont {W.}~\bibnamefont {De~Roeck}},\
  and\ \bibinfo {author} {\bibfnamefont {M.}~\bibnamefont {Fraas}},\ }\bibfield
   {title} {\bibinfo {title} {{\it Adiabatic Theorem for Quantum Spin
  Systems}},\ }\href {https://doi.org/10.1103/PhysRevLett.119.060201}
  {\bibfield  {journal} {\bibinfo  {journal} {Phys. Rev. Lett.}\ }\textbf
  {\bibinfo {volume} {119}},\ \bibinfo {pages} {060201} (\bibinfo {year}
  {2017})}\BibitemShut {NoStop}%
\bibitem [{\citenamefont {Kuwahara}\ \emph {et~al.}(2021)\citenamefont
  {Kuwahara}, \citenamefont {Alhambra},\ and\ \citenamefont
  {Anshu}}]{kuwahara2020improved}%
  \BibitemOpen
  \bibfield  {author} {\bibinfo {author} {\bibfnamefont {T.}~\bibnamefont
  {Kuwahara}}, \bibinfo {author} {\bibfnamefont {A.~M.}\ \bibnamefont
  {Alhambra}},\ and\ \bibinfo {author} {\bibfnamefont {A.}~\bibnamefont
  {Anshu}},\ }\bibfield  {title} {\bibinfo {title} {{\it Improved Thermal Area
  Law and Quasilinear Time Algorithm for Quantum Gibbs States}},\ }\href
  {https://doi.org/10.1103/PhysRevX.11.011047} {\bibfield  {journal} {\bibinfo
  {journal} {Phys. Rev. X}\ }\textbf {\bibinfo {volume} {11}},\ \bibinfo
  {pages} {011047} (\bibinfo {year} {2021})}\BibitemShut {NoStop}%
\bibitem [{\citenamefont {Maldacena}\ \emph {et~al.}(2016)\citenamefont
  {Maldacena}, \citenamefont {Shenker},\ and\ \citenamefont
  {Stanford}}]{Maldacena2016}%
  \BibitemOpen
  \bibfield  {author} {\bibinfo {author} {\bibfnamefont {J.}~\bibnamefont
  {Maldacena}}, \bibinfo {author} {\bibfnamefont {S.~H.}\ \bibnamefont
  {Shenker}},\ and\ \bibinfo {author} {\bibfnamefont {D.}~\bibnamefont
  {Stanford}},\ }\bibfield  {title} {\bibinfo {title} {{\it A bound on
  chaos}},\ }\href {https://doi.org/10.1007/JHEP08(2016)106} {\bibfield
  {journal} {\bibinfo  {journal} {Journal of High Energy Physics}\ }\textbf
  {\bibinfo {volume} {2016}},\ \bibinfo {pages} {106} (\bibinfo {year}
  {2016})}\BibitemShut {NoStop}%
\bibitem [{\citenamefont {Roberts}\ and\ \citenamefont
  {Swingle}(2016)}]{PhysRevLett.117.091602}%
  \BibitemOpen
  \bibfield  {author} {\bibinfo {author} {\bibfnamefont {D.~A.}\ \bibnamefont
  {Roberts}}\ and\ \bibinfo {author} {\bibfnamefont {B.}~\bibnamefont
  {Swingle}},\ }\bibfield  {title} {\bibinfo {title} {{\it Lieb-Robinson Bound
  and the Butterfly Effect in Quantum Field Theories}},\ }\href
  {https://doi.org/10.1103/PhysRevLett.117.091602} {\bibfield  {journal}
  {\bibinfo  {journal} {Phys. Rev. Lett.}\ }\textbf {\bibinfo {volume} {117}},\
  \bibinfo {pages} {091602} (\bibinfo {year} {2016})}\BibitemShut {NoStop}%
\bibitem [{\citenamefont {Joshi}\ \emph {et~al.}(2020)\citenamefont {Joshi},
  \citenamefont {Elben}, \citenamefont {Vermersch}, \citenamefont {Brydges},
  \citenamefont {Maier}, \citenamefont {Zoller}, \citenamefont {Blatt},\ and\
  \citenamefont {Roos}}]{PhysRevLett.124.240505}%
  \BibitemOpen
  \bibfield  {author} {\bibinfo {author} {\bibfnamefont {M.~K.}\ \bibnamefont
  {Joshi}}, \bibinfo {author} {\bibfnamefont {A.}~\bibnamefont {Elben}},
  \bibinfo {author} {\bibfnamefont {B.}~\bibnamefont {Vermersch}}, \bibinfo
  {author} {\bibfnamefont {T.}~\bibnamefont {Brydges}}, \bibinfo {author}
  {\bibfnamefont {C.}~\bibnamefont {Maier}}, \bibinfo {author} {\bibfnamefont
  {P.}~\bibnamefont {Zoller}}, \bibinfo {author} {\bibfnamefont
  {R.}~\bibnamefont {Blatt}},\ and\ \bibinfo {author} {\bibfnamefont {C.~F.}\
  \bibnamefont {Roos}},\ }\bibfield  {title} {\bibinfo {title} {{\it Quantum
  Information Scrambling in a Trapped-Ion Quantum Simulator with Tunable Range
  Interactions}},\ }\href {https://doi.org/10.1103/PhysRevLett.124.240505}
  {\bibfield  {journal} {\bibinfo  {journal} {Phys. Rev. Lett.}\ }\textbf
  {\bibinfo {volume} {124}},\ \bibinfo {pages} {240505} (\bibinfo {year}
  {2020})}\BibitemShut {NoStop}%
\bibitem [{\citenamefont {Zhou}\ \emph {et~al.}(2020)\citenamefont {Zhou},
  \citenamefont {Xu}, \citenamefont {Chen}, \citenamefont {Guo},\ and\
  \citenamefont {Swingle}}]{PhysRevLett.124.180601}%
  \BibitemOpen
  \bibfield  {author} {\bibinfo {author} {\bibfnamefont {T.}~\bibnamefont
  {Zhou}}, \bibinfo {author} {\bibfnamefont {S.}~\bibnamefont {Xu}}, \bibinfo
  {author} {\bibfnamefont {X.}~\bibnamefont {Chen}}, \bibinfo {author}
  {\bibfnamefont {A.}~\bibnamefont {Guo}},\ and\ \bibinfo {author}
  {\bibfnamefont {B.}~\bibnamefont {Swingle}},\ }\bibfield  {title} {\bibinfo
  {title} {{\it Operator L\'evy Flight: Light Cones in Chaotic Long-Range
  Interacting Systems}},\ }\href
  {https://doi.org/10.1103/PhysRevLett.124.180601} {\bibfield  {journal}
  {\bibinfo  {journal} {Phys. Rev. Lett.}\ }\textbf {\bibinfo {volume} {124}},\
  \bibinfo {pages} {180601} (\bibinfo {year} {2020})}\BibitemShut {NoStop}%
\bibitem [{\citenamefont {Chen}\ and\ \citenamefont
  {Lucas}(2019)}]{chen2019finite}%
  \BibitemOpen
  \bibfield  {author} {\bibinfo {author} {\bibfnamefont {C.-F.}\ \bibnamefont
  {Chen}}\ and\ \bibinfo {author} {\bibfnamefont {A.}~\bibnamefont {Lucas}},\
  }\bibfield  {title} {\bibinfo {title} {{\it Finite Speed of Quantum
  Scrambling with Long Range Interactions}},\ }\href
  {https://doi.org/10.1103/PhysRevLett.123.250605} {\bibfield  {journal}
  {\bibinfo  {journal} {Phys. Rev. Lett.}\ }\textbf {\bibinfo {volume} {123}},\
  \bibinfo {pages} {250605} (\bibinfo {year} {2019})}\BibitemShut {NoStop}%
\bibitem [{\citenamefont {Kuwahara}\ and\ \citenamefont
  {Saito}(2021)}]{PhysRevLett.126.030604}%
  \BibitemOpen
  \bibfield  {author} {\bibinfo {author} {\bibfnamefont {T.}~\bibnamefont
  {Kuwahara}}\ and\ \bibinfo {author} {\bibfnamefont {K.}~\bibnamefont
  {Saito}},\ }\bibfield  {title} {\bibinfo {title} {{\it Absence of Fast
  Scrambling in Thermodynamically Stable Long-Range Interacting Systems}},\
  }\href {https://doi.org/10.1103/PhysRevLett.126.030604} {\bibfield  {journal}
  {\bibinfo  {journal} {Phys. Rev. Lett.}\ }\textbf {\bibinfo {volume} {126}},\
  \bibinfo {pages} {030604} (\bibinfo {year} {2021})}\BibitemShut {NoStop}%
\bibitem [{\citenamefont {Chen}(2021)}]{chen2021concentration}%
  \BibitemOpen
  \bibfield  {author} {\bibinfo {author} {\bibfnamefont {C.-F.}\ \bibnamefont
  {Chen}},\ }\href@noop {} {\bibinfo {title} {Concentration of otoc and
  lieb-robinson velocity in random hamiltonians}} (\bibinfo {year} {2021}),\
  \Eprint {https://arxiv.org/abs/2103.09186} {arXiv:2103.09186 [quant-ph]}
  \BibitemShut {NoStop}%
\bibitem [{\citenamefont {Hastings}(2007{\natexlab{b}})}]{PhysRevB.76.201102}%
  \BibitemOpen
  \bibfield  {author} {\bibinfo {author} {\bibfnamefont {M.~B.}\ \bibnamefont
  {Hastings}},\ }\bibfield  {title} {\bibinfo {title} {{\it Quantum belief
  propagation: An algorithm for thermal quantum systems}},\ }\href
  {https://doi.org/10.1103/PhysRevB.76.201102} {\bibfield  {journal} {\bibinfo
  {journal} {Phys. Rev. B}\ }\textbf {\bibinfo {volume} {76}},\ \bibinfo
  {pages} {201102} (\bibinfo {year} {2007}{\natexlab{b}})}\BibitemShut
  {NoStop}%
\bibitem [{\citenamefont {Eisert}\ and\ \citenamefont
  {Osborne}(2006)}]{PhysRevLett.97.150404}%
  \BibitemOpen
  \bibfield  {author} {\bibinfo {author} {\bibfnamefont {J.}~\bibnamefont
  {Eisert}}\ and\ \bibinfo {author} {\bibfnamefont {T.~J.}\ \bibnamefont
  {Osborne}},\ }\bibfield  {title} {\bibinfo {title} {{\it General Entanglement
  Scaling Laws from Time Evolution}},\ }\href
  {https://doi.org/10.1103/PhysRevLett.97.150404} {\bibfield  {journal}
  {\bibinfo  {journal} {Phys. Rev. Lett.}\ }\textbf {\bibinfo {volume} {97}},\
  \bibinfo {pages} {150404} (\bibinfo {year} {2006})}\BibitemShut {NoStop}%
\bibitem [{\citenamefont {Ge}\ \emph {et~al.}(2016)\citenamefont {Ge},
  \citenamefont {Moln\'ar},\ and\ \citenamefont
  {Cirac}}]{PhysRevLett.116.080503}%
  \BibitemOpen
  \bibfield  {author} {\bibinfo {author} {\bibfnamefont {Y.}~\bibnamefont
  {Ge}}, \bibinfo {author} {\bibfnamefont {A.}~\bibnamefont {Moln\'ar}},\ and\
  \bibinfo {author} {\bibfnamefont {J.~I.}\ \bibnamefont {Cirac}},\ }\bibfield
  {title} {\bibinfo {title} {{\it Rapid Adiabatic Preparation of Injective
  Projected Entangled Pair States and Gibbs States}},\ }\href
  {https://doi.org/10.1103/PhysRevLett.116.080503} {\bibfield  {journal}
  {\bibinfo  {journal} {Phys. Rev. Lett.}\ }\textbf {\bibinfo {volume} {116}},\
  \bibinfo {pages} {080503} (\bibinfo {year} {2016})}\BibitemShut {NoStop}%
\bibitem [{\citenamefont {Deshpande}\ \emph {et~al.}(2018)\citenamefont
  {Deshpande}, \citenamefont {Fefferman}, \citenamefont {Tran}, \citenamefont
  {Foss-Feig},\ and\ \citenamefont {Gorshkov}}]{PhysRevLett.121.030501}%
  \BibitemOpen
  \bibfield  {author} {\bibinfo {author} {\bibfnamefont {A.}~\bibnamefont
  {Deshpande}}, \bibinfo {author} {\bibfnamefont {B.}~\bibnamefont
  {Fefferman}}, \bibinfo {author} {\bibfnamefont {M.~C.}\ \bibnamefont {Tran}},
  \bibinfo {author} {\bibfnamefont {M.}~\bibnamefont {Foss-Feig}},\ and\
  \bibinfo {author} {\bibfnamefont {A.~V.}\ \bibnamefont {Gorshkov}},\
  }\bibfield  {title} {\bibinfo {title} {{\it Dynamical Phase Transitions in
  Sampling Complexity}},\ }\href
  {https://doi.org/10.1103/PhysRevLett.121.030501} {\bibfield  {journal}
  {\bibinfo  {journal} {Phys. Rev. Lett.}\ }\textbf {\bibinfo {volume} {121}},\
  \bibinfo {pages} {030501} (\bibinfo {year} {2018})}\BibitemShut {NoStop}%
\bibitem [{\citenamefont {Childs}\ \emph {et~al.}(2021)\citenamefont {Childs},
  \citenamefont {Su}, \citenamefont {Tran}, \citenamefont {Wiebe},\ and\
  \citenamefont {Zhu}}]{PhysRevX.11.011020}%
  \BibitemOpen
  \bibfield  {author} {\bibinfo {author} {\bibfnamefont {A.~M.}\ \bibnamefont
  {Childs}}, \bibinfo {author} {\bibfnamefont {Y.}~\bibnamefont {Su}}, \bibinfo
  {author} {\bibfnamefont {M.~C.}\ \bibnamefont {Tran}}, \bibinfo {author}
  {\bibfnamefont {N.}~\bibnamefont {Wiebe}},\ and\ \bibinfo {author}
  {\bibfnamefont {S.}~\bibnamefont {Zhu}},\ }\bibfield  {title} {\bibinfo
  {title} {{\it Theory of Trotter Error with Commutator Scaling}},\ }\href
  {https://doi.org/10.1103/PhysRevX.11.011020} {\bibfield  {journal} {\bibinfo
  {journal} {Phys. Rev. X}\ }\textbf {\bibinfo {volume} {11}},\ \bibinfo
  {pages} {011020} (\bibinfo {year} {2021})}\BibitemShut {NoStop}%
\bibitem [{\citenamefont {Nachtergaele}\ \emph {et~al.}(2006)\citenamefont
  {Nachtergaele}, \citenamefont {Ogata},\ and\ \citenamefont
  {Sims}}]{Nachtergaele2006}%
  \BibitemOpen
  \bibfield  {author} {\bibinfo {author} {\bibfnamefont {B.}~\bibnamefont
  {Nachtergaele}}, \bibinfo {author} {\bibfnamefont {Y.}~\bibnamefont
  {Ogata}},\ and\ \bibinfo {author} {\bibfnamefont {R.}~\bibnamefont {Sims}},\
  }\bibfield  {title} {\bibinfo {title} {{\it Propagation of Correlations in
  Quantum Lattice Systems}},\ }\href
  {https://doi.org/10.1007/s10955-006-9143-6} {\bibfield  {journal} {\bibinfo
  {journal} {Journal of Statistical Physics}\ }\textbf {\bibinfo {volume}
  {124}},\ \bibinfo {pages} {1} (\bibinfo {year} {2006})}\BibitemShut {NoStop}%
\bibitem [{\citenamefont {Nachtergaele}\ and\ \citenamefont
  {Sims}(2010)}]{nachtergaele2010lieb}%
  \BibitemOpen
  \bibfield  {author} {\bibinfo {author} {\bibfnamefont {B.}~\bibnamefont
  {Nachtergaele}}\ and\ \bibinfo {author} {\bibfnamefont {R.}~\bibnamefont
  {Sims}},\ }\bibfield  {title} {\bibinfo {title} {{\it Lieb-Robinson bounds in
  quantum many-body physics}},\ }\href@noop {} {\bibfield  {journal} {\bibinfo
  {journal} {Contemp. Math}\ }\textbf {\bibinfo {volume} {529}},\ \bibinfo
  {pages} {141} (\bibinfo {year} {2010})},\ \Eprint
  {https://arxiv.org/abs/arXiv:1004.2086} {arXiv:1004.2086} \BibitemShut
  {NoStop}%
\bibitem [{\citenamefont {Eisert}\ \emph {et~al.}(2013)\citenamefont {Eisert},
  \citenamefont {van~den Worm}, \citenamefont {Manmana},\ and\ \citenamefont
  {Kastner}}]{PhysRevLett.111.260401}%
  \BibitemOpen
  \bibfield  {author} {\bibinfo {author} {\bibfnamefont {J.}~\bibnamefont
  {Eisert}}, \bibinfo {author} {\bibfnamefont {M.}~\bibnamefont {van~den
  Worm}}, \bibinfo {author} {\bibfnamefont {S.~R.}\ \bibnamefont {Manmana}},\
  and\ \bibinfo {author} {\bibfnamefont {M.}~\bibnamefont {Kastner}},\
  }\bibfield  {title} {\bibinfo {title} {{\it Breakdown of Quasilocality in
  Long-Range Quantum Lattice Models}},\ }\href
  {https://doi.org/10.1103/PhysRevLett.111.260401} {\bibfield  {journal}
  {\bibinfo  {journal} {Phys. Rev. Lett.}\ }\textbf {\bibinfo {volume} {111}},\
  \bibinfo {pages} {260401} (\bibinfo {year} {2013})}\BibitemShut {NoStop}%
\bibitem [{\citenamefont {Foss-Feig}\ \emph {et~al.}(2015)\citenamefont
  {Foss-Feig}, \citenamefont {Gong}, \citenamefont {Clark},\ and\ \citenamefont
  {Gorshkov}}]{PhysRevLett.114.157201}%
  \BibitemOpen
  \bibfield  {author} {\bibinfo {author} {\bibfnamefont {M.}~\bibnamefont
  {Foss-Feig}}, \bibinfo {author} {\bibfnamefont {Z.-X.}\ \bibnamefont {Gong}},
  \bibinfo {author} {\bibfnamefont {C.~W.}\ \bibnamefont {Clark}},\ and\
  \bibinfo {author} {\bibfnamefont {A.~V.}\ \bibnamefont {Gorshkov}},\
  }\bibfield  {title} {\bibinfo {title} {{\it Nearly Linear Light Cones in
  Long-Range Interacting Quantum Systems}},\ }\href
  {https://doi.org/10.1103/PhysRevLett.114.157201} {\bibfield  {journal}
  {\bibinfo  {journal} {Phys. Rev. Lett.}\ }\textbf {\bibinfo {volume} {114}},\
  \bibinfo {pages} {157201} (\bibinfo {year} {2015})}\BibitemShut {NoStop}%
\bibitem [{\citenamefont {Tran}\ \emph {et~al.}(2020)\citenamefont {Tran},
  \citenamefont {Chen}, \citenamefont {Ehrenberg}, \citenamefont {Guo},
  \citenamefont {Deshpande}, \citenamefont {Hong}, \citenamefont {Gong},
  \citenamefont {Gorshkov},\ and\ \citenamefont {Lucas}}]{PhysRevX.10.031009}%
  \BibitemOpen
  \bibfield  {author} {\bibinfo {author} {\bibfnamefont {M.~C.}\ \bibnamefont
  {Tran}}, \bibinfo {author} {\bibfnamefont {C.-F.}\ \bibnamefont {Chen}},
  \bibinfo {author} {\bibfnamefont {A.}~\bibnamefont {Ehrenberg}}, \bibinfo
  {author} {\bibfnamefont {A.~Y.}\ \bibnamefont {Guo}}, \bibinfo {author}
  {\bibfnamefont {A.}~\bibnamefont {Deshpande}}, \bibinfo {author}
  {\bibfnamefont {Y.}~\bibnamefont {Hong}}, \bibinfo {author} {\bibfnamefont
  {Z.-X.}\ \bibnamefont {Gong}}, \bibinfo {author} {\bibfnamefont {A.~V.}\
  \bibnamefont {Gorshkov}},\ and\ \bibinfo {author} {\bibfnamefont
  {A.}~\bibnamefont {Lucas}},\ }\bibfield  {title} {\bibinfo {title} {{\it
  Hierarchy of Linear Light Cones with Long-Range Interactions}},\ }\href
  {https://doi.org/10.1103/PhysRevX.10.031009} {\bibfield  {journal} {\bibinfo
  {journal} {Phys. Rev. X}\ }\textbf {\bibinfo {volume} {10}},\ \bibinfo
  {pages} {031009} (\bibinfo {year} {2020})}\BibitemShut {NoStop}%
\bibitem [{\citenamefont {Kuwahara}\ and\ \citenamefont
  {Saito}(2020)}]{PhysRevX.10.031010}%
  \BibitemOpen
  \bibfield  {author} {\bibinfo {author} {\bibfnamefont {T.}~\bibnamefont
  {Kuwahara}}\ and\ \bibinfo {author} {\bibfnamefont {K.}~\bibnamefont
  {Saito}},\ }\bibfield  {title} {\bibinfo {title} {{\it Strictly Linear Light
  Cones in Long-Range Interacting Systems of Arbitrary Dimensions}},\ }\href
  {https://doi.org/10.1103/PhysRevX.10.031010} {\bibfield  {journal} {\bibinfo
  {journal} {Phys. Rev. X}\ }\textbf {\bibinfo {volume} {10}},\ \bibinfo
  {pages} {031010} (\bibinfo {year} {2020})}\BibitemShut {NoStop}%
\bibitem [{\citenamefont {Tran}\ \emph {et~al.}(2021)\citenamefont {Tran},
  \citenamefont {Guo}, \citenamefont {Deshpande}, \citenamefont {Lucas},\ and\
  \citenamefont {Gorshkov}}]{tran2020optimal}%
  \BibitemOpen
  \bibfield  {author} {\bibinfo {author} {\bibfnamefont {M.~C.}\ \bibnamefont
  {Tran}}, \bibinfo {author} {\bibfnamefont {A.~Y.}\ \bibnamefont {Guo}},
  \bibinfo {author} {\bibfnamefont {A.}~\bibnamefont {Deshpande}}, \bibinfo
  {author} {\bibfnamefont {A.}~\bibnamefont {Lucas}},\ and\ \bibinfo {author}
  {\bibfnamefont {A.~V.}\ \bibnamefont {Gorshkov}},\ }\bibfield  {title}
  {\bibinfo {title} {{\it Optimal State Transfer and Entanglement Generation in
  Power-Law Interacting Systems}},\ }\href
  {https://doi.org/10.1103/PhysRevX.11.031016} {\bibfield  {journal} {\bibinfo
  {journal} {Phys. Rev. X}\ }\textbf {\bibinfo {volume} {11}},\ \bibinfo
  {pages} {031016} (\bibinfo {year} {2021})}\BibitemShut {NoStop}%
\bibitem [{\citenamefont {Cramer}\ \emph
  {et~al.}(2008{\natexlab{a}})\citenamefont {Cramer}, \citenamefont
  {Serafini},\ and\ \citenamefont {Eisert}}]{cramer2008locality}%
  \BibitemOpen
  \bibfield  {author} {\bibinfo {author} {\bibfnamefont {M.}~\bibnamefont
  {Cramer}}, \bibinfo {author} {\bibfnamefont {A.}~\bibnamefont {Serafini}},\
  and\ \bibinfo {author} {\bibfnamefont {J.}~\bibnamefont {Eisert}},\
  }\href@noop {} {\bibinfo {title} {{\it Locality of dynamics in general
  harmonic quantum systems}}} (\bibinfo {year} {2008}{\natexlab{a}}),\ \Eprint
  {https://arxiv.org/abs/0803.0890} {arXiv:0803.0890 [quant-ph]} \BibitemShut
  {NoStop}%
\bibitem [{\citenamefont {Nachtergaele}\ \emph {et~al.}(2009)\citenamefont
  {Nachtergaele}, \citenamefont {Raz}, \citenamefont {Schlein},\ and\
  \citenamefont {Sims}}]{Nachtergaele2009}%
  \BibitemOpen
  \bibfield  {author} {\bibinfo {author} {\bibfnamefont {B.}~\bibnamefont
  {Nachtergaele}}, \bibinfo {author} {\bibfnamefont {H.}~\bibnamefont {Raz}},
  \bibinfo {author} {\bibfnamefont {B.}~\bibnamefont {Schlein}},\ and\ \bibinfo
  {author} {\bibfnamefont {R.}~\bibnamefont {Sims}},\ }\bibfield  {title}
  {\bibinfo {title} {{\it Lieb-Robinson Bounds for Harmonic and Anharmonic
  Lattice Systems}},\ }\href {https://doi.org/10.1007/s00220-008-0630-2}
  {\bibfield  {journal} {\bibinfo  {journal} {Communications in Mathematical
  Physics}\ }\textbf {\bibinfo {volume} {286}},\ \bibinfo {pages} {1073}
  (\bibinfo {year} {2009})}\BibitemShut {NoStop}%
\bibitem [{\citenamefont {Raz}\ and\ \citenamefont {Sims}(2009)}]{Raz2009}%
  \BibitemOpen
  \bibfield  {author} {\bibinfo {author} {\bibfnamefont {H.}~\bibnamefont
  {Raz}}\ and\ \bibinfo {author} {\bibfnamefont {R.}~\bibnamefont {Sims}},\
  }\bibfield  {title} {\bibinfo {title} {{\it Estimating the Lieb-Robinson
  Velocity for Classical Anharmonic Lattice Systems}},\ }\href
  {https://doi.org/10.1007/s10955-009-9839-5} {\bibfield  {journal} {\bibinfo
  {journal} {Journal of Statistical Physics}\ }\textbf {\bibinfo {volume}
  {137}},\ \bibinfo {pages} {79} (\bibinfo {year} {2009})}\BibitemShut
  {NoStop}%
\bibitem [{\citenamefont {Nachtergaele}\ \emph {et~al.}(2010)\citenamefont
  {Nachtergaele}, \citenamefont {Schlein}, \citenamefont {Sims}, \citenamefont
  {Starr},\ and\ \citenamefont {Valentin}}]{doi:10.1142/S0129055X1000393X}%
  \BibitemOpen
  \bibfield  {author} {\bibinfo {author} {\bibfnamefont {B.}~\bibnamefont
  {Nachtergaele}}, \bibinfo {author} {\bibfnamefont {B.}~\bibnamefont
  {Schlein}}, \bibinfo {author} {\bibfnamefont {R.}~\bibnamefont {Sims}},
  \bibinfo {author} {\bibfnamefont {S.}~\bibnamefont {Starr}},\ and\ \bibinfo
  {author} {\bibfnamefont {Z.}~\bibnamefont {Valentin}},\ }\bibfield  {title}
  {\bibinfo {title} {{\it On the Existence of the Dynamics for Anharmonic
  Quantum Oscillator Systems}},\ }\href
  {https://doi.org/10.1142/S0129055X1000393X} {\bibfield  {journal} {\bibinfo
  {journal} {Reviews in Mathematical Physics}\ }\textbf {\bibinfo {volume}
  {22}},\ \bibinfo {pages} {207} (\bibinfo {year} {2010})}\BibitemShut
  {NoStop}%
\bibitem [{\citenamefont {Nachtergaele}\ and\ \citenamefont
  {Sims}(2014)}]{nachtergaele2014dynamics}%
  \BibitemOpen
  \bibfield  {author} {\bibinfo {author} {\bibfnamefont {B.}~\bibnamefont
  {Nachtergaele}}\ and\ \bibinfo {author} {\bibfnamefont {R.}~\bibnamefont
  {Sims}},\ }\href@noop {} {\bibinfo {title} {{\it On the dynamics of lattice
  systems with unbounded on-site terms in the Hamiltonian}}} (\bibinfo {year}
  {2014}),\ \Eprint {https://arxiv.org/abs/1410.8174} {arXiv:1410.8174
  [math-ph]} \BibitemShut {NoStop}%
\bibitem [{\citenamefont {J\"unemann}\ \emph {et~al.}(2013)\citenamefont
  {J\"unemann}, \citenamefont {Cadarso}, \citenamefont {P\'erez-Garc\'{\i}a},
  \citenamefont {Bermudez},\ and\ \citenamefont
  {Garc\'{\i}a-Ripoll}}]{PhysRevLett.111.230404}%
  \BibitemOpen
  \bibfield  {author} {\bibinfo {author} {\bibfnamefont {J.}~\bibnamefont
  {J\"unemann}}, \bibinfo {author} {\bibfnamefont {A.}~\bibnamefont {Cadarso}},
  \bibinfo {author} {\bibfnamefont {D.}~\bibnamefont {P\'erez-Garc\'{\i}a}},
  \bibinfo {author} {\bibfnamefont {A.}~\bibnamefont {Bermudez}},\ and\
  \bibinfo {author} {\bibfnamefont {J.~J.}\ \bibnamefont
  {Garc\'{\i}a-Ripoll}},\ }\bibfield  {title} {\bibinfo {title} {{\it
  Lieb-Robinson Bounds for Spin-Boson Lattice Models and Trapped Ions}},\
  }\href {https://doi.org/10.1103/PhysRevLett.111.230404} {\bibfield  {journal}
  {\bibinfo  {journal} {Phys. Rev. Lett.}\ }\textbf {\bibinfo {volume} {111}},\
  \bibinfo {pages} {230404} (\bibinfo {year} {2013})}\BibitemShut {NoStop}%
\bibitem [{\citenamefont {Woods}\ \emph {et~al.}(2015)\citenamefont {Woods},
  \citenamefont {Cramer},\ and\ \citenamefont
  {Plenio}}]{PhysRevLett.115.130401}%
  \BibitemOpen
  \bibfield  {author} {\bibinfo {author} {\bibfnamefont {M.~P.}\ \bibnamefont
  {Woods}}, \bibinfo {author} {\bibfnamefont {M.}~\bibnamefont {Cramer}},\ and\
  \bibinfo {author} {\bibfnamefont {M.~B.}\ \bibnamefont {Plenio}},\ }\bibfield
   {title} {\bibinfo {title} {{\it Simulating Bosonic Baths with Error Bars}},\
  }\href {https://doi.org/10.1103/PhysRevLett.115.130401} {\bibfield  {journal}
  {\bibinfo  {journal} {Phys. Rev. Lett.}\ }\textbf {\bibinfo {volume} {115}},\
  \bibinfo {pages} {130401} (\bibinfo {year} {2015})}\BibitemShut {NoStop}%
\bibitem [{\citenamefont {Woods}\ and\ \citenamefont
  {Plenio}(2016)}]{doi:10.1063/1.4940436}%
  \BibitemOpen
  \bibfield  {author} {\bibinfo {author} {\bibfnamefont {M.~P.}\ \bibnamefont
  {Woods}}\ and\ \bibinfo {author} {\bibfnamefont {M.~B.}\ \bibnamefont
  {Plenio}},\ }\bibfield  {title} {\bibinfo {title} {{\it Dynamical error
  bounds for continuum discretisation via Gauss quadrature rules—A
  Lieb-Robinson bound approach}},\ }\href {https://doi.org/10.1063/1.4940436}
  {\bibfield  {journal} {\bibinfo  {journal} {Journal of Mathematical Physics}\
  }\textbf {\bibinfo {volume} {57}},\ \bibinfo {pages} {022105} (\bibinfo
  {year} {2016})}\BibitemShut {NoStop}%
\bibitem [{\citenamefont {Eisert}\ and\ \citenamefont
  {Gross}(2009)}]{PhysRevLett.102.240501}%
  \BibitemOpen
  \bibfield  {author} {\bibinfo {author} {\bibfnamefont {J.}~\bibnamefont
  {Eisert}}\ and\ \bibinfo {author} {\bibfnamefont {D.}~\bibnamefont {Gross}},\
  }\bibfield  {title} {\bibinfo {title} {{\it Supersonic Quantum
  Communication}},\ }\href {https://doi.org/10.1103/PhysRevLett.102.240501}
  {\bibfield  {journal} {\bibinfo  {journal} {Phys. Rev. Lett.}\ }\textbf
  {\bibinfo {volume} {102}},\ \bibinfo {pages} {240501} (\bibinfo {year}
  {2009})}\BibitemShut {NoStop}%
\bibitem [{\citenamefont {Bloch}\ \emph {et~al.}(2008)\citenamefont {Bloch},
  \citenamefont {Dalibard},\ and\ \citenamefont {Zwerger}}]{RevModPhys.80.885}%
  \BibitemOpen
  \bibfield  {author} {\bibinfo {author} {\bibfnamefont {I.}~\bibnamefont
  {Bloch}}, \bibinfo {author} {\bibfnamefont {J.}~\bibnamefont {Dalibard}},\
  and\ \bibinfo {author} {\bibfnamefont {W.}~\bibnamefont {Zwerger}},\
  }\bibfield  {title} {\bibinfo {title} {{\it Many-body physics with ultracold
  gases}},\ }\href {https://doi.org/10.1103/RevModPhys.80.885} {\bibfield
  {journal} {\bibinfo  {journal} {Rev. Mod. Phys.}\ }\textbf {\bibinfo {volume}
  {80}},\ \bibinfo {pages} {885} (\bibinfo {year} {2008})}\BibitemShut
  {NoStop}%
\bibitem [{\citenamefont {Sherson}\ \emph {et~al.}(2010)\citenamefont
  {Sherson}, \citenamefont {Weitenberg}, \citenamefont {Endres}, \citenamefont
  {Cheneau}, \citenamefont {Bloch},\ and\ \citenamefont {Kuhr}}]{Sherson2010}%
  \BibitemOpen
  \bibfield  {author} {\bibinfo {author} {\bibfnamefont {J.~F.}\ \bibnamefont
  {Sherson}}, \bibinfo {author} {\bibfnamefont {C.}~\bibnamefont {Weitenberg}},
  \bibinfo {author} {\bibfnamefont {M.}~\bibnamefont {Endres}}, \bibinfo
  {author} {\bibfnamefont {M.}~\bibnamefont {Cheneau}}, \bibinfo {author}
  {\bibfnamefont {I.}~\bibnamefont {Bloch}},\ and\ \bibinfo {author}
  {\bibfnamefont {S.}~\bibnamefont {Kuhr}},\ }\bibfield  {title} {\bibinfo
  {title} {{\it Single-atom-resolved fluorescence imaging of an atomic Mott
  insulator}},\ }\href {https://doi.org/10.1038/nature09378} {\bibfield
  {journal} {\bibinfo  {journal} {Nature}\ }\textbf {\bibinfo {volume} {467}},\
  \bibinfo {pages} {68} (\bibinfo {year} {2010})}\BibitemShut {NoStop}%
\bibitem [{\citenamefont {Bakr}\ \emph {et~al.}(2010)\citenamefont {Bakr},
  \citenamefont {Peng}, \citenamefont {Tai}, \citenamefont {Ma}, \citenamefont
  {Simon}, \citenamefont {Gillen}, \citenamefont {F{\"o}lling}, \citenamefont
  {Pollet},\ and\ \citenamefont {Greiner}}]{Bakr547}%
  \BibitemOpen
  \bibfield  {author} {\bibinfo {author} {\bibfnamefont {W.~S.}\ \bibnamefont
  {Bakr}}, \bibinfo {author} {\bibfnamefont {A.}~\bibnamefont {Peng}}, \bibinfo
  {author} {\bibfnamefont {M.~E.}\ \bibnamefont {Tai}}, \bibinfo {author}
  {\bibfnamefont {R.}~\bibnamefont {Ma}}, \bibinfo {author} {\bibfnamefont
  {J.}~\bibnamefont {Simon}}, \bibinfo {author} {\bibfnamefont {J.~I.}\
  \bibnamefont {Gillen}}, \bibinfo {author} {\bibfnamefont {S.}~\bibnamefont
  {F{\"o}lling}}, \bibinfo {author} {\bibfnamefont {L.}~\bibnamefont
  {Pollet}},\ and\ \bibinfo {author} {\bibfnamefont {M.}~\bibnamefont
  {Greiner}},\ }\bibfield  {title} {\bibinfo {title} {{\it Probing the
  Superfluid{\textendash}to{\textendash}Mott Insulator Transition at the
  Single-Atom Level}},\ }\href {https://doi.org/10.1126/science.1192368}
  {\bibfield  {journal} {\bibinfo  {journal} {Science}\ }\textbf {\bibinfo
  {volume} {329}},\ \bibinfo {pages} {547} (\bibinfo {year}
  {2010})}\BibitemShut {NoStop}%
\bibitem [{\citenamefont {Cheneau}\ \emph {et~al.}(2012)\citenamefont
  {Cheneau}, \citenamefont {Barmettler}, \citenamefont {Poletti}, \citenamefont
  {Endres}, \citenamefont {Schau{\ss}}, \citenamefont {Fukuhara}, \citenamefont
  {Gross}, \citenamefont {Bloch}, \citenamefont {Kollath},\ and\ \citenamefont
  {Kuhr}}]{cheneau2012light}%
  \BibitemOpen
  \bibfield  {author} {\bibinfo {author} {\bibfnamefont {M.}~\bibnamefont
  {Cheneau}}, \bibinfo {author} {\bibfnamefont {P.}~\bibnamefont {Barmettler}},
  \bibinfo {author} {\bibfnamefont {D.}~\bibnamefont {Poletti}}, \bibinfo
  {author} {\bibfnamefont {M.}~\bibnamefont {Endres}}, \bibinfo {author}
  {\bibfnamefont {P.}~\bibnamefont {Schau{\ss}}}, \bibinfo {author}
  {\bibfnamefont {T.}~\bibnamefont {Fukuhara}}, \bibinfo {author}
  {\bibfnamefont {C.}~\bibnamefont {Gross}}, \bibinfo {author} {\bibfnamefont
  {I.}~\bibnamefont {Bloch}}, \bibinfo {author} {\bibfnamefont
  {C.}~\bibnamefont {Kollath}},\ and\ \bibinfo {author} {\bibfnamefont
  {S.}~\bibnamefont {Kuhr}},\ }\bibfield  {title} {\bibinfo {title} {{\it
  Light-cone-like spreading of correlations in a quantum many-body system}},\
  }\href {https://doi.org/10.1038/nature10748} {\bibfield  {journal} {\bibinfo
  {journal} {Nature}\ }\textbf {\bibinfo {volume} {481}},\ \bibinfo {pages}
  {484} (\bibinfo {year} {2012})}\BibitemShut {NoStop}%
\bibitem [{\citenamefont {Langen}\ \emph {et~al.}(2013)\citenamefont {Langen},
  \citenamefont {Geiger}, \citenamefont {Kuhnert}, \citenamefont {Rauer},\ and\
  \citenamefont {Schmiedmayer}}]{langen2013local}%
  \BibitemOpen
  \bibfield  {author} {\bibinfo {author} {\bibfnamefont {T.}~\bibnamefont
  {Langen}}, \bibinfo {author} {\bibfnamefont {R.}~\bibnamefont {Geiger}},
  \bibinfo {author} {\bibfnamefont {M.}~\bibnamefont {Kuhnert}}, \bibinfo
  {author} {\bibfnamefont {B.}~\bibnamefont {Rauer}},\ and\ \bibinfo {author}
  {\bibfnamefont {J.}~\bibnamefont {Schmiedmayer}},\ }\bibfield  {title}
  {\bibinfo {title} {{\it Local emergence of thermal correlations in an
  isolated quantum many-body system}},\ }\href
  {https://doi.org/10.1038/nphys2739} {\bibfield  {journal} {\bibinfo
  {journal} {Nature Physics}\ }\textbf {\bibinfo {volume} {9}},\ \bibinfo
  {pages} {640} (\bibinfo {year} {2013})}\BibitemShut {NoStop}%
\bibitem [{\citenamefont {Braun}\ \emph {et~al.}(2015)\citenamefont {Braun},
  \citenamefont {Friesdorf}, \citenamefont {Hodgman}, \citenamefont
  {Schreiber}, \citenamefont {Ronzheimer}, \citenamefont {Riera}, \citenamefont
  {del Rey}, \citenamefont {Bloch}, \citenamefont {Eisert},\ and\ \citenamefont
  {Schneider}}]{Braun3641}%
  \BibitemOpen
  \bibfield  {author} {\bibinfo {author} {\bibfnamefont {S.}~\bibnamefont
  {Braun}}, \bibinfo {author} {\bibfnamefont {M.}~\bibnamefont {Friesdorf}},
  \bibinfo {author} {\bibfnamefont {S.~S.}\ \bibnamefont {Hodgman}}, \bibinfo
  {author} {\bibfnamefont {M.}~\bibnamefont {Schreiber}}, \bibinfo {author}
  {\bibfnamefont {J.~P.}\ \bibnamefont {Ronzheimer}}, \bibinfo {author}
  {\bibfnamefont {A.}~\bibnamefont {Riera}}, \bibinfo {author} {\bibfnamefont
  {M.}~\bibnamefont {del Rey}}, \bibinfo {author} {\bibfnamefont
  {I.}~\bibnamefont {Bloch}}, \bibinfo {author} {\bibfnamefont
  {J.}~\bibnamefont {Eisert}},\ and\ \bibinfo {author} {\bibfnamefont
  {U.}~\bibnamefont {Schneider}},\ }\bibfield  {title} {\bibinfo {title} {{\it
  Emergence of coherence and the dynamics of quantum phase transitions}},\
  }\href {https://doi.org/10.1073/pnas.1408861112} {\bibfield  {journal}
  {\bibinfo  {journal} {Proceedings of the National Academy of Sciences}\
  }\textbf {\bibinfo {volume} {112}},\ \bibinfo {pages} {3641} (\bibinfo {year}
  {2015})}\BibitemShut {NoStop}%
\bibitem [{\citenamefont {Islam}\ \emph {et~al.}(2015)\citenamefont {Islam},
  \citenamefont {Ma}, \citenamefont {Preiss}, \citenamefont {Eric~Tai},
  \citenamefont {Lukin}, \citenamefont {Rispoli},\ and\ \citenamefont
  {Greiner}}]{Islam2015}%
  \BibitemOpen
  \bibfield  {author} {\bibinfo {author} {\bibfnamefont {R.}~\bibnamefont
  {Islam}}, \bibinfo {author} {\bibfnamefont {R.}~\bibnamefont {Ma}}, \bibinfo
  {author} {\bibfnamefont {P.~M.}\ \bibnamefont {Preiss}}, \bibinfo {author}
  {\bibfnamefont {M.}~\bibnamefont {Eric~Tai}}, \bibinfo {author}
  {\bibfnamefont {A.}~\bibnamefont {Lukin}}, \bibinfo {author} {\bibfnamefont
  {M.}~\bibnamefont {Rispoli}},\ and\ \bibinfo {author} {\bibfnamefont
  {M.}~\bibnamefont {Greiner}},\ }\bibfield  {title} {\bibinfo {title} {{\it
  Measuring entanglement entropy in a quantum many-body system}},\ }\href
  {https://doi.org/10.1038/nature15750} {\bibfield  {journal} {\bibinfo
  {journal} {Nature}\ }\textbf {\bibinfo {volume} {528}},\ \bibinfo {pages}
  {77} (\bibinfo {year} {2015})}\BibitemShut {NoStop}%
\bibitem [{\citenamefont {Choi}\ \emph {et~al.}(2016)\citenamefont {Choi},
  \citenamefont {Hild}, \citenamefont {Zeiher}, \citenamefont {Schau{\ss}},
  \citenamefont {Rubio-Abadal}, \citenamefont {Yefsah}, \citenamefont
  {Khemani}, \citenamefont {Huse}, \citenamefont {Bloch},\ and\ \citenamefont
  {Gross}}]{Choi1547}%
  \BibitemOpen
  \bibfield  {author} {\bibinfo {author} {\bibfnamefont {J.-y.}\ \bibnamefont
  {Choi}}, \bibinfo {author} {\bibfnamefont {S.}~\bibnamefont {Hild}}, \bibinfo
  {author} {\bibfnamefont {J.}~\bibnamefont {Zeiher}}, \bibinfo {author}
  {\bibfnamefont {P.}~\bibnamefont {Schau{\ss}}}, \bibinfo {author}
  {\bibfnamefont {A.}~\bibnamefont {Rubio-Abadal}}, \bibinfo {author}
  {\bibfnamefont {T.}~\bibnamefont {Yefsah}}, \bibinfo {author} {\bibfnamefont
  {V.}~\bibnamefont {Khemani}}, \bibinfo {author} {\bibfnamefont {D.~A.}\
  \bibnamefont {Huse}}, \bibinfo {author} {\bibfnamefont {I.}~\bibnamefont
  {Bloch}},\ and\ \bibinfo {author} {\bibfnamefont {C.}~\bibnamefont {Gross}},\
  }\bibfield  {title} {\bibinfo {title} {{\it Exploring the many-body
  localization transition in two dimensions}},\ }\href
  {https://doi.org/10.1126/science.aaf8834} {\bibfield  {journal} {\bibinfo
  {journal} {Science}\ }\textbf {\bibinfo {volume} {352}},\ \bibinfo {pages}
  {1547} (\bibinfo {year} {2016})}\BibitemShut {NoStop}%
\bibitem [{\citenamefont {Meinert}\ \emph {et~al.}(2016)\citenamefont
  {Meinert}, \citenamefont {Mark}, \citenamefont {Lauber}, \citenamefont
  {Daley},\ and\ \citenamefont {N\"agerl}}]{PhysRevLett.116.205301}%
  \BibitemOpen
  \bibfield  {author} {\bibinfo {author} {\bibfnamefont {F.}~\bibnamefont
  {Meinert}}, \bibinfo {author} {\bibfnamefont {M.~J.}\ \bibnamefont {Mark}},
  \bibinfo {author} {\bibfnamefont {K.}~\bibnamefont {Lauber}}, \bibinfo
  {author} {\bibfnamefont {A.~J.}\ \bibnamefont {Daley}},\ and\ \bibinfo
  {author} {\bibfnamefont {H.-C.}\ \bibnamefont {N\"agerl}},\ }\bibfield
  {title} {\bibinfo {title} {{\it Floquet Engineering of Correlated Tunneling
  in the Bose-Hubbard Model with Ultracold Atoms}},\ }\href
  {https://doi.org/10.1103/PhysRevLett.116.205301} {\bibfield  {journal}
  {\bibinfo  {journal} {Phys. Rev. Lett.}\ }\textbf {\bibinfo {volume} {116}},\
  \bibinfo {pages} {205301} (\bibinfo {year} {2016})}\BibitemShut {NoStop}%
\bibitem [{\citenamefont {Baier}\ \emph {et~al.}(2016)\citenamefont {Baier},
  \citenamefont {Mark}, \citenamefont {Petter}, \citenamefont {Aikawa},
  \citenamefont {Chomaz}, \citenamefont {Cai}, \citenamefont {Baranov},
  \citenamefont {Zoller},\ and\ \citenamefont {Ferlaino}}]{Baier201}%
  \BibitemOpen
  \bibfield  {author} {\bibinfo {author} {\bibfnamefont {S.}~\bibnamefont
  {Baier}}, \bibinfo {author} {\bibfnamefont {M.~J.}\ \bibnamefont {Mark}},
  \bibinfo {author} {\bibfnamefont {D.}~\bibnamefont {Petter}}, \bibinfo
  {author} {\bibfnamefont {K.}~\bibnamefont {Aikawa}}, \bibinfo {author}
  {\bibfnamefont {L.}~\bibnamefont {Chomaz}}, \bibinfo {author} {\bibfnamefont
  {Z.}~\bibnamefont {Cai}}, \bibinfo {author} {\bibfnamefont {M.}~\bibnamefont
  {Baranov}}, \bibinfo {author} {\bibfnamefont {P.}~\bibnamefont {Zoller}},\
  and\ \bibinfo {author} {\bibfnamefont {F.}~\bibnamefont {Ferlaino}},\
  }\bibfield  {title} {\bibinfo {title} {{\it Extended Bose-Hubbard models with
  ultracold magnetic atoms}},\ }\href {https://doi.org/10.1126/science.aac9812}
  {\bibfield  {journal} {\bibinfo  {journal} {Science}\ }\textbf {\bibinfo
  {volume} {352}},\ \bibinfo {pages} {201} (\bibinfo {year}
  {2016})}\BibitemShut {NoStop}%
\bibitem [{\citenamefont {Ye}\ \emph {et~al.}(2019)\citenamefont {Ye},
  \citenamefont {Ge}, \citenamefont {Wu}, \citenamefont {Wang}, \citenamefont
  {Gong}, \citenamefont {Zhang}, \citenamefont {Zhu}, \citenamefont {Yang},
  \citenamefont {Li}, \citenamefont {Liang}, \citenamefont {Lin}, \citenamefont
  {Xu}, \citenamefont {Guo}, \citenamefont {Sun}, \citenamefont {Cheng},
  \citenamefont {Ma}, \citenamefont {Meng}, \citenamefont {Deng}, \citenamefont
  {Rong}, \citenamefont {Lu}, \citenamefont {Peng}, \citenamefont {Fan},
  \citenamefont {Zhu},\ and\ \citenamefont {Pan}}]{PhysRevLett.123.050502}%
  \BibitemOpen
  \bibfield  {author} {\bibinfo {author} {\bibfnamefont {Y.}~\bibnamefont
  {Ye}}, \bibinfo {author} {\bibfnamefont {Z.-Y.}\ \bibnamefont {Ge}}, \bibinfo
  {author} {\bibfnamefont {Y.}~\bibnamefont {Wu}}, \bibinfo {author}
  {\bibfnamefont {S.}~\bibnamefont {Wang}}, \bibinfo {author} {\bibfnamefont
  {M.}~\bibnamefont {Gong}}, \bibinfo {author} {\bibfnamefont {Y.-R.}\
  \bibnamefont {Zhang}}, \bibinfo {author} {\bibfnamefont {Q.}~\bibnamefont
  {Zhu}}, \bibinfo {author} {\bibfnamefont {R.}~\bibnamefont {Yang}}, \bibinfo
  {author} {\bibfnamefont {S.}~\bibnamefont {Li}}, \bibinfo {author}
  {\bibfnamefont {F.}~\bibnamefont {Liang}}, \bibinfo {author} {\bibfnamefont
  {J.}~\bibnamefont {Lin}}, \bibinfo {author} {\bibfnamefont {Y.}~\bibnamefont
  {Xu}}, \bibinfo {author} {\bibfnamefont {C.}~\bibnamefont {Guo}}, \bibinfo
  {author} {\bibfnamefont {L.}~\bibnamefont {Sun}}, \bibinfo {author}
  {\bibfnamefont {C.}~\bibnamefont {Cheng}}, \bibinfo {author} {\bibfnamefont
  {N.}~\bibnamefont {Ma}}, \bibinfo {author} {\bibfnamefont {Z.~Y.}\
  \bibnamefont {Meng}}, \bibinfo {author} {\bibfnamefont {H.}~\bibnamefont
  {Deng}}, \bibinfo {author} {\bibfnamefont {H.}~\bibnamefont {Rong}}, \bibinfo
  {author} {\bibfnamefont {C.-Y.}\ \bibnamefont {Lu}}, \bibinfo {author}
  {\bibfnamefont {C.-Z.}\ \bibnamefont {Peng}}, \bibinfo {author}
  {\bibfnamefont {H.}~\bibnamefont {Fan}}, \bibinfo {author} {\bibfnamefont
  {X.}~\bibnamefont {Zhu}},\ and\ \bibinfo {author} {\bibfnamefont {J.-W.}\
  \bibnamefont {Pan}},\ }\bibfield  {title} {\bibinfo {title} {{\it Propagation
  and Localization of Collective Excitations on a 24-Qubit Superconducting
  Processor}},\ }\href {https://doi.org/10.1103/PhysRevLett.123.050502}
  {\bibfield  {journal} {\bibinfo  {journal} {Phys. Rev. Lett.}\ }\textbf
  {\bibinfo {volume} {123}},\ \bibinfo {pages} {050502} (\bibinfo {year}
  {2019})}\BibitemShut {NoStop}%
\bibitem [{\citenamefont {Yan}\ \emph {et~al.}(2019)\citenamefont {Yan},
  \citenamefont {Zhang}, \citenamefont {Gong}, \citenamefont {Wu},
  \citenamefont {Zheng}, \citenamefont {Li}, \citenamefont {Wang},
  \citenamefont {Liang}, \citenamefont {Lin}, \citenamefont {Xu}, \citenamefont
  {Guo}, \citenamefont {Sun}, \citenamefont {Peng}, \citenamefont {Xia},
  \citenamefont {Deng}, \citenamefont {Rong}, \citenamefont {You},
  \citenamefont {Nori}, \citenamefont {Fan}, \citenamefont {Zhu},\ and\
  \citenamefont {Pan}}]{Yan753}%
  \BibitemOpen
  \bibfield  {author} {\bibinfo {author} {\bibfnamefont {Z.}~\bibnamefont
  {Yan}}, \bibinfo {author} {\bibfnamefont {Y.-R.}\ \bibnamefont {Zhang}},
  \bibinfo {author} {\bibfnamefont {M.}~\bibnamefont {Gong}}, \bibinfo {author}
  {\bibfnamefont {Y.}~\bibnamefont {Wu}}, \bibinfo {author} {\bibfnamefont
  {Y.}~\bibnamefont {Zheng}}, \bibinfo {author} {\bibfnamefont
  {S.}~\bibnamefont {Li}}, \bibinfo {author} {\bibfnamefont {C.}~\bibnamefont
  {Wang}}, \bibinfo {author} {\bibfnamefont {F.}~\bibnamefont {Liang}},
  \bibinfo {author} {\bibfnamefont {J.}~\bibnamefont {Lin}}, \bibinfo {author}
  {\bibfnamefont {Y.}~\bibnamefont {Xu}}, \bibinfo {author} {\bibfnamefont
  {C.}~\bibnamefont {Guo}}, \bibinfo {author} {\bibfnamefont {L.}~\bibnamefont
  {Sun}}, \bibinfo {author} {\bibfnamefont {C.-Z.}\ \bibnamefont {Peng}},
  \bibinfo {author} {\bibfnamefont {K.}~\bibnamefont {Xia}}, \bibinfo {author}
  {\bibfnamefont {H.}~\bibnamefont {Deng}}, \bibinfo {author} {\bibfnamefont
  {H.}~\bibnamefont {Rong}}, \bibinfo {author} {\bibfnamefont {J.~Q.}\
  \bibnamefont {You}}, \bibinfo {author} {\bibfnamefont {F.}~\bibnamefont
  {Nori}}, \bibinfo {author} {\bibfnamefont {H.}~\bibnamefont {Fan}}, \bibinfo
  {author} {\bibfnamefont {X.}~\bibnamefont {Zhu}},\ and\ \bibinfo {author}
  {\bibfnamefont {J.-W.}\ \bibnamefont {Pan}},\ }\bibfield  {title} {\bibinfo
  {title} {{\it Strongly correlated quantum walks with a 12-qubit
  superconducting processor}},\ }\href
  {https://doi.org/10.1126/science.aaw1611} {\bibfield  {journal} {\bibinfo
  {journal} {Science}\ }\textbf {\bibinfo {volume} {364}},\ \bibinfo {pages}
  {753} (\bibinfo {year} {2019})}\BibitemShut {NoStop}%
\bibitem [{\citenamefont {Rubio-Abadal}\ \emph {et~al.}(2020)\citenamefont
  {Rubio-Abadal}, \citenamefont {Ippoliti}, \citenamefont {Hollerith},
  \citenamefont {Wei}, \citenamefont {Rui}, \citenamefont {Sondhi},
  \citenamefont {Khemani}, \citenamefont {Gross},\ and\ \citenamefont
  {Bloch}}]{PhysRevX.10.021044}%
  \BibitemOpen
  \bibfield  {author} {\bibinfo {author} {\bibfnamefont {A.}~\bibnamefont
  {Rubio-Abadal}}, \bibinfo {author} {\bibfnamefont {M.}~\bibnamefont
  {Ippoliti}}, \bibinfo {author} {\bibfnamefont {S.}~\bibnamefont {Hollerith}},
  \bibinfo {author} {\bibfnamefont {D.}~\bibnamefont {Wei}}, \bibinfo {author}
  {\bibfnamefont {J.}~\bibnamefont {Rui}}, \bibinfo {author} {\bibfnamefont
  {S.~L.}\ \bibnamefont {Sondhi}}, \bibinfo {author} {\bibfnamefont
  {V.}~\bibnamefont {Khemani}}, \bibinfo {author} {\bibfnamefont
  {C.}~\bibnamefont {Gross}},\ and\ \bibinfo {author} {\bibfnamefont
  {I.}~\bibnamefont {Bloch}},\ }\bibfield  {title} {\bibinfo {title} {{\it
  Floquet Prethermalization in a Bose-Hubbard System}},\ }\href
  {https://doi.org/10.1103/PhysRevX.10.021044} {\bibfield  {journal} {\bibinfo
  {journal} {Phys. Rev. X}\ }\textbf {\bibinfo {volume} {10}},\ \bibinfo
  {pages} {021044} (\bibinfo {year} {2020})}\BibitemShut {NoStop}%
\bibitem [{\citenamefont {Yang}\ \emph {et~al.}(2020)\citenamefont {Yang},
  \citenamefont {Sun}, \citenamefont {Ott}, \citenamefont {Wang}, \citenamefont
  {Zache}, \citenamefont {Halimeh}, \citenamefont {Yuan}, \citenamefont
  {Hauke},\ and\ \citenamefont {Pan}}]{Yang2020}%
  \BibitemOpen
  \bibfield  {author} {\bibinfo {author} {\bibfnamefont {B.}~\bibnamefont
  {Yang}}, \bibinfo {author} {\bibfnamefont {H.}~\bibnamefont {Sun}}, \bibinfo
  {author} {\bibfnamefont {R.}~\bibnamefont {Ott}}, \bibinfo {author}
  {\bibfnamefont {H.-Y.}\ \bibnamefont {Wang}}, \bibinfo {author}
  {\bibfnamefont {T.~V.}\ \bibnamefont {Zache}}, \bibinfo {author}
  {\bibfnamefont {J.~C.}\ \bibnamefont {Halimeh}}, \bibinfo {author}
  {\bibfnamefont {Z.-S.}\ \bibnamefont {Yuan}}, \bibinfo {author}
  {\bibfnamefont {P.}~\bibnamefont {Hauke}},\ and\ \bibinfo {author}
  {\bibfnamefont {J.-W.}\ \bibnamefont {Pan}},\ }\bibfield  {title} {\bibinfo
  {title} {{\it Observation of gauge invariance in a 71-site Bose--Hubbard
  quantum simulator}},\ }\href {https://doi.org/10.1038/s41586-020-2910-8}
  {\bibfield  {journal} {\bibinfo  {journal} {Nature}\ }\textbf {\bibinfo
  {volume} {587}},\ \bibinfo {pages} {392} (\bibinfo {year}
  {2020})}\BibitemShut {NoStop}%
\bibitem [{\citenamefont {Takasu}\ \emph {et~al.}(2020)\citenamefont {Takasu},
  \citenamefont {Yagami}, \citenamefont {Asaka}, \citenamefont {Fukushima},
  \citenamefont {Nagao}, \citenamefont {Goto}, \citenamefont {Danshita},\ and\
  \citenamefont {Takahashi}}]{Takasueaba9255}%
  \BibitemOpen
  \bibfield  {author} {\bibinfo {author} {\bibfnamefont {Y.}~\bibnamefont
  {Takasu}}, \bibinfo {author} {\bibfnamefont {T.}~\bibnamefont {Yagami}},
  \bibinfo {author} {\bibfnamefont {H.}~\bibnamefont {Asaka}}, \bibinfo
  {author} {\bibfnamefont {Y.}~\bibnamefont {Fukushima}}, \bibinfo {author}
  {\bibfnamefont {K.}~\bibnamefont {Nagao}}, \bibinfo {author} {\bibfnamefont
  {S.}~\bibnamefont {Goto}}, \bibinfo {author} {\bibfnamefont {I.}~\bibnamefont
  {Danshita}},\ and\ \bibinfo {author} {\bibfnamefont {Y.}~\bibnamefont
  {Takahashi}},\ }\bibfield  {title} {\bibinfo {title} {{\it Energy
  redistribution and spatiotemporal evolution of correlations after a sudden
  quench of the Bose-Hubbard model}},\ }\bibfield  {journal} {\bibinfo
  {journal} {Science Advances}\ }\textbf {\bibinfo {volume} {6}},\ \href
  {https://doi.org/10.1126/sciadv.aba9255} {10.1126/sciadv.aba9255} (\bibinfo
  {year} {2020})\BibitemShut {NoStop}%
\bibitem [{\citenamefont {Schuch}\ \emph {et~al.}(2011)\citenamefont {Schuch},
  \citenamefont {Harrison}, \citenamefont {Osborne},\ and\ \citenamefont
  {Eisert}}]{PhysRevA.84.032309}%
  \BibitemOpen
  \bibfield  {author} {\bibinfo {author} {\bibfnamefont {N.}~\bibnamefont
  {Schuch}}, \bibinfo {author} {\bibfnamefont {S.~K.}\ \bibnamefont
  {Harrison}}, \bibinfo {author} {\bibfnamefont {T.~J.}\ \bibnamefont
  {Osborne}},\ and\ \bibinfo {author} {\bibfnamefont {J.}~\bibnamefont
  {Eisert}},\ }\bibfield  {title} {\bibinfo {title} {{\it Information
  propagation for interacting-particle systems}},\ }\href
  {https://doi.org/10.1103/PhysRevA.84.032309} {\bibfield  {journal} {\bibinfo
  {journal} {Phys. Rev. A}\ }\textbf {\bibinfo {volume} {84}},\ \bibinfo
  {pages} {032309} (\bibinfo {year} {2011})}\BibitemShut {NoStop}%
\bibitem [{\citenamefont {Wang}\ and\ \citenamefont
  {Hazzard}(2020)}]{PRXQuantum.1.010303}%
  \BibitemOpen
  \bibfield  {author} {\bibinfo {author} {\bibfnamefont {Z.}~\bibnamefont
  {Wang}}\ and\ \bibinfo {author} {\bibfnamefont {K.~R.}\ \bibnamefont
  {Hazzard}},\ }\bibfield  {title} {\bibinfo {title} {{\it Tightening the
  Lieb-Robinson Bound in Locally Interacting Systems}},\ }\href
  {https://doi.org/10.1103/PRXQuantum.1.010303} {\bibfield  {journal} {\bibinfo
   {journal} {PRX Quantum}\ }\textbf {\bibinfo {volume} {1}},\ \bibinfo {pages}
  {010303} (\bibinfo {year} {2020})}\BibitemShut {NoStop}%
\bibitem [{\citenamefont {Läuchli}\ and\ \citenamefont
  {Kollath}(2008)}]{L_uchli_2008}%
  \BibitemOpen
  \bibfield  {author} {\bibinfo {author} {\bibfnamefont {A.~M.}\ \bibnamefont
  {Läuchli}}\ and\ \bibinfo {author} {\bibfnamefont {C.}~\bibnamefont
  {Kollath}},\ }\bibfield  {title} {\bibinfo {title} {{\it Spreading of
  correlations and entanglement after a quench in the one-dimensional
  Bose{\textendash}Hubbard model}},\ }\href
  {https://doi.org/10.1088/1742-5468/2008/05/p05018} {\bibfield  {journal}
  {\bibinfo  {journal} {Journal of Statistical Mechanics: Theory and
  Experiment}\ }\textbf {\bibinfo {volume} {2008}},\ \bibinfo {pages} {P05018}
  (\bibinfo {year} {2008})}\BibitemShut {NoStop}%
\bibitem [{\citenamefont {Barmettler}\ \emph {et~al.}(2012)\citenamefont
  {Barmettler}, \citenamefont {Poletti}, \citenamefont {Cheneau},\ and\
  \citenamefont {Kollath}}]{PhysRevA.85.053625}%
  \BibitemOpen
  \bibfield  {author} {\bibinfo {author} {\bibfnamefont {P.}~\bibnamefont
  {Barmettler}}, \bibinfo {author} {\bibfnamefont {D.}~\bibnamefont {Poletti}},
  \bibinfo {author} {\bibfnamefont {M.}~\bibnamefont {Cheneau}},\ and\ \bibinfo
  {author} {\bibfnamefont {C.}~\bibnamefont {Kollath}},\ }\bibfield  {title}
  {\bibinfo {title} {{\it Propagation front of correlations in an interacting
  Bose gas}},\ }\href {https://doi.org/10.1103/PhysRevA.85.053625} {\bibfield
  {journal} {\bibinfo  {journal} {Phys. Rev. A}\ }\textbf {\bibinfo {volume}
  {85}},\ \bibinfo {pages} {053625} (\bibinfo {year} {2012})}\BibitemShut
  {NoStop}%
\bibitem [{\citenamefont {Natu}\ and\ \citenamefont
  {Mueller}(2013)}]{PhysRevA.87.063616}%
  \BibitemOpen
  \bibfield  {author} {\bibinfo {author} {\bibfnamefont {S.~S.}\ \bibnamefont
  {Natu}}\ and\ \bibinfo {author} {\bibfnamefont {E.~J.}\ \bibnamefont
  {Mueller}},\ }\bibfield  {title} {\bibinfo {title} {{\it Dynamics of
  correlations in shallow optical lattices}},\ }\href
  {https://doi.org/10.1103/PhysRevA.87.063616} {\bibfield  {journal} {\bibinfo
  {journal} {Phys. Rev. A}\ }\textbf {\bibinfo {volume} {87}},\ \bibinfo
  {pages} {063616} (\bibinfo {year} {2013})}\BibitemShut {NoStop}%
\bibitem [{\citenamefont {Carleo}\ \emph {et~al.}(2014)\citenamefont {Carleo},
  \citenamefont {Becca}, \citenamefont {Sanchez-Palencia}, \citenamefont
  {Sorella},\ and\ \citenamefont {Fabrizio}}]{PhysRevA.89.031602}%
  \BibitemOpen
  \bibfield  {author} {\bibinfo {author} {\bibfnamefont {G.}~\bibnamefont
  {Carleo}}, \bibinfo {author} {\bibfnamefont {F.}~\bibnamefont {Becca}},
  \bibinfo {author} {\bibfnamefont {L.}~\bibnamefont {Sanchez-Palencia}},
  \bibinfo {author} {\bibfnamefont {S.}~\bibnamefont {Sorella}},\ and\ \bibinfo
  {author} {\bibfnamefont {M.}~\bibnamefont {Fabrizio}},\ }\bibfield  {title}
  {\bibinfo {title} {{\it Light-cone effect and supersonic correlations in one-
  and two-dimensional bosonic superfluids}},\ }\href
  {https://doi.org/10.1103/PhysRevA.89.031602} {\bibfield  {journal} {\bibinfo
  {journal} {Phys. Rev. A}\ }\textbf {\bibinfo {volume} {89}},\ \bibinfo
  {pages} {031602} (\bibinfo {year} {2014})}\BibitemShut {NoStop}%
\bibitem [{\citenamefont {Lo~Gullo}\ and\ \citenamefont
  {Dell'Anna}(2015)}]{PhysRevA.92.063619}%
  \BibitemOpen
  \bibfield  {author} {\bibinfo {author} {\bibfnamefont {N.}~\bibnamefont
  {Lo~Gullo}}\ and\ \bibinfo {author} {\bibfnamefont {L.}~\bibnamefont
  {Dell'Anna}},\ }\bibfield  {title} {\bibinfo {title} {{\it Spreading of
  correlations and Loschmidt echo after quantum quenches of a Bose gas in the
  Aubry-Andr\'e potential}},\ }\href
  {https://doi.org/10.1103/PhysRevA.92.063619} {\bibfield  {journal} {\bibinfo
  {journal} {Phys. Rev. A}\ }\textbf {\bibinfo {volume} {92}},\ \bibinfo
  {pages} {063619} (\bibinfo {year} {2015})}\BibitemShut {NoStop}%
\bibitem [{\citenamefont {Bernier}\ \emph {et~al.}(2018)\citenamefont
  {Bernier}, \citenamefont {Tan}, \citenamefont {Bonnes}, \citenamefont {Guo},
  \citenamefont {Poletti},\ and\ \citenamefont
  {Kollath}}]{PhysRevLett.120.020401}%
  \BibitemOpen
  \bibfield  {author} {\bibinfo {author} {\bibfnamefont {J.-S.}\ \bibnamefont
  {Bernier}}, \bibinfo {author} {\bibfnamefont {R.}~\bibnamefont {Tan}},
  \bibinfo {author} {\bibfnamefont {L.}~\bibnamefont {Bonnes}}, \bibinfo
  {author} {\bibfnamefont {C.}~\bibnamefont {Guo}}, \bibinfo {author}
  {\bibfnamefont {D.}~\bibnamefont {Poletti}},\ and\ \bibinfo {author}
  {\bibfnamefont {C.}~\bibnamefont {Kollath}},\ }\bibfield  {title} {\bibinfo
  {title} {{\it Light-Cone and Diffusive Propagation of Correlations in a
  Many-Body Dissipative System}},\ }\href
  {https://doi.org/10.1103/PhysRevLett.120.020401} {\bibfield  {journal}
  {\bibinfo  {journal} {Phys. Rev. Lett.}\ }\textbf {\bibinfo {volume} {120}},\
  \bibinfo {pages} {020401} (\bibinfo {year} {2018})}\BibitemShut {NoStop}%
\bibitem [{\citenamefont {Fitzpatrick}\ and\ \citenamefont
  {Kennett}(2018)}]{PhysRevA.98.053618}%
  \BibitemOpen
  \bibfield  {author} {\bibinfo {author} {\bibfnamefont {M.~R.~C.}\
  \bibnamefont {Fitzpatrick}}\ and\ \bibinfo {author} {\bibfnamefont {M.~P.}\
  \bibnamefont {Kennett}},\ }\bibfield  {title} {\bibinfo {title} {{\it
  Light-cone-like spreading of single-particle correlations in the Bose-Hubbard
  model after a quantum quench in the strong-coupling regime}},\ }\href
  {https://doi.org/10.1103/PhysRevA.98.053618} {\bibfield  {journal} {\bibinfo
  {journal} {Phys. Rev. A}\ }\textbf {\bibinfo {volume} {98}},\ \bibinfo
  {pages} {053618} (\bibinfo {year} {2018})}\BibitemShut {NoStop}%
\bibitem [{\citenamefont {Mokhtari-Jazi}\ \emph {et~al.}(2021)\citenamefont
  {Mokhtari-Jazi}, \citenamefont {Fitzpatrick},\ and\ \citenamefont
  {Kennett}}]{PhysRevA.103.023334}%
  \BibitemOpen
  \bibfield  {author} {\bibinfo {author} {\bibfnamefont {A.}~\bibnamefont
  {Mokhtari-Jazi}}, \bibinfo {author} {\bibfnamefont {M.~R.~C.}\ \bibnamefont
  {Fitzpatrick}},\ and\ \bibinfo {author} {\bibfnamefont {M.~P.}\ \bibnamefont
  {Kennett}},\ }\bibfield  {title} {\bibinfo {title} {{\it Phase and group
  velocities for correlation spreading in the Mott phase of the Bose-Hubbard
  model in dimensions greater than one}},\ }\href
  {https://doi.org/10.1103/PhysRevA.103.023334} {\bibfield  {journal} {\bibinfo
   {journal} {Phys. Rev. A}\ }\textbf {\bibinfo {volume} {103}},\ \bibinfo
  {pages} {023334} (\bibinfo {year} {2021})}\BibitemShut {NoStop}%
\bibitem [{\citenamefont {Kollath}\ \emph {et~al.}(2007)\citenamefont
  {Kollath}, \citenamefont {L\"auchli},\ and\ \citenamefont
  {Altman}}]{PhysRevLett.98.180601}%
  \BibitemOpen
  \bibfield  {author} {\bibinfo {author} {\bibfnamefont {C.}~\bibnamefont
  {Kollath}}, \bibinfo {author} {\bibfnamefont {A.~M.}\ \bibnamefont
  {L\"auchli}},\ and\ \bibinfo {author} {\bibfnamefont {E.}~\bibnamefont
  {Altman}},\ }\bibfield  {title} {\bibinfo {title} {{\it Quench Dynamics and
  Nonequilibrium Phase Diagram of the Bose-Hubbard Model}},\ }\href
  {https://doi.org/10.1103/PhysRevLett.98.180601} {\bibfield  {journal}
  {\bibinfo  {journal} {Phys. Rev. Lett.}\ }\textbf {\bibinfo {volume} {98}},\
  \bibinfo {pages} {180601} (\bibinfo {year} {2007})}\BibitemShut {NoStop}%
\bibitem [{\citenamefont {Cramer}\ \emph
  {et~al.}(2008{\natexlab{b}})\citenamefont {Cramer}, \citenamefont {Dawson},
  \citenamefont {Eisert},\ and\ \citenamefont
  {Osborne}}]{PhysRevLett.100.030602}%
  \BibitemOpen
  \bibfield  {author} {\bibinfo {author} {\bibfnamefont {M.}~\bibnamefont
  {Cramer}}, \bibinfo {author} {\bibfnamefont {C.~M.}\ \bibnamefont {Dawson}},
  \bibinfo {author} {\bibfnamefont {J.}~\bibnamefont {Eisert}},\ and\ \bibinfo
  {author} {\bibfnamefont {T.~J.}\ \bibnamefont {Osborne}},\ }\bibfield
  {title} {\bibinfo {title} {{\it Exact Relaxation in a Class of Nonequilibrium
  Quantum Lattice Systems}},\ }\href
  {https://doi.org/10.1103/PhysRevLett.100.030602} {\bibfield  {journal}
  {\bibinfo  {journal} {Phys. Rev. Lett.}\ }\textbf {\bibinfo {volume} {100}},\
  \bibinfo {pages} {030602} (\bibinfo {year} {2008}{\natexlab{b}})}\BibitemShut
  {NoStop}%
\bibitem [{\citenamefont {Cramer}\ \emph
  {et~al.}(2008{\natexlab{c}})\citenamefont {Cramer}, \citenamefont {Flesch},
  \citenamefont {McCulloch}, \citenamefont {Schollw\"ock},\ and\ \citenamefont
  {Eisert}}]{PhysRevLett.101.063001}%
  \BibitemOpen
  \bibfield  {author} {\bibinfo {author} {\bibfnamefont {M.}~\bibnamefont
  {Cramer}}, \bibinfo {author} {\bibfnamefont {A.}~\bibnamefont {Flesch}},
  \bibinfo {author} {\bibfnamefont {I.~P.}\ \bibnamefont {McCulloch}}, \bibinfo
  {author} {\bibfnamefont {U.}~\bibnamefont {Schollw\"ock}},\ and\ \bibinfo
  {author} {\bibfnamefont {J.}~\bibnamefont {Eisert}},\ }\bibfield  {title}
  {\bibinfo {title} {{\it Exploring Local Quantum Many-Body Relaxation by Atoms
  in Optical Superlattices}},\ }\href
  {https://doi.org/10.1103/PhysRevLett.101.063001} {\bibfield  {journal}
  {\bibinfo  {journal} {Phys. Rev. Lett.}\ }\textbf {\bibinfo {volume} {101}},\
  \bibinfo {pages} {063001} (\bibinfo {year} {2008}{\natexlab{c}})}\BibitemShut
  {NoStop}%
\bibitem [{\citenamefont {Roux}(2009)}]{PhysRevA.79.021608}%
  \BibitemOpen
  \bibfield  {author} {\bibinfo {author} {\bibfnamefont {G.}~\bibnamefont
  {Roux}},\ }\bibfield  {title} {\bibinfo {title} {{\it Quenches in quantum
  many-body systems: One-dimensional Bose-Hubbard model reexamined}},\ }\href
  {https://doi.org/10.1103/PhysRevA.79.021608} {\bibfield  {journal} {\bibinfo
  {journal} {Phys. Rev. A}\ }\textbf {\bibinfo {volume} {79}},\ \bibinfo
  {pages} {021608} (\bibinfo {year} {2009})}\BibitemShut {NoStop}%
\bibitem [{\citenamefont {Navez}\ and\ \citenamefont
  {Sch\"utzhold}(2010)}]{PhysRevA.82.063603}%
  \BibitemOpen
  \bibfield  {author} {\bibinfo {author} {\bibfnamefont {P.}~\bibnamefont
  {Navez}}\ and\ \bibinfo {author} {\bibfnamefont {R.}~\bibnamefont
  {Sch\"utzhold}},\ }\bibfield  {title} {\bibinfo {title} {{\it Emergence of
  coherence in the Mott-insulator--superfluid quench of the Bose-Hubbard
  model}},\ }\href {https://doi.org/10.1103/PhysRevA.82.063603} {\bibfield
  {journal} {\bibinfo  {journal} {Phys. Rev. A}\ }\textbf {\bibinfo {volume}
  {82}},\ \bibinfo {pages} {063603} (\bibinfo {year} {2010})}\BibitemShut
  {NoStop}%
\bibitem [{\citenamefont {Enss}\ and\ \citenamefont
  {Sirker}(2012)}]{Enss_2012}%
  \BibitemOpen
  \bibfield  {author} {\bibinfo {author} {\bibfnamefont {T.}~\bibnamefont
  {Enss}}\ and\ \bibinfo {author} {\bibfnamefont {J.}~\bibnamefont {Sirker}},\
  }\bibfield  {title} {\bibinfo {title} {{\it Light cone renormalization and
  quantum quenches in one-dimensional Hubbard models}},\ }\href
  {https://doi.org/10.1088/1367-2630/14/2/023008} {\bibfield  {journal}
  {\bibinfo  {journal} {New Journal of Physics}\ }\textbf {\bibinfo {volume}
  {14}},\ \bibinfo {pages} {023008} (\bibinfo {year} {2012})}\BibitemShut
  {NoStop}%
\bibitem [{\citenamefont {Bernier}\ \emph {et~al.}(2012)\citenamefont
  {Bernier}, \citenamefont {Poletti}, \citenamefont {Barmettler}, \citenamefont
  {Roux},\ and\ \citenamefont {Kollath}}]{PhysRevA.85.033641}%
  \BibitemOpen
  \bibfield  {author} {\bibinfo {author} {\bibfnamefont {J.-S.}\ \bibnamefont
  {Bernier}}, \bibinfo {author} {\bibfnamefont {D.}~\bibnamefont {Poletti}},
  \bibinfo {author} {\bibfnamefont {P.}~\bibnamefont {Barmettler}}, \bibinfo
  {author} {\bibfnamefont {G.}~\bibnamefont {Roux}},\ and\ \bibinfo {author}
  {\bibfnamefont {C.}~\bibnamefont {Kollath}},\ }\bibfield  {title} {\bibinfo
  {title} {{\it Slow quench dynamics of Mott-insulating regions in a trapped
  Bose gas}},\ }\href {https://doi.org/10.1103/PhysRevA.85.033641} {\bibfield
  {journal} {\bibinfo  {journal} {Phys. Rev. A}\ }\textbf {\bibinfo {volume}
  {85}},\ \bibinfo {pages} {033641} (\bibinfo {year} {2012})}\BibitemShut
  {NoStop}%
\bibitem [{\citenamefont {Sorg}\ \emph {et~al.}(2014)\citenamefont {Sorg},
  \citenamefont {Vidmar}, \citenamefont {Pollet},\ and\ \citenamefont
  {Heidrich-Meisner}}]{PhysRevA.90.033606}%
  \BibitemOpen
  \bibfield  {author} {\bibinfo {author} {\bibfnamefont {S.}~\bibnamefont
  {Sorg}}, \bibinfo {author} {\bibfnamefont {L.}~\bibnamefont {Vidmar}},
  \bibinfo {author} {\bibfnamefont {L.}~\bibnamefont {Pollet}},\ and\ \bibinfo
  {author} {\bibfnamefont {F.}~\bibnamefont {Heidrich-Meisner}},\ }\bibfield
  {title} {\bibinfo {title} {{\it Relaxation and thermalization in the
  one-dimensional Bose-Hubbard model: A case study for the interaction quantum
  quench from the atomic limit}},\ }\href
  {https://doi.org/10.1103/PhysRevA.90.033606} {\bibfield  {journal} {\bibinfo
  {journal} {Phys. Rev. A}\ }\textbf {\bibinfo {volume} {90}},\ \bibinfo
  {pages} {033606} (\bibinfo {year} {2014})}\BibitemShut {NoStop}%
\bibitem [{\citenamefont {Bernier}\ \emph {et~al.}(2014)\citenamefont
  {Bernier}, \citenamefont {Citro}, \citenamefont {Kollath},\ and\
  \citenamefont {Orignac}}]{PhysRevLett.112.065301}%
  \BibitemOpen
  \bibfield  {author} {\bibinfo {author} {\bibfnamefont {J.-S.}\ \bibnamefont
  {Bernier}}, \bibinfo {author} {\bibfnamefont {R.}~\bibnamefont {Citro}},
  \bibinfo {author} {\bibfnamefont {C.}~\bibnamefont {Kollath}},\ and\ \bibinfo
  {author} {\bibfnamefont {E.}~\bibnamefont {Orignac}},\ }\bibfield  {title}
  {\bibinfo {title} {{\it Correlation Dynamics During a Slow Interaction Quench
  in a One-Dimensional Bose Gas}},\ }\href
  {https://doi.org/10.1103/PhysRevLett.112.065301} {\bibfield  {journal}
  {\bibinfo  {journal} {Phys. Rev. Lett.}\ }\textbf {\bibinfo {volume} {112}},\
  \bibinfo {pages} {065301} (\bibinfo {year} {2014})}\BibitemShut {NoStop}%
\bibitem [{\citenamefont {Geiger}\ \emph {et~al.}(2014)\citenamefont {Geiger},
  \citenamefont {Langen}, \citenamefont {Mazets},\ and\ \citenamefont
  {Schmiedmayer}}]{Geiger_2014}%
  \BibitemOpen
  \bibfield  {author} {\bibinfo {author} {\bibfnamefont {R.}~\bibnamefont
  {Geiger}}, \bibinfo {author} {\bibfnamefont {T.}~\bibnamefont {Langen}},
  \bibinfo {author} {\bibfnamefont {I.~E.}\ \bibnamefont {Mazets}},\ and\
  \bibinfo {author} {\bibfnamefont {J.}~\bibnamefont {Schmiedmayer}},\
  }\bibfield  {title} {\bibinfo {title} {{\it Local relaxation and
  light-cone-like propagation of correlations in a trapped one-dimensional Bose
  gas}},\ }\href {https://doi.org/10.1088/1367-2630/16/5/053034} {\bibfield
  {journal} {\bibinfo  {journal} {New Journal of Physics}\ }\textbf {\bibinfo
  {volume} {16}},\ \bibinfo {pages} {053034} (\bibinfo {year}
  {2014})}\BibitemShut {NoStop}%
\bibitem [{\citenamefont {Krutitsky}\ \emph {et~al.}(2014)\citenamefont
  {Krutitsky}, \citenamefont {Navez}, \citenamefont {Queisser},\ and\
  \citenamefont {Sch{\"u}tzhold}}]{Krutitsky2014}%
  \BibitemOpen
  \bibfield  {author} {\bibinfo {author} {\bibfnamefont {K.~V.}\ \bibnamefont
  {Krutitsky}}, \bibinfo {author} {\bibfnamefont {P.}~\bibnamefont {Navez}},
  \bibinfo {author} {\bibfnamefont {F.}~\bibnamefont {Queisser}},\ and\
  \bibinfo {author} {\bibfnamefont {R.}~\bibnamefont {Sch{\"u}tzhold}},\
  }\bibfield  {title} {\bibinfo {title} {{\it Propagation of quantum
  correlations after a quench in the Mott-insulator regime of the Bose-Hubbard
  model}},\ }\href {https://doi.org/10.1140/epjqt12} {\bibfield  {journal}
  {\bibinfo  {journal} {EPJ Quantum Technology}\ }\textbf {\bibinfo {volume}
  {1}},\ \bibinfo {pages} {12} (\bibinfo {year} {2014})}\BibitemShut {NoStop}%
\bibitem [{\citenamefont {Andraschko}\ and\ \citenamefont
  {Sirker}(2015)}]{PhysRevB.91.235132}%
  \BibitemOpen
  \bibfield  {author} {\bibinfo {author} {\bibfnamefont {F.}~\bibnamefont
  {Andraschko}}\ and\ \bibinfo {author} {\bibfnamefont {J.}~\bibnamefont
  {Sirker}},\ }\bibfield  {title} {\bibinfo {title} {{\it Propagation of a
  single-hole defect in the one-dimensional Bose-Hubbard model}},\ }\href
  {https://doi.org/10.1103/PhysRevB.91.235132} {\bibfield  {journal} {\bibinfo
  {journal} {Phys. Rev. B}\ }\textbf {\bibinfo {volume} {91}},\ \bibinfo
  {pages} {235132} (\bibinfo {year} {2015})}\BibitemShut {NoStop}%
\bibitem [{\citenamefont {Shen}\ \emph {et~al.}(2017)\citenamefont {Shen},
  \citenamefont {Zhang}, \citenamefont {Fan},\ and\ \citenamefont
  {Zhai}}]{PhysRevB.96.054503}%
  \BibitemOpen
  \bibfield  {author} {\bibinfo {author} {\bibfnamefont {H.}~\bibnamefont
  {Shen}}, \bibinfo {author} {\bibfnamefont {P.}~\bibnamefont {Zhang}},
  \bibinfo {author} {\bibfnamefont {R.}~\bibnamefont {Fan}},\ and\ \bibinfo
  {author} {\bibfnamefont {H.}~\bibnamefont {Zhai}},\ }\bibfield  {title}
  {\bibinfo {title} {{\it Out-of-time-order correlation at a quantum phase
  transition}},\ }\href {https://doi.org/10.1103/PhysRevB.96.054503} {\bibfield
   {journal} {\bibinfo  {journal} {Phys. Rev. B}\ }\textbf {\bibinfo {volume}
  {96}},\ \bibinfo {pages} {054503} (\bibinfo {year} {2017})}\BibitemShut
  {NoStop}%
\bibitem [{\citenamefont {Liu}\ \emph {et~al.}(2018)\citenamefont {Liu},
  \citenamefont {Garrison}, \citenamefont {Deng}, \citenamefont {Gong},\ and\
  \citenamefont {Gorshkov}}]{PhysRevLett.121.250404}%
  \BibitemOpen
  \bibfield  {author} {\bibinfo {author} {\bibfnamefont {F.}~\bibnamefont
  {Liu}}, \bibinfo {author} {\bibfnamefont {J.~R.}\ \bibnamefont {Garrison}},
  \bibinfo {author} {\bibfnamefont {D.-L.}\ \bibnamefont {Deng}}, \bibinfo
  {author} {\bibfnamefont {Z.-X.}\ \bibnamefont {Gong}},\ and\ \bibinfo
  {author} {\bibfnamefont {A.~V.}\ \bibnamefont {Gorshkov}},\ }\bibfield
  {title} {\bibinfo {title} {{\it Asymmetric Particle Transport and Light-Cone
  Dynamics Induced by Anyonic Statistics}},\ }\href
  {https://doi.org/10.1103/PhysRevLett.121.250404} {\bibfield  {journal}
  {\bibinfo  {journal} {Phys. Rev. Lett.}\ }\textbf {\bibinfo {volume} {121}},\
  \bibinfo {pages} {250404} (\bibinfo {year} {2018})}\BibitemShut {NoStop}%
\bibitem [{\citenamefont {Pietraszewicz}\ \emph {et~al.}(2019)\citenamefont
  {Pietraszewicz}, \citenamefont {Stobi\ifmmode~\acute{n}\else \'{n}\fi{}ska},\
  and\ \citenamefont {Deuar}}]{PhysRevA.99.023620}%
  \BibitemOpen
  \bibfield  {author} {\bibinfo {author} {\bibfnamefont {J.}~\bibnamefont
  {Pietraszewicz}}, \bibinfo {author} {\bibfnamefont {M.}~\bibnamefont
  {Stobi\ifmmode~\acute{n}\else \'{n}\fi{}ska}},\ and\ \bibinfo {author}
  {\bibfnamefont {P.}~\bibnamefont {Deuar}},\ }\bibfield  {title} {\bibinfo
  {title} {Correlation evolution in dilute bose-einstein condensates after
  quantum quenches},\ }\href {https://doi.org/10.1103/PhysRevA.99.023620}
  {\bibfield  {journal} {\bibinfo  {journal} {Phys. Rev. A}\ }\textbf {\bibinfo
  {volume} {99}},\ \bibinfo {pages} {023620} (\bibinfo {year}
  {2019})}\BibitemShut {NoStop}%
\bibitem [{\citenamefont {Despres}\ \emph {et~al.}(2019)\citenamefont
  {Despres}, \citenamefont {Villa},\ and\ \citenamefont
  {Sanchez-Palencia}}]{Despres2019}%
  \BibitemOpen
  \bibfield  {author} {\bibinfo {author} {\bibfnamefont {J.}~\bibnamefont
  {Despres}}, \bibinfo {author} {\bibfnamefont {L.}~\bibnamefont {Villa}},\
  and\ \bibinfo {author} {\bibfnamefont {L.}~\bibnamefont {Sanchez-Palencia}},\
  }\bibfield  {title} {\bibinfo {title} {{\it Twofold correlation spreading in
  a strongly correlated lattice Bose gas}},\ }\href
  {https://doi.org/10.1038/s41598-019-40679-3} {\bibfield  {journal} {\bibinfo
  {journal} {Scientific Reports}\ }\textbf {\bibinfo {volume} {9}},\ \bibinfo
  {pages} {4135} (\bibinfo {year} {2019})}\BibitemShut {NoStop}%
\bibitem [{\citenamefont {Villa}\ \emph {et~al.}(2020)\citenamefont {Villa},
  \citenamefont {Despres}, \citenamefont {Thomson},\ and\ \citenamefont
  {Sanchez-Palencia}}]{PhysRevA.102.033337}%
  \BibitemOpen
  \bibfield  {author} {\bibinfo {author} {\bibfnamefont {L.}~\bibnamefont
  {Villa}}, \bibinfo {author} {\bibfnamefont {J.}~\bibnamefont {Despres}},
  \bibinfo {author} {\bibfnamefont {S.~J.}\ \bibnamefont {Thomson}},\ and\
  \bibinfo {author} {\bibfnamefont {L.}~\bibnamefont {Sanchez-Palencia}},\
  }\bibfield  {title} {\bibinfo {title} {{\it Local quench spectroscopy of
  many-body quantum systems}},\ }\href
  {https://doi.org/10.1103/PhysRevA.102.033337} {\bibfield  {journal} {\bibinfo
   {journal} {Phys. Rev. A}\ }\textbf {\bibinfo {volume} {102}},\ \bibinfo
  {pages} {033337} (\bibinfo {year} {2020})}\BibitemShut {NoStop}%
\bibitem [{Sup()}]{Supplement_boson}%
  \BibitemOpen
  \href@noop {} {\bibinfo  {journal} {See Supplemental Material for the details
  of the rigorous proof of the main statements, which includes
  Refs.~\cite{Arad_connecting_global_local_dist, Cramer_2006}.
  \cite{Arad_connecting_global_local_dist} [1st reference in Supplemental
  Material not already in paper], \cite{Cramer_2006} [8th reference in
  Supplemental Material not already in paper]}\ }\BibitemShut {NoStop}%
\bibitem [{\citenamefont {Kuwahara}(2016)}]{Kuwahara_2016_njp}%
  \BibitemOpen
\bibfield  {journal} {  }\bibfield  {author} {\bibinfo {author} {\bibfnamefont
  {T.}~\bibnamefont {Kuwahara}},\ }\bibfield  {title} {\bibinfo {title} {{\it
  Exponential bound on information spreading induced by quantum many-body
  dynamics with long-range interactions}},\ }\href
  {https://doi.org/10.1088/1367-2630/18/5/053034} {\bibfield  {journal}
  {\bibinfo  {journal} {New Journal of Physics}\ }\textbf {\bibinfo {volume}
  {18}},\ \bibinfo {pages} {053034} (\bibinfo {year} {2016})}\BibitemShut
  {NoStop}%
\bibitem [{\citenamefont {Yin}\ and\ \citenamefont
  {Lucas}(2021)}]{yin2021finite}%
  \BibitemOpen
  \bibfield  {author} {\bibinfo {author} {\bibfnamefont {C.}~\bibnamefont
  {Yin}}\ and\ \bibinfo {author} {\bibfnamefont {A.}~\bibnamefont {Lucas}},\
  }\href@noop {} {\bibinfo {title} {Finite speed of quantum information in
  models of interacting bosons at finite density}} (\bibinfo {year} {2021}),\
  \Eprint {https://arxiv.org/abs/2106.09726} {arXiv:2106.09726 [quant-ph]}
  \BibitemShut {NoStop}%
\bibitem [{\citenamefont {Arad}\ \emph {et~al.}(2016)\citenamefont {Arad},
  \citenamefont {Kuwahara},\ and\ \citenamefont
  {Landau}}]{Arad_connecting_global_local_dist}%
  \BibitemOpen
  \bibfield  {author} {\bibinfo {author} {\bibfnamefont {I.}~\bibnamefont
  {Arad}}, \bibinfo {author} {\bibfnamefont {T.}~\bibnamefont {Kuwahara}},\
  and\ \bibinfo {author} {\bibfnamefont {Z.}~\bibnamefont {Landau}},\
  }\bibfield  {title} {\bibinfo {title} {Connecting global and local energy
  distributions in quantum spin models on a lattice},\ }\href
  {https://doi.org/10.1088/1742-5468/2016/03/033301} {\bibfield  {journal}
  {\bibinfo  {journal} {Journal of Statistical Mechanics: Theory and
  Experiment}\ }\textbf {\bibinfo {volume} {2016}},\ \bibinfo {pages} {033301}
  (\bibinfo {year} {2016})}\BibitemShut {NoStop}%
\bibitem [{\citenamefont {Cramer}\ and\ \citenamefont
  {Eisert}(2006)}]{Cramer_2006}%
  \BibitemOpen
  \bibfield  {author} {\bibinfo {author} {\bibfnamefont {M.}~\bibnamefont
  {Cramer}}\ and\ \bibinfo {author} {\bibfnamefont {J.}~\bibnamefont
  {Eisert}},\ }\bibfield  {title} {\bibinfo {title} {Correlations, spectral gap
  and entanglement in harmonic quantum systems on generic lattices},\ }\href
  {https://doi.org/10.1088/1367-2630/8/5/071} {\bibfield  {journal} {\bibinfo
  {journal} {New Journal of Physics}\ }\textbf {\bibinfo {volume} {8}},\
  \bibinfo {pages} {71} (\bibinfo {year} {2006})}\BibitemShut {NoStop}%
\end{thebibliography}%

\renewcommand\thefootnote{*\arabic{footnote}}

\clearpage
\newpage

%\counterwithout{section}{section}
\addtocounter{section}{0}

%\counterwithout{equation}{section}
\addtocounter{equation}{-19}

\renewcommand{\theequation}{S.\arabic{equation}}

\renewcommand{\thesection}{S.\Roman{section}}
\begin{widetext}

\begin{center}
{\large \bf Supplementary Material for  ``Lieb-Robinson bound and almost-linear light-cone in interacting boson systems''}\\
\vspace*{0.3cm}
Tomotaka Kuwahara$^{1}$, Keiji Saito$^{2}$ \\
\vspace*{0.1cm}
$^{1}${\small \it Mathematical Science Team, RIKEN Center for Advanced Intelligence Project (AIP),1-4-1 Nihonbashi, Chuo-ku, Tokyo 103-0027, Japan}\\
%$^{2}$Department of Mathematics, Faculty of Science and Technology, Keio University, 3-14-1 Hiyoshi, Kouhoku-ku, Yokohama 223-8522, Japan \protect \\
%$^{2}${\small \it Interdisciplinary Theoretical \& Mathematical Sciences Program (iTHEMS) RIKEN 2-1, Hirosawa, Wako, Saitama 351-0198, Japan} \\
$^{2}${\small \it Department of Physics, Keio University, Yokohama 223-8522, Japan} 
\end{center}

\tableofcontents

%%%%%%%%%%%%%%%%%%%%%%%%%%%%%%%%%%%%%%%%%%%%%%%%%%%%%%%%%%%%%%%%%%%%%%%%%%

\section{Set up}
We first describe the setup, which is described in less detail in the main text.

We consider a quantum system on a $D$-dimensional lattice (graph),
% with $n$ sites, 
where bosons interact with each other. 
An unbounded number of bosons occupy each site, and hence the local Hilbert dimension is infinitely large.
We define $d_G$ as the maximum degree of the lattice (graph). 
%The total Hilbert space dimension is now given by $(N+1)^n$.
We denote the set of total sites by $\Lambda$. For an arbitrary partial set $X\subseteq \Lambda$, we denote the cardinality (i.e., the number of sites contained in $X$) by $|X|$.

For arbitrary subsets $X, Y \subseteq \Lambda$, we define $\dist_{X,Y}$ as the shortest path length on the graph that connects $X$ and $Y$; that is, if $X\cap Y \neq \emptyset$, $\dist_{X,Y}=0$. 
When $X$ contains only one element (i.e., $X=\{i\}$), we abbreviate $\dist_{\{i\},Y}$ as $\dist_{i,Y}$ for simplicity.
We also denote the complementary subset and surface subset of $X$ by $X^\co := \Lambda\setminus X$ and $\partial X:=\{ i\in X| \dist_{i,X^\co}=1\}$, respectively.
We also define $\diam(X)$ as follows: 
\begin{align}
\diam(X):  =1+ \max_{i,j\in X} (\dist_{i,j}).
\end{align}
For a subset $X\subseteq \Lambda$, we define the extended subset $\bal{X}{r}$ as
\begin{align}
\bal{X}{r}:= \{i\in \Lambda| \dist_{X,i} \le r \} , \label{def:bal_X_r}
\end{align}
where $\bal{X}{0}=X$, and $r$ is an arbitrary positive number (i.e., $r\in \mathbb{R}^+$).
From the definition, $i[r]$ is given by a ball region with radius $r$ centered at the site $i$. 
We introduce a geometric parameter $\gamma$, which is determined only by the lattice structure.
We define $\gamma \ge 1$ as a constant of $\orderof{1}$ that satisfies the following inequalities:
\begin{align}
|\bal{i}{r}| \le \gamma  r^D \quad  (r\ge 1) .\label{def:parameter_gamma}  
\end{align}

We define the constant $\lambda_0$ as  follows:
\begin{align}
\label{def_gamma_0}
\lambda_0:= \max_{i\in \Lambda} \sum_{j\in \Lambda}  e^{-\dist_{i,j}} .
\end{align}
The parameter $\lambda_0$ depends on $\gamma$ and $D$ because, when the parameter $\gamma$ is used, it is upper-bounded as 
\begin{align}
\sum_{j\in \Lambda}  e^{-\dist_{i,j}} &=1+ \sum_{x=1}^\infty \sum_{j: \dist_{i,j}=x} e^{-x} \notag \\
&\le 1 +  \gamma \sum_{x=1}^\infty x^D e^{-x}  \le1 + \gamma  \int_0^\infty x^D e^{-x+1}dx = 1+ e \gamma D! ,
\end{align}
where we use the inequality $\# \{j: \dist_{i,j}=x\} \le \# \{j: \dist_{i,j}\le x\} = |i[x]| \le \gamma x^D$.

\subsection{Boson operators}

We define $b_i$ and $b_i^\dagger$ as the boson annihilation and creation operators, respectively.
We also define $\nb_i$ as the boson number operator on site $i$, $\nb_i:=b_i^\dagger b_i$.
We denote the boson number on a subset $X$ by $\nb_X$ as follows: 
\begin{align}
\nb_X = \sum_{i\in X} \nb_i .
\end{align}
For an arbitrary subset $X\subseteq \Lambda$, we define $\Pi_{X,q}$ as the projection onto the eigenspace of $\nb_X$ with eigenvalue $q$:
\begin{align}
\label{def_Pi_X_q}
\nb_X \Pi_{X,q} = q \Pi_{X,q}.
\end{align}
When $X$ includes only one site (i.e., $X=\{i\}$), we denote $\Pi_{\{i\},q}$ by $\Pi_{i,q}$ for simplicity. 
We also define $\Pi_{X, \ge q}$ as $\sum_{q'=q}^\infty \Pi_{X,q'}$.

\subsection{Bose-Hubbard type Hamiltonian}

We consider a Hamiltonian of the form
\begin{align}
&H= H_0+ V , \notag \\
&H_0:= \sum_{\langle i, j \rangle} J_{i,j} (b_i b_j^\dagger +{\rm h.c.} ), \quad   V:=  \sum_{Z \subset \Lambda: |Z|\le k} v_Z ,\notag \\ 
&|J_{i,j}| \le \bar{J} , \label{def:Ham}
\end{align}
where $\sum_{\langle i, j \rangle}$ represents the summation of all the pairs of adjacent sites $\{i,j\}$ on the lattice, 
and $v_Z$ represents boson--boson interactions in subset $Z$.
We assume that $v_Z$ is now given by a function of the number operators $\{\nb_i\}_{i\in Z}$.
For example, for $Z=\{i,j\}$, $v_Z$ includes interactions such as $\nb_i^4$, $\nb_i^2e^{\nb_j}$, $e^{\nb_i^2\nb_j^3}$, and so on.
The simplest example is the Bose--Hubbard model: 
\begin{align}
H= \sum_{\langle i, j \rangle}J (b_i b_j^\dagger +{\rm h.c.} ) + \frac{U}{2} \sum_{i\in \Lambda} \nb_i(\nb_i-1)
-\mu\sum_{i\in \Lambda} \nb_i  , \label{def:Ham_BH}
\end{align}
where $U$ and $\mu$ are $\orderof{1}$ constants.

For an arbitrary subset $X \subseteq \Lambda$, we define the subset Hamiltonians $H_{0,X}$, $V_X$, and $H_X$ as follows:
\begin{align}
&H_X= H_{0,X}+ V_X , \notag \\
&H_{0,X}:= \sum_{i,j \in X} [ J_{i,j} (b_i b_j^\dagger +{\rm h.c.} )], \quad   V_X:=  \sum_{Z \subseteq X: |Z|\le k} v_Z .
 \label{def:Ham_subset}
\end{align}
Note that they are supported on the subset $X$.

For an arbitrary operator $O$, we denote the time evolution by an operator $A$ as $O(A,t)$ as follows:
\begin{align}
O(A,t) := e^{iAt} O e^{-iA t}.
\end{align}
In particular, when $A=H$, we often abbreviate $O(H,t)$ as $O(t)$ for simplicity. 

We define $t_0$ as the unit of time, which is an $\orderof{1}$ constant; for example, we can choose $t_0=1$.

\subsection{Initial condition for the boson density}

We here define the condition of low boson density as follows.
\begin{assump}[Low boson density] \label{only_the_assumption_initial}
For a quantum state $\sigma$, we say that the state $\sigma$ satisfies the low-boson-density condition if 
there exist $\orderof{1}$ constants $c_{0}$ and $\bar{q}$ such that 
\begin{align}
\label{condition_for_moment_generating}
\max_{i\in \Lambda} \tr (e^{c_{0} (\nb_i -\bar{q}) } \sigma) \le  1 \quad (c_{0} \le 1).
\end{align}
\end{assump}

This condition ensures that the probability that many bosons are concentrated on one site is exponentially small.
Indeed, the probability that $\sigma$ has more than $q$ ($q\in \mathbb{N}$) bosons on a site $i$ is upper-bounded by
\begin{align}
\label{ineq:condition_for_moment_generating}
\tr (\Pi_{i,\ge q} \rho_0) 
&= \tr \br{ \Pi_{i,\ge q}e^{-c_{0} (\nb_i -\bar{q})/2}   e^{c_{0} (\nb_i -\bar{q})/2} \rho_0 e^{c_{0} (\nb_i -\bar{q})/2} e^{-c_{0} (\nb_i -\bar{q})/2} \Pi_{i,\ge q} } \notag \\   
&\le \| \Pi_{i,\ge q}e^{-c_{0} (\nb_i -\bar{q})/2}\|^2 \cdot \tr (e^{c_{0} (\nb_i -\bar{q})}\rho_0) \le e^{-c_{0} (q -\bar{q})}  .
\end{align}
In the first inequality, we use the inequality $\tr (O^\dagger A O ) \le  \|O\|^2\tr (A)$ for an arbitrary positive operator $A$. 
Therefore, the probability decays exponentially beyond $q\approx \bar{q}$.

{\bf Remark.} 
This condition is expected to hold in real experimental setups, although it would be difficult to prove rigorously in general. 
Quantum Gibbs states are among the candidates that satisfy the condition; 
as a trivial example, our theory can be applied to the infinite-temperature state.  
If we specify a setup (e.g., that there exists a repulsive force between bosons), we could prove the condition~\eqref{condition_for_moment_generating} for low-energy states by employing the techniques in~\cite{Arad_connecting_global_local_dist}.

%The condition is expected to hold in real experimental setups although it would be difficult to rigorously prove in general.
%If we restrict a specific setup (e.g., there exists a force of repulsion between bosons), we would be able to prove the inequality~\eqref{condition_for_moment_generating} for low-energy states by employing the techniques in Ref.~\cite{Arad_connecting_global_local_dist}.   
%
%\begin{align}
%N_0 := \max_{i\in \Lambda} \tr (\nb_i \rho_0 )  
%\end{align}
%Note that $n_0$ has a value of $\bar{q}+\orderof{1/c_{0}}$.
%

\section{Main result: Lieb--Robinson bound for interacting bosons}

Let $\rho_0$ be a time-independent quantum state, $[\rho_0,H]=0$.
We then consider the propagation of a perturbation to $\rho_0$ as 
\begin{align}
\rho = O_{X_0} \rho_0 O_{X_0}^\dagger ,\quad X\subseteq i_0[r_0].
\end{align}
We are now interested in how fast this perturbation propagates. 
After time evolution, $\rho(t)$ is given mathematically by $O_{X_0}(t) \rho_0 O_{X_0}(t)^\dagger$, 
and hence we need to estimate the approximation error of 
\begin{align}
O_{X_0}(t) \rho_0 \approx O^{(t)}_{i_0[R]} \rho_0  ,
\end{align}
where $O^{(t)}_{i_0[R]}$ is an appropriate operator supported on the subset $i_0[R]$. 
We aim to upper-bound the error as a function of $R$ (Fig.~\ref{fig_main_theorem}).

In deriving the Lieb--Robinson bound, we need to assume that boson creation by $O_{X_0}$ is not infinitely large; thus, we adopt the following condition:
\begin{align}
\label{cond_u_X_0_spect}
\| \Pi_{X_0,q}O_{X_0}\Pi_{X_0,q'} \| =0 \for q'> q+q_0 
\end{align}
with $q_0=\orderof{1}$,
where the projection $\Pi_{X_0,q}$ has been defined by Eq.~\eqref{def_Pi_X_q}. 
The above condition implies that the number of bosons created by $O_{X_0}$ is less than or equal to $q_0$.
We notice that the above condition also implies 
\begin{align}
\| \Pi_{X,q}O_{X_0}\Pi_{X,q'} \| =0 \for q'> q+q_0
\end{align}
for an arbitrary subset $X\supseteq X_0$\footnote{If we consider a subset $X_1\subset X_0$, it may break down. 
For example, for $X_0= X_1 \sqcup X_2$, let us consider an eigenstate $\ket{q_{X_1}, q_{X_2}}$, where $\nb_{X_1}\ket{q_{X_1}, q_{X_2}}=q_{X_1}\ket{q_{X_1}, q_{X_2}}$, $\nb_{X_2}\ket{q_{X_1}, q_{X_2}}=q_{X_2}\ket{q_{X_1}, q_{X_2}}$, and $q_{X_1}+q_{X_2}=q$. If we have $O_{X_0} \ket{q_{X_1}, q_{X_2}} =\ket{q_{X_1}+q_1, q_{X_2}-q_2}$ with $q_1-q_2=q_0$, the operator $O_{X_0}$ satisfies Eq.~\eqref{cond_u_X_0_spect}. 
However, for $q_2>0$, the equation 
\begin{align}
\| \Pi_{X_1,q}O_{X_0}\Pi_{X_1,q'} \| =0 \for q'> q+q_0
\end{align}
does not hold.}.

Our main result gives the Lieb--Robinson bound for an arbitrary quantum state $\rho_0$ satisfying the low-boson-density condition~\eqref{only_the_assumption_initial}.
\begin{theorem} \label{main_theorem_long_time_LR}
Let $O_{X_0}$ be an arbitrary operator supported on a subset $X_0\subseteq i_0[r_0]$ ($i_0\in \Lambda$).
We assume that the number of bosons created by $O_{X_0}$ is finite, as in Eq.~\eqref{cond_u_X_0_spect}. 
Then, for an arbitrary steady quantum state $\rho_0$ satisfying the assumption~\eqref{only_the_assumption_initial}, 
the operator $O_{X_0}(t) \rho_0$ is approximated by using another operator $O^{(t)}_{i_0[R]}$ supported on $i_0[R]$ with the following approximation error:
\begin{align}
\label{main_theorem_short_time_LR_main_ineq}
&\left \| \br{ O_{X_0}(t)-O^{(t)}_{i_0[R]} }\rho_0 \right\|_1 \le \| O_{X_0}\| \exp\br{c_0\bar{q}- C_1 \frac{(R-r_0)}{t\log (R)} + C_2 \log(R)} \quad (t\ge 1) ,
\end{align}
where $C_1$ and $C_2$ are constants of $\orderof{1}$ which are independent of $\bar{q}$ and depend only on the details of the system. 
\end{theorem}

 \begin{figure}[tt]
\centering
\includegraphics[clip, scale=0.5]{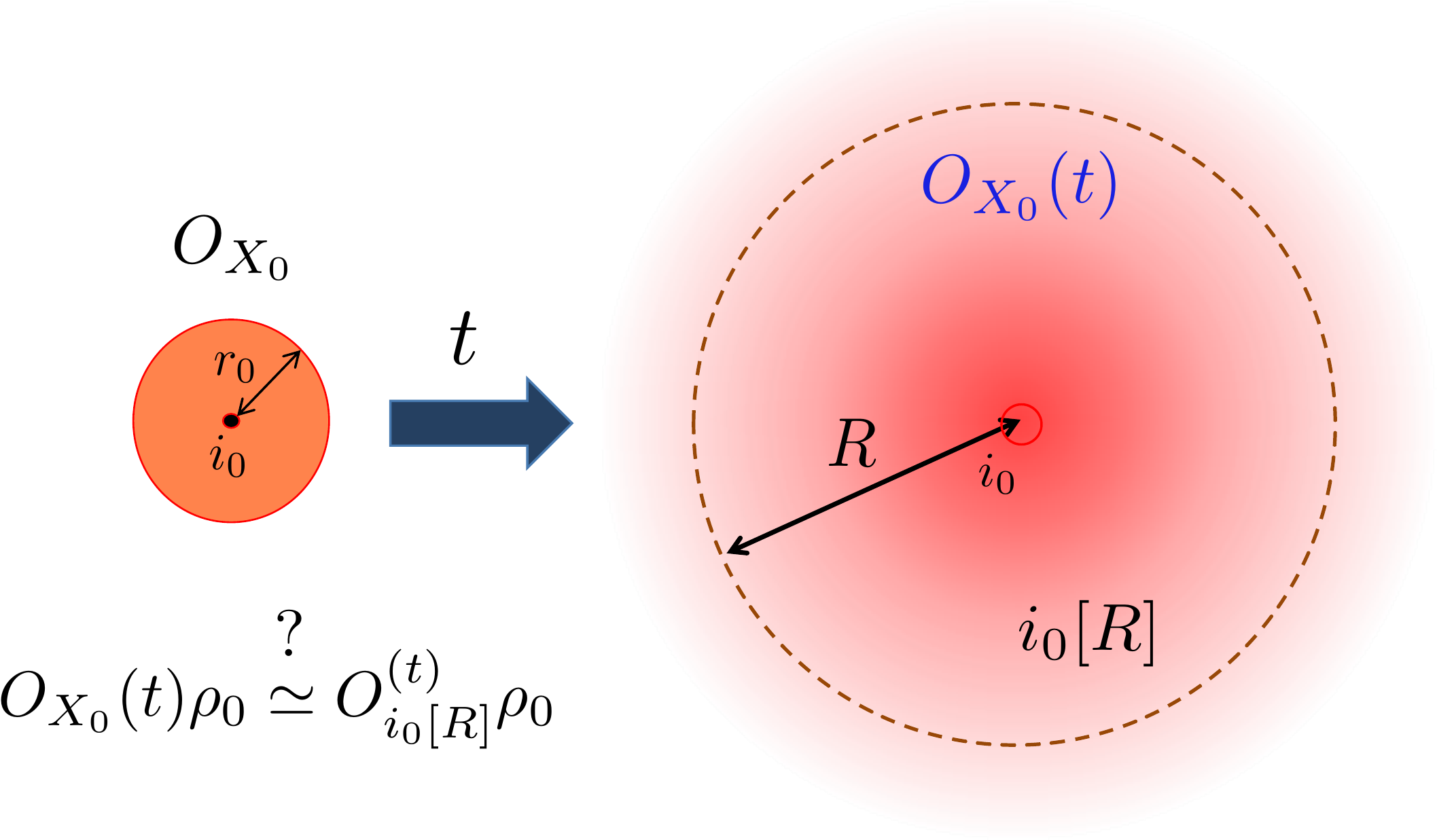}
\caption{Schematic illustration of Theorem~\ref{main_theorem_long_time_LR}.
}
\label{fig_main_theorem}
\end{figure}

{\bf Remark.}
From the above result, we can identify the shape of the effective light cone. 
We here assume $t\ge e$ and $X_0=i_0$ ($r_0=0$) for simplicity.
As shown in the main text, the velocity $ v_{t,\delta}$ to characterize the effective light cone has been defined by the following inequality:
 \begin{align}
&\| [O_{i_0}(t)-O^{(t)}_{i_0[R]} ]\|\le \delta \norm{O_{i_0}} \for R\ge v_{t,\delta} |t|. \label{def:linear_light_cone_supp}
\end{align}
In order to achieve the above inequality for a particular choices of $t$ and $\delta$, we need to choose $R$ such that 
 \begin{align}
&\exp\br{c_0\bar{q}- C_1 \frac{R}{t\log (R)} + C_2 \log(R)} \le \delta  \notag \\
&\longrightarrow C_1 \frac{R}{t\log(R)} -C_2 \log(R) \ge c_0 \bar{q} + \log(1/\delta) \longrightarrow \frac{R}{t\log(R)} - \tilde{C}_2 \log(R) \ge Q ,
\end{align}
where we define $\tilde{C}_2:=C_2/C_1$ and $Q=[c_0 \bar{q} + \log(1/\delta)]/C_1$. 
We now choose $R=\Gamma t \log^2 (t)$ by using a parameter $\Gamma$ ($\ge 1$). 
Then, from $\Gamma t \le R \le \Gamma t^2$ for $t\ge e$, we obtain the condition of
 \begin{align}
 \label{cond_Gamma_cone}
\Gamma \log (t) - 2\tilde{C}_2 \log(t) - \tilde{C}_2 \log(\Gamma) \ge
 Q.
\end{align}
Let $\Gamma_c$ be a constant depending on $\tilde{C}_2$ such that $(1/2)\Gamma \ge \tilde{C}_2 \log(\Gamma)$ for $\Gamma \ge \Gamma_c$. 
Then, for $\Gamma \ge \Gamma_c$, the condition~\eqref{cond_Gamma_cone} reduces to
 \begin{align}
\Gamma (\log (t) -1/2)   \ge 2\tilde{C}_2 \log(t) +Q  
\longrightarrow \Gamma \ge \frac{2\tilde{C}_2 \log(t) +Q }{\log (t) -1/2}
\longrightarrow \Gamma \ge 4\tilde{C}_2 +2Q ,
\end{align}
where the last inequality results from $t\ge e$. 
Therefore, by choosing 
 \begin{align}
 \label{cond_Gamma_cone}
R=t \log^2 (t) \max(\Gamma_c, 4\tilde{C}_2 +2Q) =t \log^2 (t) \max \left(\Gamma_c , 4C_2/C_1+2[c_0 \bar{q} + \log(1/\delta)]/C_1 \right),
\end{align}
the inequality~\eqref{def:linear_light_cone_supp} is satisfied.
We thus conclude that $v_{t,\delta} \propto \log^2 (t) [c_0\bar{q} + \log(1/\delta)]$.

\section{Clustering theorem for gapped ground states}

As a direct application of Theorem~\ref{main_theorem_long_time_LR}, we can discuss the clustering property of the bipartite correlations in gapped ground states.
Let $\{\ket{E_j}\}_{j\ge 0}$ be eigenstates of the Hamiltonian $H$. 
We set $E_0=0$ and define $\Delta E:=E_1$ as the spectral gap between the ground and first excited energies. 
We can prove the following corollary.
\begin{corol} \label{clustering_boson_corol}
Let $O_X$ and $O_Y$ be arbitrary operators satisfying the condition~\eqref{cond_u_X_0_spect} for a fixed $q_0$. 
Also, we here assume that $\bar{q}=\orderof{1}$. 
Then, for the ground states $\ket{E_0}$ with low boson density according to Eq.~\eqref{only_the_assumption_initial}, the bipartite correlation between $O_X$ and $O_Y$ satisfies the inequality 
\begin{align}
\label{ineq:clustering_boson_corol}
{\rm Cor}(O_X,O_Y):= \bra{E_0}O_XO_Y\ket{E_0} - \bra{E_0}O_X\ket{E_0} \bra{E_0}O_Y\ket{E_0} \le 
C_3 \cdot\|O_X\| \cdot \|O_Y\| \cdot \exp\br{-\sqrt{\frac{C_3' \Delta E}{ \log(R)}R}} 
\end{align}
for $R\ge {\rm const.} \times (1/\Delta E) \log^3(1/\Delta E)$.
Here, $C_3$ and $C_3'$ are $\orderof{1}$ constants.
\end{corol}

{\bf Remark.} 
According to this corollary, the correlation decays sub-exponentially with $R$ that is larger than $\tilde{\mathcal{O}}(1/\Delta E)$.
This corollary is weaker than the standard clustering theorem~\cite{ref:Hastings2004-Markov,ref:Hastings2006-ExpDec,ref:Nachtergaele2006-LR}, which yields exponential decay of the bipartite correlations as $e^{-\orderof{\Delta E R}}$. 
The primary reason is that the asymptotic form of the Lieb--Robinson bound in Theorem~\ref{main_theorem_long_time_LR} is given by 
$e^{-\orderof{R/(t\log R)}}$ instead of $e^{-\orderof{R- vt}}$.
This point is reflected in the choice of $T$ in Eqs.~\eqref{start_clustering_Lieb_Robisnon_4}, \eqref{start_clustering_Lieb_Robisnon_5}, and \eqref{start_clustering_Lieb_Robisnon_6} below.
For example, if we could improve the present upper bound to $e^{-\orderof{R/(\log R) -vt}}$, 
we would be able to obtain nearly exponential decay of the bipartite correlations as $e^{-\orderof{\Delta E R/(\log R)}}$.

\subsection{Proof of Corollary~\ref{clustering_boson_corol}}
The proof relies on the method in Refs.~\cite{ref:Hastings2004-Markov,ref:Hastings2006-ExpDec,ref:Nachtergaele2006-LR}. 
We start with the equation 
\begin{align}
\bra{E_0} [O_X(t), O_Y] \ket{E_0}
&= \sum_{s\ge 0}\br{ e^{-iE_s t}\bra{E_0} O_X \ket{E_s}\bra{E_s} O_Y \ket{E_0} -e^{iE_st}\bra{E_0} O_Y \ket{E_s} \bra{E_s} O_X \ket{E_0}}   \notag \\
&=\sum_{s\ge 1} \br{ e^{-iE_s t}\bra{E_0} O_X \ket{E_s}\bra{E_s} O_Y \ket{E_0}  - e^{iE_s t}\bra{E_0} O_Y \ket{E_s} \bra{E_s} O_X \ket{E_0}}.  
\label{start_clustering_Lieb_Robisnon}
\end{align}
Using the function $K(t)$, where
\begin{align}
K(t)= \frac{i}{2\pi} \lim_{\epsilon\to +0} \frac{e^{-\alpha t^2}}{t+i\epsilon} ,
\end{align}
we have, from Ref.~\cite{ref:Hastings2006-ExpDec},
\begin{align}
\int_{-\infty}^\infty e^{-i\omega t} K(t) dt =\begin{cases} 
 {\rm const} \cdot e^{-\omega^2/(4\alpha)} &\for \omega>0,\notag \\
 1+  {\rm const} \cdot e^{-\omega^2/(4\alpha)} &\for \omega<0. \end{cases} 
\end{align}
Here, $\alpha$ is a free parameter. 
Because $K(t)$ decays as $e^{-\alpha t^2}$, we can obtain 
\begin{align}
\int_{-T}^Te^{-i\omega t} K(t) dt &=\begin{cases} 
 {\rm const} \cdot \br{ e^{-\omega^2/(4\alpha)} +c_{\alpha,T} e^{-\alpha T^2}} &\for \omega>0,  \\
 1+  {\rm const} \cdot \br{ e^{-\omega^2/(4\alpha)} + c_{\alpha,T}  e^{-\alpha T^2}}  &\for \omega<0, \end{cases} \notag \\
 &=\begin{cases} 
 {\rm const} \cdot ( e^{- T\omega^2/(2\Delta E) } + c_{\Delta E,T} e^{-T\Delta E /2}  )&\for \omega>0,  \\
 1+  {\rm const} \cdot ( e^{- T\omega^2/(2\Delta E) } +c_{\Delta E,T}  e^{-T\Delta E /2}  )  &\for \omega<0, \end{cases}  
 \label{Eq_K_t_integral_T}
\end{align}
where $c_{\alpha,T}$ and $c_{\Delta E,T}$ are  appropriate constants, and we choose $\alpha=\Delta E/(2T)$.
We note that $|c_{\Delta E,T}| \le {\rm const.} /\sqrt{\alpha} = {\rm const.} \sqrt{2T/\Delta E}$.

By applying Eq.~\eqref{Eq_K_t_integral_T} to Eq.~\eqref{start_clustering_Lieb_Robisnon}, we obtain 
\begin{align}
&\int_{-T}^T K(t) \bra{E_0} [O_X(t), O_Y] \ket{E_0} dt   \notag \\
&= \int_{-T}^T  K(t) \sum_{s\ge 1} \br{ e^{-iE_s t}\bra{E_0} O_X \ket{E_s}\bra{E_s} O_Y \ket{E_0}  - e^{iE_s t}\bra{E_0} O_Y \ket{E_s} \bra{E_s} O_X \ket{E_0}}  dt 
\notag \\
&=\sum_{s\ge 1} \left[ K_{E_s,\Delta E}   \bra{E_0} O_X \ket{E_s}\bra{E_s} O_Y \ket{E_0} 
+(1+ K_{E_s,\Delta E}  ) \bra{E_0} O_Y \ket{E_s} \bra{E_s} O_X \ket{E_0}  \right] \notag \\
&= ( \bra{E_0} O_X Q_T O_Y \ket{E_0} + {\rm c.c.})  
+\sum_{s\ge 1} \bra{E_0} O_Y \ket{E_s} \bra{E_s} O_X \ket{E_0},
\label{start_clustering_Lieb_Robisnon_2}  
\end{align}
where we set $K_{\omega,\Delta E} := {\rm const} \cdot ( e^{- T\omega^2/(2\Delta E) } +c_{\Delta E,T}  e^{-T\Delta E /2}  )$ and define $Q_T$ as 
\begin{align}
\label{start_clustering_Lieb_Robisnon_3}
Q_T := \sum_{s\ge 1} K_{E_s,\Delta E} \ket{E_s}\bra{E_s} . 
\end{align}
By combining Eqs.~\eqref{Eq_K_t_integral_T} and \eqref{start_clustering_Lieb_Robisnon_2}, we obtain
 \begin{align}
 \label{start_clustering_Lieb_Robisnon_4}
\left | {\rm Cor}(O_X,O_Y)\right | 
&\le  \int_{-T}^T K(t) |\bra{E_0} [O_X(t), O_Y] \ket{E_0} |dt   + {\rm const} \cdot c_{\Delta E,T}\|O_X\| \cdot \|O_Y\| \cdot e^{-T\Delta E/2} ,
\end{align}
where we use $\sum_{s\ge 1} \bra{E_0} O_Y \ket{E_s} \bra{E_s} O_X \ket{E_0}={\rm Cor}(O_X,O_Y)$ and $\|Q_T\| \le K_{E_1,\Delta E}= {\rm const} \cdot c_{\Delta E,T} e^{-T\Delta E/2}$ from $\Delta E=E_1$.
From Theorem~\ref{main_theorem_long_time_LR}, we have 
 \begin{align}
| \bra{E_0} [O_X(t), O_Y] \ket{E_0} | \le  \| [O_X(t), O_Y] \|\le  \|O_X\| \cdot \| O_Y\|  \cdot \exp\br{- {\rm const} \cdot  \frac{R}{t\log (R)}} 
\end{align}
for $R\ge {\rm const} \cdot t \log^2 (t)$.  
Hence, for $R\ge {\rm const}\cdot T\log^2(T)$, we obtain
\begin{align}
\label{start_clustering_Lieb_Robisnon_5}
\int_{-T}^T K(t) |\bra{E_0} [O_X(t), O_Y] \ket{E_0}| dt \le  {\rm const} \cdot \|O_X\| \cdot \| O_Y\|  \cdot T \exp\br{- {\rm const} \cdot  \frac{R}{T\log (R)}}  .
\end{align}
We thus choose $T$ as 
 \begin{align}
 \label{start_clustering_Lieb_Robisnon_6}
T= \sqrt{\frac{{\rm const} \cdot R}{\Delta E \log(R)}} ,
\end{align}
and we obtain the main inequality~\eqref{ineq:clustering_boson_corol}. 
Finally, the condition $R\ge {\rm const}\cdot T\log^2(T)$ reduces to
\begin{align}
R\log R \ge {\rm const} \cdot \frac{1}{\Delta E} [ \log^4(R) + \log^4(1/\Delta E) ],
\end{align}
which is satisfied if $R\ge {\rm const} \cdot (1/\Delta E) \log^3(1/\Delta E)$.
This completes the proof. $\square$

\section{Proof outline of the main theorem (Fig.~\ref{fig_Proof_flow})}

\subsection{Lieb--Robinson bound for short-time evolution}

The key ingredient in our proof is the Lieb--Robinson bound for short-time evolution.
We consider a quantum state $\tilde{\rho}$, which is defined using an operator $O_X$ supported on a subset $X\subset \Lambda$: 
 \begin{align}
 \label{tilde_rho_definition/}
\tilde{\rho}:= O_X \rho_0 O_X^\dagger , \quad X \in i_0[r]   \quad (i_0\in \Lambda,\ r\ge 3) ,
\end{align}
where $O_X$ is given in the form
\begin{align}
 \label{O_X_unitary_definition/}
 O_X = U_X^\dagger O_{X_0} U_X, \quad [U_X,\nb_X] = 0, 
\end{align}
where $U_X$ is a unitary operator that commutes with $\hat{n}_X$.
Because $O_{X_0}$ satisfies the condition~\eqref{cond_u_X_0_spect}, and $U_X$ does not change the total number of bosons on $X$, we obtain 
\begin{align}
\label{cond_O_X_spect}
\| \Pi_{X,q} O_X \Pi_{X,q'} \| =0 \for q'> q+q_0.
\end{align}

We are now interested in the approximation 
\begin{align}
 O_X(t) \rho_0 \approx 
 (U_{X[\ell]}^\dagger O_X U_{X[\ell]}) \rho_0
\end{align}
for a sufficiently small $t=\orderof{1}$, 
where $U_{X[\ell]}$ is an appropriate unitary operator defined on the subset $X[\ell]$.
We can prove the following statement about the approximation.
\begin{subtheorem} \label{main_theorem_short_time_LR}
 Let $O_X$ be an arbitrary operator as defined in Eq.~\eqref{O_X_unitary_definition/}, which is supported on the subset $X\subseteq i_0[r]$ ($i_0\in \Lambda$).
 Then, for a length $\ell$ that satisfies 
 \begin{align}
 \label{condition_for_lenfth_R}
\ell \ge  C_0\log^2(r),
\end{align}
with $C_0=\orderof{1}$ which does not depend on $\bar{q}$, 
we can find a unitary operator $U_{X[\ell]}$ that does not depend on the form of $O_X$ such that 
\begin{align}
[ U_{X[\ell]}, \nb_{X[\ell]}]=0
\end{align}
and
\begin{align}
\label{ineq:main_theorem_short_time_LR}
&\left \| \left( O_X (t) - U_{X[\ell]}^\dagger O_X U_{X[\ell]} \right) \rho_0  \right\|_1
\le \| O_X\| e^{c_0\bar{q} -\ell/\log (r)} 
\end{align}
for $t\le \Delta t_0$, 
where $\Delta t_0=\orderof{1}$ and $C_0$ are appropriately chosen. 
We can combine condition~\eqref{condition_for_lenfth_R} with Ineq.~\eqref{ineq:main_theorem_short_time_LR} as follows:
\begin{align}
\label{ineq:main_theorem_short_time_LR_2}
&\left \| \left( O_X (t) - U_{X[\ell]}^\dagger O_X U_{X[\ell]} \right) \rho_0  \right\|_1
\le \| O_X\| e^{c_0\bar{q}  -\ell/\log (r) + C_0\log(r)} ,
\end{align}
where the inequality holds trivially for $\ell \le  C_0\log^2(r)$.
\end{subtheorem}

{\bf Remark.} In this subtheorem, we do not need to assume the time-independence of the state $\rho_0$.
Hence, only the assumption~\ref{only_the_assumption_initial} is used.
When $\rho_0$ does not satisfy $[\rho_0,H]=0$, Ineq.~\eqref{ineq:main_theorem_short_time_LR} is replaced by
\begin{align}
\label{ineq:main_theorem_short_time_LR_time_dependent}
&\left \| \left( O_X (t) - U_{X[\ell]}^\dagger O_X U_{X[\ell]} \right) \rho_0(-t)  \right\|_1
\le \| O_X\| e^{c_0\bar{q} -\ell/\log (r)} .
\end{align}

By contrast, when we prove the main theorem, we need to connect the short-time evolutions, 
and to perform this procedure, the time-independence of $\rho_0$ is required [see Eqs.~\eqref{time_dependence_why_1} and \eqref{time_dependence_why_2}].

It is necessary to generalize Eq.~\eqref{ineq:main_theorem_short_time_LR_time_dependent} when applying the subtheorem to analyze the quench dynamics (see Sec.~\ref{Sec:LR_Quench}).

\subsection{Proof of Theorem~\ref{main_theorem_long_time_LR} based on Subtheorem~\ref{main_theorem_short_time_LR}}
\label{sec:proof:Lieb--Robinson bound for short-time evolution}

For the convenience of readers, we present the approach to the proof in the main text again with additional explanations.  
We use the connection of unitary time evolutions addressed in Refs.~\cite{Kuwahara_2016_njp,PhysRevLett.126.030604}, which assumes the time-independence of the initial state (i.e., $\rho_0(t)=\rho_0$).

We introduce $\Delta t \le \Delta t_0$ and the following decompositions of the time $t$ to $t/\Delta t$ pieces:
\begin{align}
t := m_t \Delta t \quad  {\rm with} \quad \Delta t \ge \min(t, \Delta t_0/2).
\end{align}
For fixed $R$, we define the subset $X_m$ as follows: 
\begin{align}
\label{Choice_Delta_r_X_m}
X_m:= i_0[r_0+ m\Delta r] = X_0[m\Delta r], \quad \Delta r = \left \lfloor \frac{R-r_0}{m_t} \right \rfloor . 
\end{align}
Note that $X_{m_t} \subseteq i_0[R]$.
%We define $\bar{m}$ such that $ i_0[\bar{s}\Delta r]=i_0[R]$. 
%we define the set of $\{s_0,s_1,\ldots, s_{m_t}\}$ such that 
%\begin{align}
%\label{def_s_0_bar_s}
%s_0=  \lceil r_0/ \Delta r \rceil , \quad s_{m_t} = \bar{s}, \quad 
%s_m= m \lfloor (\bar{s} - s_0)/m_t \rfloor \quad (1\le m \le m_t)
%\end{align}

Then, assuming that $\rho_0$ is time-invariant, we can derive the following inequality~\cite{PhysRevLett.126.030604}: 
 \begin{align}
 \label{unitary_connect_upper_bound}
\norm{\left [O_{X_0} (m_t\Delta t) - O_{X_{m_t}}^{(m_t)}\right]\rho_0 }_1& \le 
 \sum_{m=1}^{m_t} \norm{\left [ O_{X_{m-1}}^{(m-1)} (\Delta t) - O_{X_{m}}^{(m)} \right]\rho_0 }_1,
\end{align}
where $O_{X_0}^{(0)} = O_{X_0}$, and $O_{X_{m}}^{(m)}$ is recursively defined by approximating $O_{X_{m-1}}^{(m-1)} (\Delta t)$.
To see why the time-invariance of $\rho_0$ is essential, let us look at the derivation of Ineq.~\eqref{unitary_connect_upper_bound} for $m_t=2$.
For $m=1$, we define 
\begin{align}
O_{X_1}^{(1)} := U_{X_1}^{(1)\dagger} O_{X_0} U_{X_1}^{(1)} ,
\end{align}
where we choose the unitary operator $U_{X_1} $ by following Subtheorem~\ref{main_theorem_short_time_LR}. 
Note that $O_{X_1}^{(1)}$ is now supported on the subset $X_1$.
For $m=2$, we consider the approximation $O_{X_1}^{(1)}(\Delta t)$ by
\begin{align}
O_{X_2}^{(2)}:= U_{X_2}^{(2)\dagger} O_{X_1}^{(1)} U_{X_2}^{(2)}   , 
\end{align}
where the unitary operator $U_{X_2}^{(2)}$ is chosen according to Subtheorem~\ref{main_theorem_short_time_LR}. 
We then connect the two approximations:
\begin{align}
O_{X_0}(\Delta t) \xrightarrow{\rm approximate} O_{X_1}^{(1)} ,\quad O_{X_1}^{(1)} (\Delta t) \xrightarrow{\rm approximate} O_{X_2}^{(2)} .
\end{align}
To obtain the approximation error, we need to consider 
\begin{align}
\label{time_dependence_why_1}
\norm{ \left [ O_{X_0}(2\Delta t) - O_{X_2}^{(2)} \right]\rho_0}_1
& \le  \norm{ \left [ O_{X_0}(2\Delta t) - O_{X_1}^{(1)}(\Delta t) +O_{X_1}^{(1)}(\Delta t) - O_{X_2}^{(2)} \right]\rho_0  }_1      \notag \\
&\le \norm{ \left [ O_{X_0}(2\Delta t) - O_{X_1}^{(1)}(\Delta t) \right]\rho_0  }_1 
+\norm{ \left [ O_{X_1}^{(1)}(\Delta t) - O_{X_2}^{(2)} \right]\rho_0  }_1. 
\end{align}
The second term is upper-bounded according to Subtheorem~\ref{main_theorem_short_time_LR}.  
The first term is given by
\begin{align}
\label{time_dependence_why_2}
\norm{ \left [ O_{X_0}(2\Delta t) - O_{X_1}^{(1)}(\Delta t) \right]\rho_0  }_1 = 
\norm{ \left [ O_{X_0}(\Delta t) - O_{X_1}^{(1)} \right]\rho_0(-\Delta t)  }_1  .
\end{align}
If $\rho_0(-\Delta t) =\rho_0$, we can upper-bound the above quantity using Subtheorem~\ref{main_theorem_short_time_LR}.  
However, when $\rho_0$ is time-dependent, the condition~\eqref{only_the_assumption_initial} for low boson density may not be satisfied for $\rho_0(-\Delta t)$. 
Therefore, to prove the main theorem for generic $\rho_0$, we need to prove 
the low-boson-density condition for $\rho_0(-m\Delta t)$ ($m\le m_t-1$).

We return to Ineq.~\eqref{unitary_connect_upper_bound}.
According to Subtheorem~\ref{main_theorem_short_time_LR}, we can find $O_{X_{m}}^{(m)}$ such that 
\begin{align}
O_{X_{m}}^{(m)} :=U_{X_{m}}^{(m)\dagger} O_{X_{m-1}}^{(m-1)} U_{X_{m}}^{(m)}   
\end{align}
for each $m=1,2,\ldots,m_t$. 
From Ineq.~\eqref{ineq:main_theorem_short_time_LR_2}, the unitary operator $U_{X_{m}}^{(m)}$ satisfies
\begin{align}
\label{theorem_main_inequality_1_apply}
&\norm { \br{ U_{X_{m}}^{(m)\dagger} O_{X_{m-1}}^{(m-1)} U_{X_{m}}^{(m)}  -  O_{X_{m-1}}^{(m-1)} (\Delta t)} \rho_0}_1 
\le \|O_{X_0}\| e^{c_0\bar{q} -\Delta r/\log(R)+C_0\log(R)} ,
\end{align} 
where we use $X_m \subseteq i_0[R]$ for all $m$ in applying Subtheorem~\ref{main_theorem_short_time_LR}.
We thus obtain 
 \begin{align}
 \label{proof_of_theorem_main_last}
\norm{\left [O_{X_0} (m_t\Delta t) - O_{X_{m_t}}^{(m_t)}\right]\rho_0 }_1
& \le m_t  \|O_{X_0}\| e^{c_0\bar{q} -\Delta r/\log(R)+C_0\log(R)}  \notag \\
& \le  \|O_{X_0}\| \exp\left(c_0\bar{q} - \frac{\Delta t (R-r_0)}{t\log(R)}+\frac{1}{\log(R)} +(C_0+1) \log(R) \right) ,
\end{align}
for $R\ge 2$,
where in the second inequality we use $\Delta r \ge 1$ [or $m_t \le (R-r_0)$]\footnote{Otherwise, the upper bound is worse than the trivial inequality, i.e., $\|  [O_{X_0} (m_t\Delta t) - O_{X_{m_t}}^{(m_t)} ]\rho_0 \|_1\le \|O_{X_0}\|$.}.
Thus, because of $1/\log(R) \le 3\log(R)$ for $R\ge 2$, by choosing $C_1=\Delta t$ and $C_2=C_0+4$, 
the inequality~\eqref{proof_of_theorem_main_last} reduces to Ineq.~\eqref{main_theorem_short_time_LR_main_ineq}. 
This completes the proof of Theorem~\ref{main_theorem_long_time_LR}. $\square$

In the following sections, we give the full proof of Subtheorem~\ref{main_theorem_short_time_LR}. 
Note that the following proof repeats some of the explanations in the main text.  
We show the outline of the proof in Fig.~\ref{fig_Proof_flow}.

 \begin{figure}[tt]
\centering
\includegraphics[clip, scale=0.5]{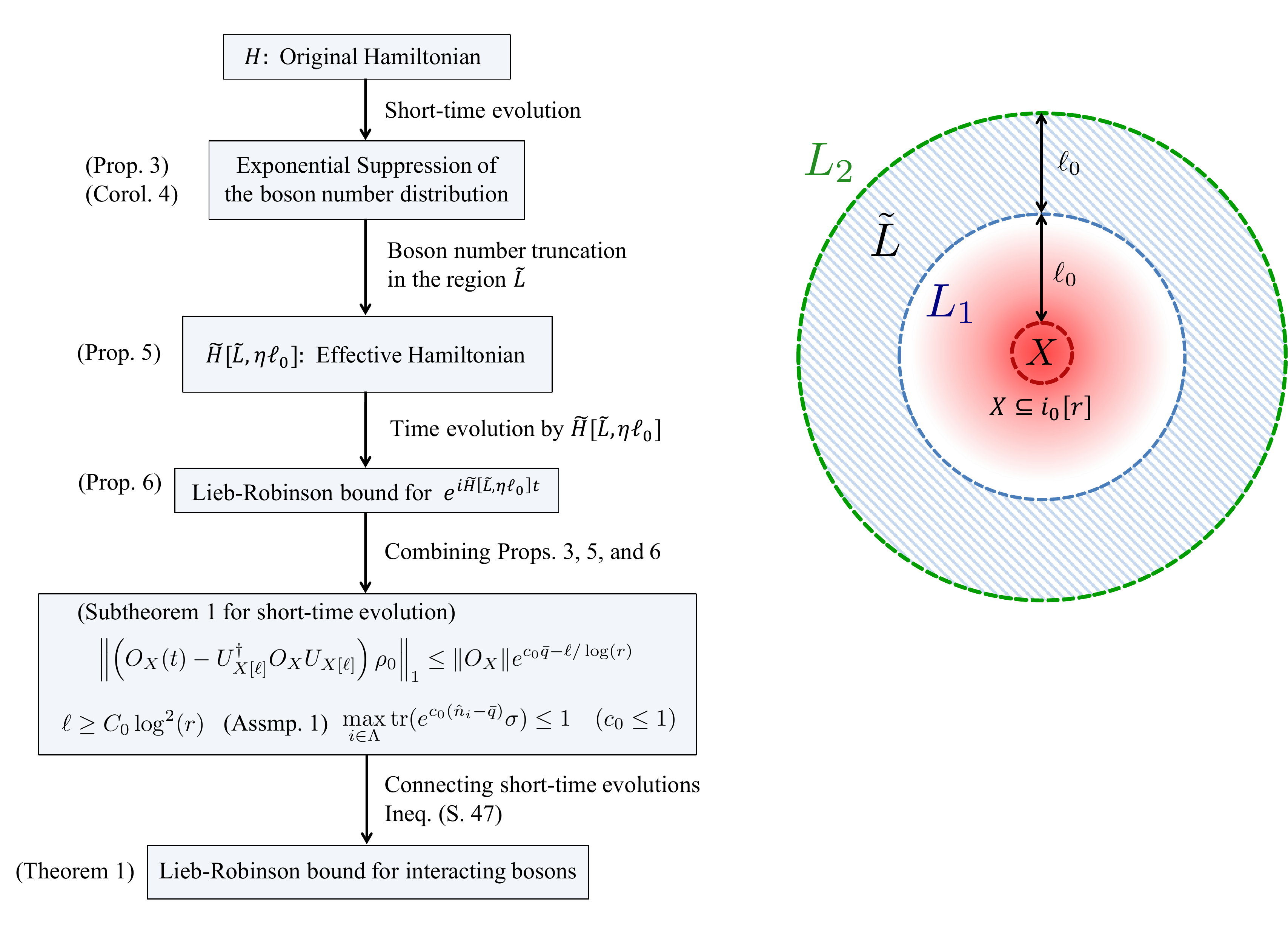}
\caption{Schematic illustration of the proof of Theorem~\ref{main_theorem_long_time_LR}.
}
\label{fig_Proof_flow}
\end{figure}

%
%This inequality gives an upper bound in the form of $e^{- \frac{\orderof{1}}{t} [R/\log(R) - vt \log(R)]}$ and slightly weaker than the inequality~\eqref{main_theorem_short_time_LR_main_ineq}, which gives 
%the bound in the form of $e^{-R/\log(R) + vt \log(R)}$. 
%In the next subsection, we refine the above analyses to derive the inequality~\eqref{main_theorem_short_time_LR_main_ineq} without assuming the time independence of $\rho_0$. 
%
%
%
%

\section{Proof of Subtheorem~\ref{main_theorem_short_time_LR}}

Here, we derive the inequality~\eqref{ineq:main_theorem_short_time_LR_time_dependent} in which the time-independence of $\rho_0$ is not satisfied.
Throughout the proof, we denote $\|O_{X_0}\|$ by $\zeta_0$. 
Because of the definition~\eqref{O_X_unitary_definition/} of $O_X$, we have 
\begin{align}
\label{norm_O_X_def_zeta_0}
\|O_X\|= \zeta_0.
\end{align}

We first show that the following simple analysis \textit{does not work} in proving the subtheorem.
If short-time evolution is considered, the following simple Taylor expansion is expected to work: 
\begin{align}
\label{BCH_expansion}
\| [O_X(t),u_i] \rho_0 \|_1 =
\| [O_X,u_i(-t)] \rho_0 \|_1 =
\norm{ \sum_{s=0}^\infty \frac{(-t)^s}{s!} [\ad_{H}^s (u_i) , O_X] \rho_0}_1 
%\stackrel{\rm{?}}{\approx}
% \overbrace{\approx}^? 
\le  \sum_{s=0}^{\bar{s}}\frac{t^s}{s!}  \norm{ [\ad_{H}^s (u_i) , O_X] \rho_0}_1 ,
\end{align}
where $u_i$ is an arbitrary unitary operator acting on a site $i\in \Lambda$.
Because the Hamiltonian is short-range, we have $[\ad_{H}^s (u_i) , O_X]=0$ for $s\le \bar{s}$, where 
$\bar{s}=\orderof{\dist_{i,X}}$. 
Hence, we have 
\begin{align}
\label{BCH_expansion2}
\| [O_X(t),u_i] \rho_0 \|_1 \le 
 \sum_{s=\bar{s}+1}^\infty \frac{t^s}{s!}\norm{ [\ad_{H}^s (u_i) , O_X] \rho_0}_1. 
\end{align}

In standard spin models with bounded local energy, we have 
\begin{align}
\| \ad_{H}^s (u_i) \| \lesssim s! ,
\end{align}
and hence we can ensure the exponential convergence of the expansion~\eqref{BCH_expansion} for $t=\orderof{1}$.
Unfortunately, this simple estimation cannot be used to obtain Subtheorem~\ref{main_theorem_short_time_LR}. 
When we formally describe the Hamiltonian~\eqref{def:Ham} as  
 \begin{align}
&H= \sum_{|Z| \le k} h_Z , 
\end{align}
we need to consider the norm of 
\begin{align}
h_{Z_1} h_{Z_2} \cdots h_{Z_{s_1}} u_i h_{Z_{s_1+1}} \cdots h_{Z_s}  O_X \rho_0 ,
\end{align}
where $h_Z$ consists of the boson hoppings $b_i b_{j}^\dagger$ and the boson--boson interactions $v_Z$ in Eq.~\eqref{def:Ham}.
Because the boson number $\nb_i$ on an arbitrary site is strongly suppressed in $\rho_0$ from Assumption~\ref{only_the_assumption_initial}, 
we have $\| h_Z \rho_0\|_1 = \orderof{1}$. 
By contrast, in the state $O_X \rho_0  O_X^\dagger$, all the bosons in the region $X$ can be concentrated on one site,
which may give $\| \nb_i O_X \rho_0 \|_1 \propto |X|$ for $i\in X$.    
We thus obtain $\| h_Z O_X \rho_0 \|_1 \gtrsim |X|^{\nu}$ for $Z\cap X\neq \emptyset$, where $\nu$ is a positive integer depending on the form of $v_Z$. 
Consequently, the norm $\norm{ [\ad_{H}^s (u_i) , O_X] \rho_0}_1$ has a rather weaker upper bound: 
\begin{align}
\norm{ [\ad_{H}^s (u_i) , O_X] \rho_0}_1 \lesssim s! |X|^{\nu(s-\bar{s})} .
\end{align}
By combining the above inequality with Eq.~\eqref{BCH_expansion2}, we can ensure the convergence of the expansion only for $t=\orderof{1/|X|^\nu}$,
which is too weak to prove Theorem~\ref{main_theorem_long_time_LR}. 

In the following, we take a different route to prove  Subtheorem~\ref{main_theorem_short_time_LR} by the three steps in Secs.~\ref{Sec:Density of bosons after time evolution Proposition Schuch_boson_extend}, \ref{sec:def_Effective Hamiltonian}, and \ref{sec:Lieb--Robinson bound for the effective Hamiltonian}.

\subsection{Boson density after time evolution (Proposition~\ref{Schuch_boson_extend})} 
\label{Sec:Density of bosons after time evolution Proposition Schuch_boson_extend}

We first consider the boson number distribution after a short-time evolution. 
To this end, we need to estimate 
\begin{align}
\label{def_tilde_rho_t}
\tr \left [ \Pi_{i,\ge z}\tilde{\rho}(t) \right] ,\quad 
\tilde{\rho}(t)= e^{-iHt} O_X \rho_0 O_X^\dagger e^{iHt}= 
O_X(-t) \rho_0(t) O_X(-t)^\dagger ,
\end{align}
where $\Pi_{i,z}$ has been defined by Eq.~\eqref{def_Pi_X_q}.
In the state $\tilde{\rho}$, the boson number $\hat{n}_i$ for $i\in X$ can be as large as $\orderof{|X|}$, whereas    
the boson number $\nb_i$ ($i\in X^\co$) is exponentially suppressed, as shown in Eq.~\eqref{ineq:condition_for_moment_generating}. 
During the time evolution, the bosons concentrated on $X$ spread outside of $X$.  
We expect that after a sufficiently small time, the exponential decay of the boson number still holds for sites that are sufficiently separated from $X$.
% (see Fig.~\ref{fig_density_boson}).   

% \begin{figure}[tt]
%\centering
%\includegraphics[clip, scale=0.5]{density_boson.pdf}
%\caption{Boson density after time evolution. 
%When all the boson concentrated on a region $X$ (i.e., no bosons outside $X$), the spreading of the bosons is known to obey the Lieb--Robison bound~\cite{PhysRevA.84.032309}.
%However, if there are finite number of bosons outside $X$, the upper bound of the boson number increases exponentially with $t$. 
%This point spoils the approach in Ref.~\cite{PhysRevA.84.032309} as long as we are interested in the long-time behavior of bosons.  
%On the other hand, in our approach (Sec.~\ref{sec:proof:Lieb--Robinson bound for short-time evolution}), we only have to consider the short time $t=\orderof{1}$, where the exponential increase $e^{\orderof{t}}$ is still $\orderof{1}$. 
%We then ensure that the boson number distribution for $\nb_i$ exponentially decays if the site $i$ of interest is sufficiently separated from the region $X$ (Corollary~\ref{corol:time_distribution}).
%}
%\label{fig_density_boson}
%\end{figure}

The first proposition ensures that the boson density is not seriously affected by the time evolution for $\dist_{i,X} \gg 1$ if the time $t$ is of $\orderof{1}$ (see Sec.~\ref{Sec:Proof:density of bosons after time evolution} for the proof).
 
\begin{prop} \label{Schuch_boson_extend}
We first define the operator $M^{(s)}_i(t)$ as 
\begin{align}
M^{(s)}_i(t) := \tr \left [ \hat{n}_i^s \tilde{\rho}(t)  \right] .
\end{align}
Then, for $t\le t_0$, the following upper bound for $M^{(s)}_i(t)$ holds:
\begin{align}
\label{main_ineq_Schuch_boson_extend}
M_i^{(s)}(t) &\le  c'_{1} e^{c_{0} \bar{q}} \zeta_0^2 |X|^3 (c_{1}s |X| )^{s}   e^{-\dist_{i,X}}  +c''_{1} e^{c_{0} \bar{q}} \zeta_0^2 (c_{1}s)^{s},
\end{align}
where $c_{1}$, $c'_{1}$, and $c''_{1}$ are defined as 
\begin{align}
\label{parameters_c_c'_c''_t_0}
&c_{1} :=  e^{8\bar{J} d_G t_0}/c_{0}, \notag \\
&c'_{1} := 320 c_{0}^{-3}e^{4\bar{J} d_G t_0+ c_{0} (1+q_0/|X|)} ,  \notag \\
&c''_{1} := 80 \lambda_0c_{0}^{-1} e^{4 \bar{J} d_G t_0+c_0}  .
\end{align}
Notice that they are $\orderof{1}$ constants if $t_0=\orderof{1}$. 
\end{prop}

{\bf Remark.}
In the proof, we fully use the methods in Ref.~\cite{PhysRevA.84.032309}, which treats the case of $s=1$.
The above upper bound increases exponentially with $t$, and hence we cannot use it to upper-bound the boson density for general $t$.  
The key point of the proof in Sec.~\ref{sec:proof:Lieb--Robinson bound for short-time evolution} 
is that we need to treat only the short-time evolution in this subtheorem.
We afterward connect the short-time evolutions step by step, as in Ineq.~\eqref{unitary_connect_upper_bound}.

In this proposition, the form of the Hamiltonian, i.e., Eq.~\eqref{def:Ham}, is essential; if the Hamiltonian includes interactions such as $\nb_i \nb_j b_i b_j^\dagger$, 
the above proposition breaks down even for small $t$.

{~}\\

By using Proposition~\ref{Schuch_boson_extend}, we can immediately derive the following corollary of the boson number distribution. 

\begin{corol} \label{corol:time_distribution}
Let us define the boson number distribution on a site $i \in \Lambda$ as follows:
\begin{align}
P_{i,\ge z_0}^{(t)}  := \sum_{z\ge z_0} \tr \left [ \Pi_{i,z} \tilde{\rho}(t)  \right] , 
\end{align}
where $\Pi_{i,z}$ has been defined by Eq.~\eqref{def_Pi_X_q}.
Then, for arbitrary $i\in \Lambda$ such that
\begin{align}
\label{condition_dist_i_X_corol}
\frac{\gamma^3 c'_{1}}{c''_{1}  } r^{3D} \le  e^{\dist_{i,X}/2} \to \dist_{i,X} \ge 2 \log \left(\frac{\gamma^3 c'_{1}}{c''_{1}  }\right) + 
6D \log(r) = \orderof{\log(r)},
\end{align} 
we obtain 
\begin{align}
\label{main_ineq_dist_i_X_corol}
P_{i,\ge z_0}^{(t)}  \le 2c''_{1} e^{c_{0} \bar{q}} \zeta_0^2 \br{\frac{\tilde{c}_1 \dist_{i,X} }{z_0  }}^{\tilde{c}'_1 \dist_{i,X}/\log (r)} ,
\end{align}
where we have defined $r$ ($\ge 3$) by $X\subseteq i_0[r]$ for an appropriate choice of $i_0\in \Lambda$ in Eq.~\eqref{tilde_rho_definition/}, and 
$\tilde{c}_1$ and $\tilde{c}'_1$ are constants of $\orderof{1}$.
\end{corol}

\subsubsection{Proof of Corollary~\ref{corol:time_distribution}}

Because $X\subseteq i_0[r]$, we have from Ineq.~\eqref{def:parameter_gamma}
\begin{align}
|X| \le \gamma r^D.
\end{align}
We then choose $s$ such that 
\begin{align}
\label{cond_choise_s_corol}
c'_{1}  |X|^3 (c_{1}s |X| )^{s}   e^{-\dist_{i,X}} \le  
\gamma^3 c'_{1} r^{3D} ( \gamma c_{1}s r^D )^{s}  e^{-\dist_{i,X}} \le c''_{1} (c_{1}s)^{s}, 
\end{align}
which yields 
\begin{align}
&\frac{\gamma^3 c'_{1}}{c''_{1}  } r^{3D} ( \gamma r^D )^{s}   \le  e^{\dist_{i,X}/2} ( \gamma r^D )^{s}   \le  e^{\dist_{i,X}} \notag \\
&\longrightarrow s \le \frac{\dist_{i,X}}{2 \log (\gamma r^D)} ,
\end{align}
where we use the condition~\eqref{condition_dist_i_X_corol}.
We choose $s$ as 
\begin{align}
s =\left \lfloor \frac{\dist_{i,X}}{2 \log (\gamma r^D)} \right \rfloor \in   \left[ \tilde{c}'_1 \frac{\dist_{i,X}}{\log (r)} , \tilde{c}_1'' \frac{\dist_{i,X}}{\log (r)} \right] ,
\end{align}
where $\tilde{c}'_{1}$ and $\tilde{c}''_1$ are constants which depend only on $\gamma$ and $D$. 
By using the inequality~\eqref{cond_choise_s_corol}, we reduce the Ineq.~\eqref{main_ineq_Schuch_boson_extend} to
\begin{align}
M_i^{(s)}(t) &\le  2c''_{1} e^{c_{0} \bar{q}}\zeta_0^2 (c_{1}s)^{s} , 
\end{align}
and hence Markov's inequality yields 
\begin{align}
P_{i,\ge z_0}^{(t)} \le \frac{M_i^{(s)}(t)}{z_0^s} 
\le  2c''_{1} e^{c_{0} \bar{q}}\zeta_0^2 \br{\frac{c_{1} \tilde{c}''_1 \dist_{i,X} }{z_0 \log(r) }}^{\tilde{c}'_1 \dist_{i,X}/\log (r)} .
\end{align}
By using $\log(r) \ge 1$ from $r\ge 3$ and defining $\tilde{c}_1:=c_{1} \tilde{c}''_1$, we prove the main inequality~\eqref{main_ineq_dist_i_X_corol}.
This completes the proof. $\square$

% {~}
%
%\hrulefill{\bf [ End of Proof of Corollary~\ref{corol:time_distribution}] }
%
%{~}

\subsection{Effective Hamiltonian (Proposition~\ref{prop:error_time_evolution_effective_Ham})} \label{sec:def_Effective Hamiltonian}

\begin{figure}[tt]
\centering
\subfigure[Subsets $L_1$ and $L_2$]
{\includegraphics[clip, scale=0.4]{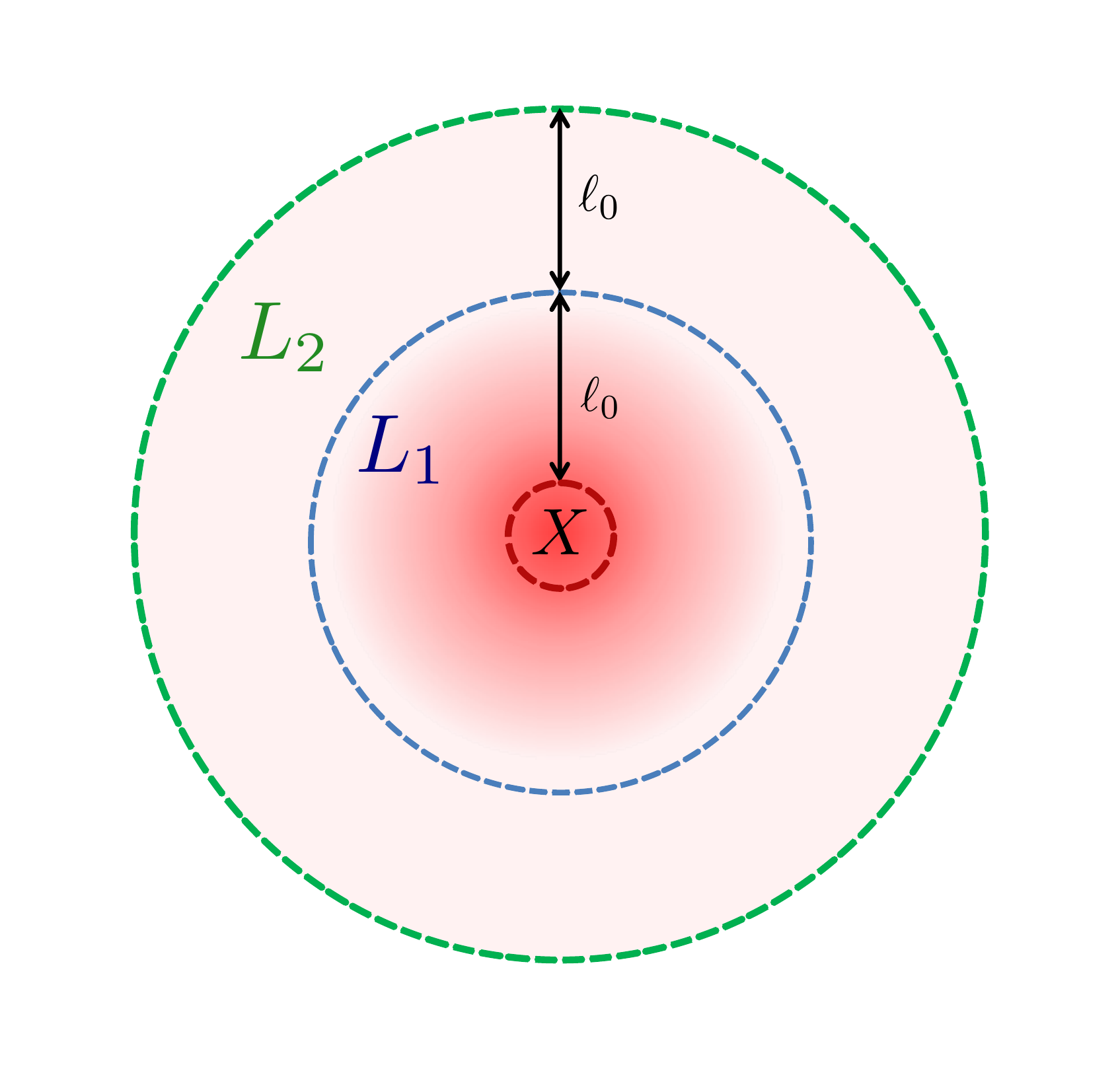}}
\subfigure[Subset $\tilde{L}$]
{\includegraphics[clip, scale=0.4]{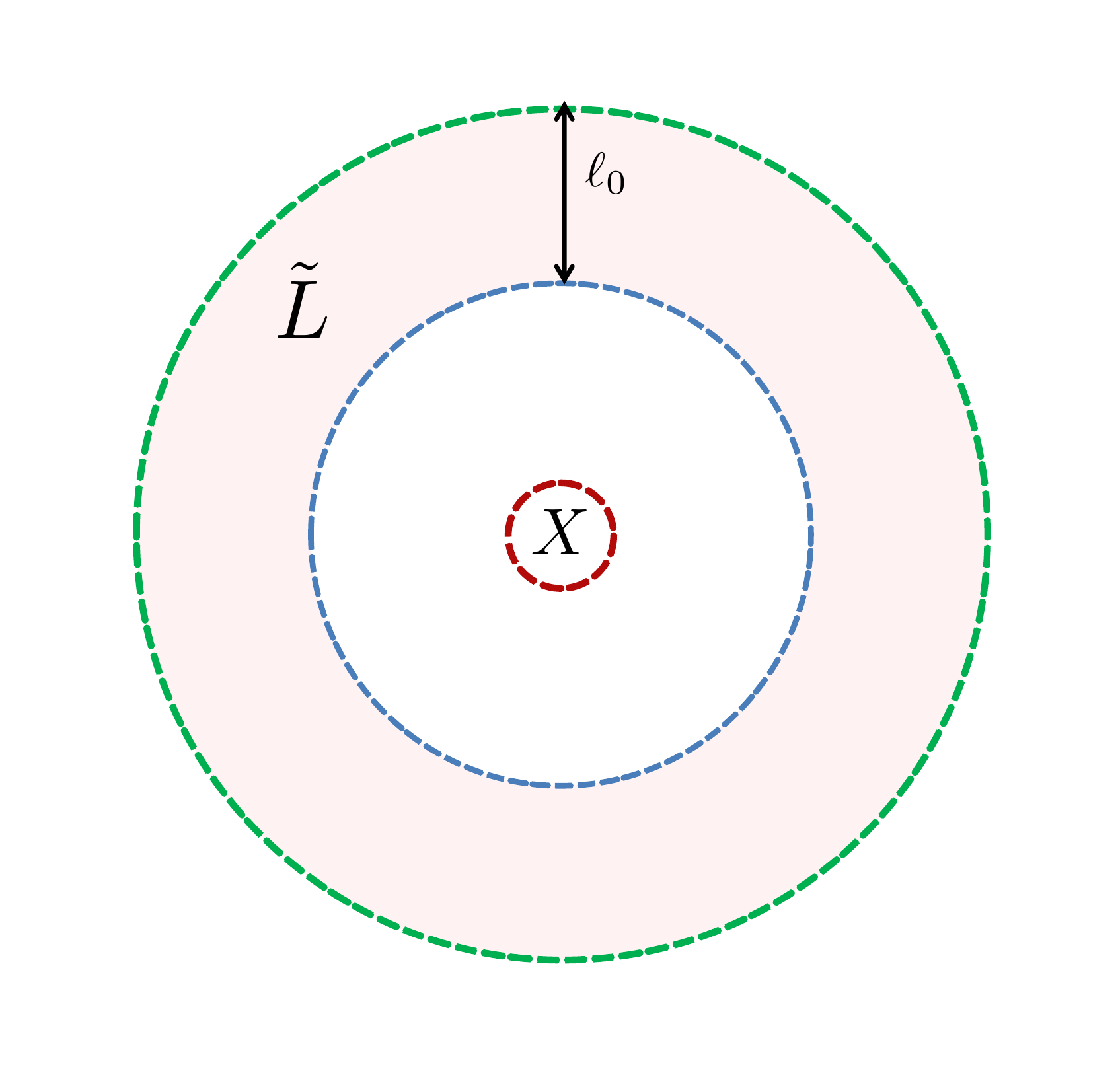}}
\caption{
Schematic illustrations of sets $L_1$, $L_2$, and $\tilde{L}$.
In the region $L_1$, we cannot ensure that the boson number distribution is exponentially localized.
By contrast, if $\ell_0$ is sufficiently large [see the condition \eqref{condition_ell_0_prop}], 
it is ensured that in the region $\tilde{L} = L_2 \setminus L_1$, the boson number can be truncated up to a finite value.
We define the effective Hamiltonian $\tilde{H}[\tilde{L},\eta\ell_0]$ using Eqs.~\eqref{def:bar_Pi_X_z} and \eqref{def:tilde_Ham_effective}, which well approximates the 
exact dynamics, as described by Ineq.~\eqref{ineq:prop:error_time_evolution_effective_Ham}.
}
\label{fig_Decomp}
\end{figure}

Proposition~\ref{Schuch_boson_extend} and Corollary~\ref{corol:time_distribution} imply that 
during the time evolution, the boson number $\nb_i$ is strongly suppressed as long as site $i$ is separated from the region $X$ by a sufficient distance. 
Hence, we expect that in the Hamiltonian $H$, the maximum boson number on one site can be truncated. 
In this subsection, we rigorously justify this procedure.   

As shown in Fig.~\ref{fig_Decomp}, we first define $L_1$ and $L_2$ as
\begin{align}
\label{def_L_1_L_2}
L_1 = X[\ell_0] , \quad L_2 = X[2\ell_0] .
\end{align}
We also define $\tilde{L}$ as 
\begin{align}
\label{def:tilde_L}
\tilde{L} : = L_2 \setminus L_1. 
\end{align}

We now define $\bar{\Pi}_{L,q}$ ($L\subseteq \Lambda$) as the projection onto the eigenspace such that for an arbitrary $i \in L$, the boson number $\nb_i$ is truncated up to $q$:
 \begin{align} \label{def:bar_Pi_X_z}
\bar{\Pi}_{L,q} := \prod_{i\in L}\Pi_{i,\le q} .
\end{align}
Note that $\|\nb_j \bar{\Pi}_{L,q}\| \le q$ for $\forall j\in L$.
During the time evolution of $\tilde{\rho}$ in Eq.~\eqref{tilde_rho_definition/}, the boson number is exponentially suppressed as long as site $i$ is sufficiently separated from $X$ (see Lemma~\ref{lemma:boson_number_concentration_pi_Z_q} below).  
We then aim to approximate the time evolution $e^{-iHt}$ by using another Hamiltonian $\tilde{H}[L,q]$ as 
 \begin{align}
 \label{def:tilde_Ham_effective}
&\tilde{H}[L,q] := \tilde{H}_0[L,q] + \tilde{V} [L,q] ,\notag \\
& \tilde{H}_0[L,q]:=\bar{\Pi}_{L,q} H_0 \bar{\Pi}_{L,q} ,\quad 
V[L,q]:=\bar{\Pi}_{L,q} V \bar{\Pi}_{L,q} ,
\end{align}
where the subset $L$ can be arbitrarily chosen in the definitions.
In this effective Hamiltonian, the boson number is truncated up to $q$ in the region $L$.

In the following, we choose $L=\tilde{L}$ in Eq.~\eqref{def:tilde_Ham_effective} and consider $\tilde{H}[\tilde{L},\eta\ell_0]$ as the effective Hamiltonian, where 
$\tilde{L}$ is given by Eq.~\eqref{def:tilde_L}, and $\eta$ is chosen appropriately (see Proposition~\ref{prop:error_time_evolution_effective_Ham} below).  
We usually cannot say that the time evolution by $H$ is approximated by $e^{-i \tilde{H}[\tilde{L},\eta\ell_0] t}$, that is, 
\begin{align}
\norm{ O_X(t) - O_X(\tilde{H}[\tilde{L},\eta\ell_0],t)  }  \approx \|O_X\| .
\end{align}
However, when the above operator acts on the state $\rho_0(-t)$, the difference can be small.   
Hence, we need to estimate the difference between 
 \begin{align}
O_X(t)  \rho_0(-t)\quad {\rm and} \quad    O_X(\tilde{H}[\tilde{L},\eta\ell_0],t)\rho_0(-t)  \quad (t\le t_0).
\end{align}
We can prove the following proposition (see Sec.~\ref{Proof_prop:error_time_evolution_effective_Ham} for the proof). 
\begin{prop} \label{prop:error_time_evolution_effective_Ham}
Let us choose $\ell_0$ so that it satisfies 
\begin{align}
\label{condition_ell_0_prop}
\ell_0 \ge c_2\log^2 (r) .
\end{align} 
Recall that $r$ was defined in Eq.~\eqref{tilde_rho_definition/}.
Then, there exists a constant $\eta$ that gives 
 \begin{align}
 \label{ineq:prop:error_time_evolution_effective_Ham}
&\left \| \left[ O_X(t) -O_X(\tilde{H}[\tilde{L},\eta\ell_0],t)\right] \rho_0(-t) \right\|_1 \le \frac{1}{2}\zeta_0 e^{c_{0} \bar{q}} e^{-2\ell_0/\log (r) }
\end{align}
for $t\le t_0$. 
Here, $c_2$ and $\eta$ are $\orderof{1}$ constants which do not depend on $\bar{q}$.
\end{prop}
 
 {~}  
 
\noindent 
From this proposition, we can see that the error decreases exponentially with the number of the boson truncations. 
In the Hamiltonian $\tilde{H}[\tilde{L},\eta\ell_0]$, the greatest obstacle, namely, the unboundedness of the interaction norms, has been removed, at least in the region $\tilde{L}$.
However, outside the region $\tilde{L}$, the norm is still unbounded. 
In the following subsection, we consider how to derive the Lieb--Robinson bound for $e^{-i \tilde{H}[\tilde{L},\eta\ell_0] t}$ only from the finiteness of the norm in the region $\tilde{L}$.

%%%%%%%%%%%%%%%%%%%%%%%%%%%%%%%%%%%%%%%%%%%%%%%%%%%%%%%%%%%%%%%%%%%%%%%%%%%%%%%%%%%%%%%%%%%%%%%%%%%%%%%%%%%%%%%%%%%%%%%%%%%%%%%%%%%%%%%%%%%%%%%%%%%%%%%%%%%%%%%%%%%%%%%%%%%%%%%%%%%%%%%%%%%%%%%%%%%%%%%%%%%%%%%%%%%%%%%%%%%%%%%%%%%%%%%%%%%%%%%%%%%%%%%%%%%%%%%%%%%%%%%%%%%%%%%%%%%%%%%%%%%%%%%%%%%%%%%%%%%%%%%%%%%%%%%%%%%%%%%%%%%%%%%%%%%%%%%%%%%%%%%%%%%%%%%%%%%%%%%%%%%%%%%%%%%%%%%%%%%%%%%%%%%%%%%%%%%%%%%%%%%%%%%%%%%%%%%%%%%%%%%%%%%%%%%%%%%%%%%%%%%%%%%%%%%%%%%%%%%%%%%%%%%%%%%%%%%%%%%%%%%%%%%%%%%%%%%%%%%%%%%%%%%%%%%%%%%%%%%%%%%%%%%%%%%%%%%%%%%%%%%%%%%%%%%%%%%%%%%%%%%%%%%%%%%%%%%%%%%%%%%%%%%%

\subsection{Lieb--Robinson bound for the effective Hamiltonian (Proposition~\ref{prop:short_time_Lieb--Robinson})}
\label{sec:Lieb--Robinson bound for the effective Hamiltonian}

We are now interested in the Lieb--Robinson bound for the effective Hamiltonian $\tilde{H}[\tilde{L},\eta\ell_0]$ defined in Proposition~\ref{prop:error_time_evolution_effective_Ham}. 
In this section, we adopt an additional condition for $\ell_0$, as follows:
 \begin{align}
 \label{new_condition_ell_0_LR}
\ell_0 \ge 8k ,
\end{align} 
where $k$ represents the maximum interaction length in $V$ [see Eq.~\eqref{def:Ham}].

We would like to calculate the norm
 \begin{align}
\label{ineq:prop:error_time_evolution_effective_Ham'}
&\left \| O_X(\tilde{H}[\tilde{L},\eta\ell_0],t) - U_{L_2}^\dagger O_X U_{L_2}    \right\|  \quad (t\le t_0),
\end{align} 
where $U_{L_2}$ is an appropriate unitary operator supported on the subset $L_2$.
To this end, we first decompose $e^{i \tilde{H}[\tilde{L},\eta\ell_0] t}$ as follows:
 \begin{align}
e^{-i \tilde{H}[\tilde{L},\eta\ell_0] t} = e^{-i \tilde{V}[\tilde{L},\eta\ell_0] t}\ \mathcal{T} \exp \br{    
-i\int_0^t e^{i \tilde{V}[\tilde{L},\eta\ell_0] x} \tilde{H}_0[\tilde{L},\eta\ell_0]  e^{-i \tilde{V}[\tilde{L},\eta\ell_0] x} dx
} ,
\end{align} 
where $\mathcal{T}$ is the time-ordering operator, and we use the definition~\eqref{def:tilde_Ham_effective}.
Because $\tilde{V}[\tilde{L},\eta\ell_0]$ consists of operators that commute with each other, the time-evolved operator 
 \begin{align}
O_X(\tilde{V}[\tilde{L},\eta\ell_0] ,\eta\ell_0],t) = e^{i \tilde{V}_{X[k]}[\tilde{L},\eta\ell_0] t} O_X e^{-i \tilde{V}_{X[k]}[\tilde{L},\eta\ell_0] t}
\end{align} 
is supported on the subset $X[k]$, 
where $\tilde{V}_{X[k]}[\tilde{L},\eta\ell_0]$ picks up all the interaction terms $v_Z[\tilde{L},\eta\ell_0]$ such that $Z\subset X[k]$ 
[see also Eqs.~\eqref{def:Ham} and \eqref{def:Ham_subset}].
Because of the condition~\eqref{new_condition_ell_0_LR}, we have $L_1=X[\ell_0] \supset X[k]$, and we write $O_X(\tilde{V}[\tilde{L},\eta\ell_0] ,\eta\ell_0],t) $ as 
 \begin{align}
 \label{def_tilde_O_L_1}
O_X(\tilde{V}[\tilde{L},\eta\ell_0] ,\eta\ell_0],t) = \tilde{O}_{L_1} \quad (\|\tilde{O}_{L_1}\|=\| O_X\|=\zeta_0) .
\end{align} 

In addition, for arbitrary $0\le \tau \le t$, $e^{-i \tilde{V}[\tilde{L},\eta\ell_0] \tau} \tilde{H}_0[\tilde{L},\eta\ell_0]  e^{i \tilde{V}[\tilde{L},\eta\ell_0] \tau}$ is formally described by
 \begin{align}
 \label{def:tilde_H_tau}
\tilde{H}_\tau := e^{i \tilde{V}[\tilde{L},\eta\ell_0] \tau } \tilde{H}_0[\tilde{L},\eta\ell_0]  e^{-i \tilde{V}[\tilde{L},\eta\ell_0] \tau} 
&=\sum_{\langle i, j\rangle } J_{i,j} e^{i \tilde{V}[\tilde{L},\eta\ell_0] \tau }\bar{\Pi}_{\tilde{L},\eta\ell_0}(b_i^\dagger b_j + {\rm h.c.}) \bar{\Pi}_{\tilde{L},\eta\ell_0}e^{-i \tilde{V}[\tilde{L},\eta\ell_0] \tau}    \notag\\
&=: \sum_{Z\subset\Lambda: \diam (Z) \le 2k} \tilde{h}_{Z,\tau} ,
\end{align} 
where we use the fact that $\tilde{V}[\tilde{L},\eta\ell_0]$ consists of interaction terms with a maximum interaction length of $k$. 
For an arbitrary time-dependent operator $A_\tau$, we adopt the following notations: 
\begin{align}
\label{notation_time_dependent_operator}
&U_{A_\tau, x\to t} = \mathcal{T} e^{-i \int_{x}^{t} A_\tau   d\tau} ,\notag \\
&O(A_\tau , x\to t) = U_{A_\tau, x\to t}^\dagger O U_{A_\tau, x\to t}.
\end{align} 
Using these notations, we obtain 
 \begin{align}
O_X(\tilde{H}[\tilde{L},\eta\ell_0],t) = \tilde{O}_{L_1} (\tilde{H}_\tau, 0\to t),
\end{align} 
where $\tilde{O}_{L_1}$ has been defined in Eq.~\eqref{def_tilde_O_L_1}. 
In the following, we approximate $\tilde{O}_{L_1} (\tilde{H}_\tau, 0\to t)$ using the subset Hamiltonian $\tilde{H}_{L,\tau}$:
 \begin{align}
 \label{approximate_L_1_H_L_tau}
&\tilde{O}_{L_1} (\tilde{H}_\tau, 0\to t) \approx \tilde{O}_{L_1} (\tilde{H}_{L,\tau} , 0\to t) , \quad \tilde{H}_{L,\tau}= \sum_{Z\subset L} \tilde{h}_{Z,\tau} ,
\end{align} 
where we choose subset $L$ appropriately as described later.
In the following proposition, we estimate the approximation error (see Sec.~\ref{Proof_prop:short_time_Lieb--Robinson} for the proof).
\begin{prop} \label{prop:short_time_Lieb--Robinson}
When $L=L_2' :=X[2\ell_0-2k]$ ($\subset L_2$) is chosen, the approximation error in Eq.~\eqref{approximate_L_1_H_L_tau} is bounded from above by
 \begin{align}
 \label{ineq:prop:short_time_Lieb--Robinson}
\left\| \tilde{O}_{L_1} (\tilde{H}_\tau, 0\to t) - \tilde{O}_{L_1} (\tilde{H}_{L'_2,\tau} , 0\to t) \right\| \le 
2e^3 \zeta_0 c_3 t  |\partial L_2'| \ell_0 e^{-\ell_0/(2k)}
\end{align}
under the condition
  \begin{align}
t\le \frac{1}{e c_3'} , 
\end{align} 
where $c_3:= 4\bar{J}  \eta \gamma (2k)^D d_G$, and $c_3' := 16ekc_3 \gamma (2k)^D$.
\end{prop}

From the above proposition, by choosing $U_{L_2}$ as 
  \begin{align}
  \label{unitary_explicit_form_L_2}
U_{L_2}= e^{-i \tilde{V}_{X[k]}[\tilde{L},\eta\ell_0] t} U_{\tilde{H}_{L'_2,\tau} , 0\to t} ,
\end{align}
we find that 
  \begin{align}
[ U_{L_2}, \nb_{L_2} ] = 0
\end{align} 
because of $[ \tilde{V}_{X[k]}[\tilde{L},\eta\ell_0], \nb_{L_2} ] =0$ and $[ \tilde{H}_{L'_2,\tau}, \nb_{L_2} ] =0$.
We upper-bound the norm~\eqref{ineq:prop:error_time_evolution_effective_Ham'} as 
 \begin{align}
 \label{ineq_effctive_L_2_unitary_error}
&\left \| O_X(\tilde{H}[\tilde{L},\eta\ell_0],t) - U_{L_2}^\dagger O_X U_{L_2}    \right\|  \le 2e^3\zeta_0 c_3 t  |\partial L_2'| \ell_0 e^{-\ell_0/(2k)} 
\le 2e^3\zeta_0 c_3 \gamma t (2\ell_0 +r)^D \ell_0 e^{-\ell_0/(2k)} ,
\end{align} 
where the last inequality is derived from $\partial L_2' \subset L_2 =X[2\ell_0] \subseteq i_0[2\ell_0 +r]$.

%
% \begin{figure}[tt]
%\centering
%\includegraphics[clip, scale=0.4]{Lieb--Robinson_setup.pdf}
%\caption{Schematic picture of the set }
%\label{fig_Lieb--Robinson_setup}
%\end{figure}

\subsection{Completing the proof}

We now have all the ingredients to prove Subtheorem~\ref{main_theorem_short_time_LR}.
First, we set the parameter $\Delta t_0$ in the statement to $\Delta t_0 = 1/(e c_3')$, which is an $\orderof{1}$ constant from the definition of $c_3'$ in Proposition~\ref{prop:short_time_Lieb--Robinson}.
By choosing $\ell_0$ such that it satisfies the conditions~\eqref{condition_ell_0_prop} and \eqref{new_condition_ell_0_LR}, we obtain 
 \begin{align}
\left \| \left [ O_X(t) -O_X(\tilde{H}[\tilde{L},\eta\ell_0],t)\right]   \rho_0(-t) \right\|_1 \le 
\frac{1}{2}e^{c_{0} \bar{q}} \zeta_0  e^{-2\ell_0/\log (r) }
\end{align}
and 
 \begin{align}
&\left \| O_X(\tilde{H}[\tilde{L},\eta\ell_0],t) - U_{L_2}^\dagger O_X U_{L_2}    \right\|  \le 2e^3\zeta_0 c_3 \gamma t (2\ell_0 +r)^D \ell_0 e^{-\ell_0/(2k)} ,
\end{align} 
from Propositions~\ref{prop:error_time_evolution_effective_Ham} and~\ref{prop:short_time_Lieb--Robinson}, respectively.
By combining them, we obtain 
 \begin{align}
 \label{approximation_e^-iHt_tilde_rho}
&\left \| \left [ O_X(t) - U_{L_2}^\dagger O_X U_{L_2}  \right]   \rho_0(-t) \right\|_1  
\le \frac{1}{2}\zeta_0 e^{c_{0} \bar{q}} e^{-2\ell_0/\log (r) } + 2e^3 \zeta_0c_3 \gamma t (2\ell_0 +r)^D \ell_0 e^{-\ell_0/(2k)} .
\end{align}
For the second term, there exists a constant $\delta c_2 = \orderof{1} $ such that for $\ell_0 \ge (c_2 + \delta c_2) \log^2 (r)$, 
 \begin{align}
 \label{cond_ell_0_error_effective}
2e^3 \zeta_0 c_3 \gamma t (2\ell_0 +r)^D \ell_0 e^{-\ell_0/(2k)} 
\le \frac{1}{2}  \zeta_0 e^{-2\ell_0/\log (r) } \le \frac{1}{2} e^{c_{0} \bar{q}} \zeta_0 e^{-2\ell_0/\log (r) },
\end{align}
which reduces Ineq.~\eqref{approximation_e^-iHt_tilde_rho} to 
 \begin{align}
&\left \| \left [ O_X(t) - U_{L_2}^\dagger O_X U_{L_2}  \right]   \rho_0(-t) \right\|_1  
\le e^{c_{0} \bar{q}} \zeta_0 e^{-2\ell_0/\log (r) } .
\end{align}
Note that $\delta c_2$ does not depend on $\bar{q}$.
By letting $2\ell_0=\ell$ in this inequality (i.e., $L_2=X[2\ell_0]=X[\ell]$), we obtain the main inequality~\eqref{ineq:main_theorem_short_time_LR}. 
All the conditions for $\ell_0=\ell/2$ can be written in the form of~\eqref{condition_for_lenfth_R} by choosing $C_0=\orderof{1}$ appropriately.

%\red{On the second inequality~\eqref{ineq2:main_theorem_short_time_LR}, 
%when we define $L_3 = X[\ell_0']$ with $\ell'_0=\ell'/2 \ge \ell_0$, we obtain the same inequality as \eqref{ineq_effctive_L_2_unitary_error}:
% \begin{align}
%&\left \| O_X(\tilde{H}[\tilde{L},\eta\ell_0],t) - U_{L_3}^\dagger O_X U_{L_3}    \right\| 
%\le 2e^3\zeta_0 c_3 \gamma t (2\ell'_0 +r)^D \ell_0 e^{-\ell_0'/(2k)} \le \frac{1}{2}\zeta_0 e^{-2\ell_0/\log (r) } 
%\end{align} 
%where the last inequality results from the condition~\eqref{cond_ell_0_error_effective} for $\ell_0$.
%Then, by using the triangle inequality, we have  
% \begin{align}
%\left \|  U_{L_3}^\dagger O_X U_{L_3} -U_{L_2}^\dagger O_X U_{L_2} \right\|
%&\le \left \| O_X(\tilde{H}[\tilde{L},\eta\ell_0],t) - U_{L_3}^\dagger O_X U_{L_3}    \right\|  +\left \| O_X(\tilde{H}[\tilde{L},\eta\ell_0],t) - U_{L_2}^\dagger O_X U_{L_2}  \right\| 
%\notag\\
%&\le \zeta_0 e^{-2\ell_0/\log (r) } .
%\end{align} 
%We thus prove the inequality~\eqref{ineq2:main_theorem_short_time_LR}. }

This completes the proof of Subtheorem~\ref{main_theorem_short_time_LR}. $\square$

\section{Proof of Proposition~\ref{Schuch_boson_extend}: boson density after time evolution}
\label{Sec:Proof:density of bosons after time evolution}

\subsection{Restatement}

{\bf Proposition~\ref{Schuch_boson_extend}.}
\textit{
We first define the operator $M^{(s)}_i(t)$ as 
\begin{align}
M^{(s)}_i(t) := \tr \left [ \hat{n}_i^s \tilde{\rho}(t)  \right] .
\end{align}
Then, for $t\le t_0$, the following upper bound for $M^{(s)}_i(t)$ holds:
\begin{align}
\label{main_ineq_Schuch_boson_extend_re}
M_i^{(s)}(t) &\le  c'_{1} e^{c_{0} \bar{q}} \zeta_0^2 |X|^3 (c_{1}s |X| )^{s}   e^{-\dist_{i,X}}  +c''_{1}e^{c_{0} \bar{q}}\zeta_0^2 (c_{1}s)^{s},
\end{align}
where $c_{1}$, $c'_{1}$, and $c''_{1}$ are defined as 
\begin{align}
\label{parameters_c_c'_c''_t_0_re}
&c_{1} :=  e^{8\bar{J} d_G t_0}/c_{0}, \notag \\
&c'_{1} := 320 c_{0}^{-3}e^{4\bar{J} d_G t_0+ c_{0} (1+q_0/|X|)} ,  \notag \\
&c''_{1} := 80 \lambda_0c_{0}^{-1} e^{4 \bar{J} d_G t_0+ c_{0}}  .
\end{align}
Note that they are $\orderof{1}$ constants if $t_0=\orderof{1}$. 
}

\subsection{Proof}

For the proof, we start from the differential equation for $M^{(s)}_i(t)$:
\begin{align}
\label{sth_moment_der}
\frac{d}{dt} M^{(s)}_i(t) &=-i \tr \left ( \hat{n}_i^s [H, \tilde{\rho}(t) ]  \right) =i\tr \left ( [H,\hat{n}_i^s]  \tilde{\rho}(t)  \right) .
\end{align}
The form of the Hamiltonian~\eqref{def:Ham} gives 
\begin{align}
\label{Ham_commute_nb_i_s}
 [H,\hat{n}_i^s ] = [H_0,\hat{n}_i^s ] =\sum_{\langle j, j' \rangle}  J_{j,j'} [b_j b_{j'}^\dagger +{\rm h.c.} , \hat{n}_i^s ] .
\end{align}
Note that $[V,\hat{n}_i^s]=0$ because $[\hat{n}_i,\hat{n}_j]=0$ for $\forall j \in \Lambda$.

% \begin{figure}[tt]
%\centering
%\includegraphics[clip, scale=0.4]{density_boson.pdf}
%\caption{Schematic picture of the set }
%\label{fig_density_boson}
%\end{figure}

For the convenience of readers, we first consider the case of $s=1$, where $M^{(1)}_i(t)$ gives the expectation value of $\nb_i$. 
This case was considered in Ref.~\cite{PhysRevA.84.032309}. 
We prove the following lemma.

\begin{lemma}\label{lem:case_s=1}
For arbitrary site $i\in \Lambda$, the first-order moment $M^{(1)}_i(t) $ is upper-bounded by
   \begin{align}
M^{(1)}_i(t) &\le 10 \tilde{N}(\dist_{i,X})  e^{3 \bar{J} d_G t}  ,\quad  \tilde{N}(\dist_{i,X}) :=  N_Xe^{-\dist_{i,X}}  + n_0 \lambda_0,
\end{align}
where we define $N_X$ and $n_0$ as $N_X:=\tr(\rho_0 \nb_X)$ and $n_0:=\max_{i\in \Lambda}\tr(\rho_0 \hat{n_i})$, respectively.
%Note that $N_X$ has been defined by Eq.~\eqref{Def_N_L} with $L=X$.
%N_Xの定義を与える。n_0の定義を与える。
\end{lemma}

%{\bf Remark.} 
%Schuch paper との比較
%N_X のみ、n_0=0、

\subsubsection{Proof of Lemma~\ref{lem:case_s=1}} \label{sec:Proof of lem:case_s=1}

Using the relation $[b_i,\nb_i] = b_i$ (or $[b^\dagger_i,\nb_i] = -b_i^\dagger$), we obtain 
\begin{align}
 [H_0,\hat{n}_i] =\sum_{j:\dist_{i, j}=1}  J_{i,j} (b_i b_j^\dagger -{\rm h.c.}),
\end{align}
which can be used to reduce Eq.~\eqref{sth_moment_der} with $s=1$ to
\begin{align}
\label{sth_moment_der_s=1}
\frac{d}{dt} M^{(1)}_i(t) &=-\sum_{j:\dist_{i, j}=1}2 J_{i,j}  {\rm Im} \tr \left[\left ( b_i b_j^\dagger \tilde{\rho}(t)  \right) \right].
\end{align}
The Cauchy--Schwarz inequality gives an upper bound of 
\begin{align}
 \left | \tr \left[\left ( b_i b_j^\dagger \tilde{\rho}(t)  \right) \right] \right| \le \sqrt{
 \tr \left[\left ( b_i^\dagger b_i \tilde{\rho}(t)  \right) \right] \cdot \tr \left[\left ( b_j^\dagger b_j \tilde{\rho}(t)  \right) \right]} \le 
 \frac{M^{(1)}_i(t) + M^{(1)}_j(t)}{2} ,
\end{align}
and hence from $|J_{i,j}| \le \bar{J}$
\begin{align}
\left| \frac{d}{dt} M^{(1)}_i(t)\right| &\le \bar{J} \sum_{j:\dist_{i, j}=1} [M^{(1)}_i(t) + M^{(1)}_j(t)].
\end{align}
We then give the upper bound of $\vec{M}^{(1)}(t)=\{M^{(1)}_i(t)\}_{i=1}^n$ as 
\begin{align}
\label{Moment_1st_Ineq_upp}
\vec{M}^{(1)}(t) \le e^{\bar{J} d_G t} e^{\bar{J} \mathcal{M} t} \vec{M}^{(1)}(0) ,
\end{align}
where $d_G$ is the maximum degree of the graph, and the matrix $\mathcal{M}$ has nonzero elements only for $\dist_{i,j}=1$ with $\mathcal{M}_{i,j} = 1$.
As shown in Ref.~\cite{Cramer_2006,PhysRevA.84.032309}, the matrix $e^{\bar{J} \mathcal{M} t} $ satisfies 
\begin{align} 
\label{parameter_ineq_moment_LR}
& [ e^{\bar{J} \mathcal{M} t}  ]_{i,j} \le C e^{\tilde{v}_0 t - \dist_{i,j}}, \notag \\
&\tilde{v}_0= \chi \bar{J} \Delta , \quad \chi\approx 3.59,\quad \Delta = \frac{\|M\|}{2} \le d_G/2 ,\quad  C = \frac{2\chi^2}{\chi-1}\approx 10.
\end{align}
Then, because $M^{(1)}_i(0)\le n_0$ for $i\notin X$, the upper bound of $M^{(1)}_i(t)$ is given by
\begin{align}
M^{(1)}_i(t) &\le e^{\bar{J} d_G t} \sum_{j\in \Lambda}   C e^{\tilde{v}_0 t - \dist_{i,j}} M^{(1)}_j(0)  \notag \\
&\le  10e^{\tilde{v}_0 t+ \bar{J} d_G t} \left( N_Xe^{-\dist_{i,X}}  +  n_0 \sum_{j\in \Lambda}  e^{-\dist_{i,j}} \right) =: 10\tilde{N}(\dist_{i,X})  e^{(\tilde{v}_0 + \bar{J} d_G) t} 
\le 10 \tilde{N}(\dist_{i,X})  e^{3 \bar{J} d_G t}  , 
\end{align}
where we define $\tilde{N}(\dist_{i,X})=  N_Xe^{-\dist_{i,X}}  + n_0 \lambda_0$ [see Eq.~\eqref{def_gamma_0} for the definition of $\lambda_0$].

 {~}

\hrulefill{\bf [ End of Proof of Lemma~\ref{lem:case_s=1}] }

{~}

For general $s$, we use a similar approach to obtain the upper bound. 
By using the relation $[b_i,\nb_i] = b_i$ (or $[b^\dagger_i,\nb_i] = -b_i^\dagger$), we first prove the following lemma.
\begin{lemma}\label{lem:commutator_boson_s_power}
For an arbitrary function $f(x)$, the commutator $[b_i,f(\nb_i)]$ is given by
\begin{align}
\label{commutator_boson_s_power}
&[b_i,f(\nb_i)] = [f(\nb_i+1) - f(\nb_i)] b_i  
\end{align}
or 
\begin{align}
\label{commutator_boson_s_power2}
&[b_i,f(\nb_i)] = b_i  [f(\nb_i) -f (\nb_i-1)] .
\end{align}
\end{lemma}

\subsubsection{Proof of Lemma~\ref{lem:commutator_boson_s_power}} \label{sec:Proof of lem:commutator_boson_s_power}

%We here prove the equation~\eqref{commutator_boson_s_power}, but the same proof is applied to Eq.~\eqref{commutator_boson_s_power2}.
For the proof, let us define $\ket{q,r}$ as an eigenstate of $n_i$ as $n_i \ket{q,r} = q\ket{q,r}$, where the index $r$ indicates the degenerate eigenstate.  
Then, we have 
\begin{align}
[b_i,f(\nb_i)] \ket{q,r} &= b_i f(q) \ket{q,r}  -  f(\nb_i) b_i \ket{q,r} \notag \\
& = f(q) \sqrt{q} \ket{q-1,r}  -  f(q-1) \sqrt{q} \ket{q-1,r}  \notag \\
& =  [f(q) - f(q-1)] \sqrt{q} \ket{q-1,r} \notag \\
& =  [f(\nb_i+1) - f(\nb_i)] \sqrt{q} \ket{q-1,r} =  [f(\nb_i+1) - f(\nb_i)] b_i \ket{q,r} .
\end{align}
This equation holds for arbitrary eigenstates $\ket{q,r}$, and we obtain Eq.~\eqref{commutator_boson_s_power}.
For the proof of Eq.~\eqref{commutator_boson_s_power2}, we take the same approach: 
\begin{align}
[b_i,f(\nb_i)] \ket{q,r} &=  [f(q) - f(q-1)] \sqrt{q} \ket{q-1,r} \notag \\
& =b_i  [f(q) - f(q-1)]   \ket{q,r} = b_i  [f(\nb_i) -f (\nb_i-1)]  \ket{q,r} .
\end{align}
We thus prove Lemma~\ref{lem:commutator_boson_s_power}. $\square$

 {~}

\hrulefill{\bf [ End of Proof of Lemma~\ref{lem:commutator_boson_s_power}] }

{~}
%We prove the lemma by induction method.  
%For $s=1$, Eq.~\eqref{commutator_boson_s_power} is trivially obtained from $[\nb_i, b_i] $.
%We assume Eq.~\eqref{commutator_boson_s_power} for general $s$ and derive the case of $s+1$.
%By using Eq.~\eqref{commutator_boson_s_power} for the given $s$, we obtain 
%\begin{align}
%[b_i, \hat{n}_i^{s+1} ] &= \hat{n}_i^s [b_i,\nb_i ] + [b_i,\hat{n}_i^s ] \nb_i   \notag \\
%&= \hat{n}_i^s b_i + \sum_{s_1=1}^{s} (-1)^{s_1-1}\binom{s}{s_1}  \hat{n}_i^{s-s_1} b_i \nb_i  \notag \\
%&= \hat{n}_i^s b_i + \sum_{s_1=1}^{s} (-1)^{s_1-1}\binom{s}{s_1}  \hat{n}_i^{s-s_1} ( \nb_i  b_i  - b_i)  .
%\label{commutator_boson_s_power_pre}
%\end{align}
%The coefficient for $\hat{n}_i^s b_i$ is now equal to $s+1$, and the coefficients for $\{\hat{n}_i^{s+1-s_1} b_i\}_{s_1=2}^s$ is given by
% \begin{align}
%(-1)^{s_1-1}\binom{s}{s_1}   -  (-1)^{s_1-2}\binom{s}{s_1-1}= (-1)^{s_1-1} \left[ \binom{s}{s_1} +\binom{s}{s_1-1} \right] = 
%(-1)^{s_1-1}  \binom{s+1}{s_1+1} ,
%\end{align}
%where we use the Pascal's rule in the last equation. 
%The above equations reduce Eq.~\eqref{commutator_boson_s_power_pre} to Eq.~\eqref{commutator_boson_s_power} with $s+1$. 
%This completes the proof. $\square$
%

To obtain the upper bounds for a higher-order moment $M^{(s)}_i(t)$, we need to consider $[b_i,\nb_i^s]$ in Eq.~\eqref{Ham_commute_nb_i_s}.
Using Lemma~\ref{lem:commutator_boson_s_power}, we obtain 
\begin{align}
&[b_i,\nb_i^s] = \sum_{s_1=0}^{s_1-1} \binom{s}{s_1} \nb_i^{s_1} b_i  .
\end{align}
From this equation, the $s$th order moment $M^{(s)}_i(t)$ depends on the moments with lower degrees, $M^{(s')}_i(t)$ ($s'<s$). 
This point complicates the analyses significantly.
 
To overcome this difficulty, let us define an $s$th-order function $f_s(x)$ that satisfies 
\begin{align}
\label{cond_f_s_x____}
f_s(x+1) - f_s(x) = sx^{s-1} ,\quad f_s(0)=0.
\end{align}
We can always find such a function by iteratively determining the coefficient for $x^{s_1}$ ($s_1\le s$) in $f_s(x)$.
From Lemma~\ref{lem:commutator_boson_s_power}, the function $f_s(x)$ satisfies
\begin{align}
\label{commutator_f_s_nb_i}
[b_i,f_s(\nb_i)] =s\nb_i^{s-1}b_i   .
\end{align}
Although the explicit form of $f_s(x)$ is not simple, as 
\begin{align}
&f_1(x)=x,\quad f_2(x)=x^2-x,\quad f_3(x)=x^3-\frac{3}{2}x^2 + \frac{1}{2}x,\quad f_4(x)=x^4-2x^3+x^2 ,\quad  f_5(x)=x^5-\frac{5}{2}x^4+\frac{5}{3}x^3 -\frac{1}{6} x^2, \notag \\
&f_6(x)=x^6-3x^5+\frac{5}{2}x^4 -\frac{1}{2} x^2, \quad 
f_7(x)=x^7-\frac{7}{2}x^6+\frac{7}{2}x^5 -\frac{7}{6} x^3+\frac{1}{6} x, \quad  \cdots,  
\end{align}
we can prove the following lemma on the properties of the function $f_s(x)$.
\begin{lemma}\label{lem:function_s}
For an arbitrary positive integer $m \in \mathbb{N}$, we prove 
\begin{align}
\label{first_part_function_s}
(m-1)^s \le f_s(m) \le m^s  .
\end{align}
In addition, the following inequality holds:
\begin{align}
\label{second_part_function_s}
f_s(m) + \frac{s^s}{4} \ge \frac{m^s}{4}  .
\end{align}
\end{lemma}

\subsubsection{Proof of Lemma~\ref{lem:function_s}} \label{sec:Proof of lem:function_s}
For the proof, we use Eq.~\eqref{cond_f_s_x____}. 
First, for $m=1$, we have $f_s(1)=0$ because 
\begin{align}
f_s(1) = f_s(0)+ s\cdot 0^{s-1} =0,
\end{align}
where we use $ f_s(0)=0$. 
For $m=2$, from $f_s(x+1) - f_s(x) = sx^{s-1}$, we have 
\begin{align}
f_s(2) =f_s(1) + s =s   .
\end{align}
Similarly, for $m=3$ and $m=4$, we have 
\begin{align}
f_s(3) =f_s(2) + s\cdot 2^{s-1} =s (1+ 2^{s-1})  
\end{align}
and
\begin{align}
f_s(4) =f_s(3) + s\cdot 3^{s-1} =s (1+ 2^{s-1}+3^{s-1})  .
\end{align}
By repeating this procedure, we obtain
\begin{align}
f_s(m) =s  \sum_{j=1}^{m-1} j^{s-1}. 
\end{align}
For $s\ge1$, when we use 
\begin{align}
&\sum_{j=1}^{m-1} j^{s-1} \le \int_1^m x^{s-1} dx= \frac{m^s -1}{s},\notag \\
&\sum_{j=1}^{m-1} j^{s-1} \ge  \int_0^{m-1} x^{s-1} dx = \frac{(m-1)^s}{s},
\end{align}
the function $f_s(m)$ is bounded from above/below by
\begin{align}
f_s(m) \le m^s -1 \le m^s 
\end{align}
and 
\begin{align}
\label{first_part_function_s_2_2}
f_s(m) \ge (m-1)^s .
\end{align}
We thus prove the first part~\eqref{first_part_function_s} of the lemma.

Next, we prove Ineq.~\eqref{second_part_function_s}.
By using Ineq.~\eqref{first_part_function_s_2_2}, we obtain the following for arbitrary $\kappa$:
\begin{align}
\label{Ineq_kappa_s_s_4_0}
f_s(m) + \kappa \ge (m-1)^s +\kappa .
\end{align}
We then prove that for $\kappa= s^s/4$, the inequality 
\begin{align}
\label{Ineq_kappa_s_s_4}
(m-1)^s +\frac{s^s}{4} \ge \frac{m^s}{4}
\end{align}
holds. 
After proving the inequality~\eqref{Ineq_kappa_s_s_4}, we can obtain the main inequality~\eqref{second_part_function_s} by using Ineq.~\eqref{Ineq_kappa_s_s_4_0} as $f_s(m) +s^s/4 \ge (m-1)^s +s^s/4 \ge m^s/4$.

In order to prove the inequality~\eqref{Ineq_kappa_s_s_4}, we first note that the inequality~\eqref{Ineq_kappa_s_s_4} holds trivially for $s\ge m$.
We thus consider the case of $m\ge s$.
The cases of $m\ge1$ and $s=1$ are trivial; hence, we need to consider the case of $m\ge s \ge 2$.
Ineq.~\eqref{Ineq_kappa_s_s_4} reduces to
\begin{align}
\frac{s^s}{4m^s} \ge \frac{1}{4} - (1- 1/m)^s.
\end{align}
Because $(1- 1/m)^s \ge 1/4$ for $m\ge s \ge 2$, the RHS of the above inequality becomes negative,  
and hence it always holds for $m\ge s \ge 2$.

This completes the proof of the lemma. $\square$

%\begin{lemma}\label{lem:function_s}
%Let $f_s(x)$ be decomposed as 
%\begin{align}
%f_s(x) = \sum_{s_1=0}^{s} a_{s_1}^{(s)} x^{s_1}.
%\end{align}
%We then obtain 
%\begin{align}
%a_{s}^{(s)}=1, \quad \sum_{s_1=0}^{s} a_{s_1}^{(s)} = 0. 
%\end{align}
%\end{lemma}
%
%\subsubsection{Proof of Lemma~\ref{lem:function_s}} \label{sec:Proof of lem:function_s}
%First of all, the coefficient for $x^{s-1}$ in the function $f_s(x+1) - f_s(x)$ is simply calculated as 
%\begin{align}
%a_{s}^{(s)} \binom{s}{s-1} =s  a_{s}^{(s)},
%\end{align}
%which yields $a_{s}^{(s)}=1$ in order to satisfy Eq.~\eqref{cond_f_s_x____}.
%Second, the coefficient for $x^{0}$ in the function $f_s(x+1) - f_s(x)$ is given by
%\begin{align}
%\sum_{s_1=1}^s a_{s_1}^{(s)} ,
%\end{align}
%Because $a_{0}^{(s)}$ does not appear in $f_s(x+1) - f_s(x)$, we can arbitrary choose $a_{0}^{(s)}$ and hence define it as $a_{0}^{(s)}=0$.
%Thus, from the condition~\eqref{cond_f_s_x____}, we obtain $\sum_{s_1=0}^s a_{s_1}^{(s)} =0$.
%This completes the proof. $\square$

{~}

\hrulefill{\bf [ End of Proof of Lemma~\ref{lem:function_s}]}

{~}

In the following, we consider the time evolution of $f_s(\nb_i)$ instead of $\nb_i^s$, which we define as 
\begin{align}
F^{(s)}_i(t) = \tr \left[ f_s(\nb_i) \tilde{\rho}(t)\right].
\end{align}
Then, using Eq.~\eqref{commutator_f_s_nb_i}, we have 
\begin{align}
\label{commutator_n_i_H}
 [H_0,f_s(\hat{n}_i) ] &=\sum_{j:\dist_{i, j}=1} s J_{i,j}  [b_i b_j^\dagger +{\rm h.c.} , f_s(\hat{n}_i) ] 
 =\sum_{j:\dist_{i, j}=1} s J_{i,j}    \left( \hat{n}_i^{s-1}   b_i   b_j^\dagger  - {\rm h.c.}  \right),
\end{align}
where we use 
\begin{align}
[ b_i^\dagger , f_s(\nb_i) ] = - \left( [ b_i, f_s(\nb_i) ]  \right)^\dagger = - s b_i^\dagger \hat{n}_i^{s-1}.
\end{align}
To estimate the upper bound of 
\begin{align}
\label{derivative_M_S_i_t}
\left| \frac{d}{dt} F^{(s)}_i(t) \right |&=  \left| \tr \bigl( [H,f_s(\nb_i)] \tilde{\rho}(t)\bigr)  \right| 
\le \sum_{j:\dist_{i, j}=1}  |J_{i,j} |  \cdot 2s\left| \tr \bigl(  \hat{n}_i^{s-1}   b_i   b_j^\dagger   \tilde{\rho}(t)\bigr)  \right|, 
\end{align}
we need to obtain the upper bound of
\begin{align}
\label{estimate_s-1_th_moment}
\left|  \tr \left (  \hat{n}_i^{s-1}  b_i  b_j^\dagger  \tilde{\rho}(t)  \right) \right |.
%=\tr \left (   b_j^\dagger   \hat{n}_i^{s''-s''_1}  \tilde{\rho}(t) \hat{n}_i^{s'} b_i   \right)  .
\end{align}
Here we derive the following lemma.

\begin{lemma}\label{lem:expectation_b_j_b_i_poly_n_i}
For arbitrary integers $s$ and $s_1$ such that $s_1\le s$, we obtain the upper bound as
\begin{align}
\left| \tr \left (  \hat{n}_i^{s-s_1}   b_i   b_j^\dagger  \tilde{\rho}(t)  \right) \right| \le \left(1-\frac{1}{2(s-s_1+1)} \right) M_i^{(s-s_1+1)}(t) + \frac{1}{2(s-s_1+1)}M_j^{(s-s_1+1)}(t) .
\label{ineq_lem:expectation_b_j_b_i_poly_n_i}
\end{align}
\end{lemma}

\subsubsection{Proof of Lemma~\ref{lem:expectation_b_j_b_i_poly_n_i}} \label{sec:Proof of lem:expectation_b_j_b_i_poly_n_i}

%We first consider the case that $s$ is given by a even integer, namely $s=2s_0$ ($s_0 \in \mathbb{N}$). 
From the Cauchy--Schwarz inequality, we have 
\begin{align}
\left|  \tr \left (  \hat{n}_i^{s-s_1} b_i    b_j^\dagger  \tilde{\rho}(t)  \right)\right| 
&=\left|  \tr \left (  \hat{n}_i^{(s-s_1)/2} b_i   \tilde{\rho}(t) b_j^\dagger  \hat{n}_i^{(s-s_1)/2} \right) \right|  \notag \\
&\le \sqrt{\tr \left ( \hat{n}_i^{(s-s_1)/2} b_j \tilde{\rho}(t) b_j^\dagger  \hat{n}_i^{(s-s_1)/2}   \right) 
\tr \left (  \hat{n}_i^{(s-s_1)/2} b_i   \tilde{\rho}(t) b_i^\dagger \hat{n}_i^{(s-s_1)/2} \right)   }  \notag \\
&\le  \frac{1}{2} \tr \left [   \hat{n}_i^{s-s_1} \hat{n}_j  \tilde{\rho}(t)  \right] 
  +\frac{1}{2} \tr \left (  b_i^\dagger  \hat{n}_i^{s-s_1}  b_i \tilde{\rho}(t)  \right)   . \label{Ineq1_for_lem:commutator_boson_s_power}
\end{align}
For the first term, we use the H\"older inequality to derive 
\begin{align}
 \tr \left [   \hat{n}_i^{s-s_1} \hat{n}_j  \tilde{\rho}(t)  \right] 
&\le  \left( \tr \left [   \hat{n}_i^{p(s-s_1)} \tilde{\rho}(t)  \right]   \right)^{1/p}\left( \tr \left [   \hat{n}_j^{q} \tilde{\rho}(t)  \right]   \right)^{1/q}    \notag \\
&=\left( \tr \left [   \hat{n}_i^{s-s_1+1} \tilde{\rho}(t)  \right]   \right)^{(s-s_1)/(s-s_1+1)} 
\left( \tr \left [   \hat{n}_j^{s-s_1+1} \tilde{\rho}(t)  \right]   \right)^{1/(s-s_1+1)}   \notag \\
&\le \frac{s-s_1}{s-s_1+1} M_i^{(s-s_1+1)}(t) + \frac{1}{s-s_1+1}M_j^{(s-s_1+1)}(t),
\label{Ineq2_for_lem:commutator_boson_s_power}
\end{align}
where we choose $p$ and $q$ such that $1/p=(s-s_1)/(s-s_1+1)$ and $1/q=1/(s-s_1+1)$, and we use the  inequality of arithmetic and geometric means as $x^t y^{1-t} \le t x + (1-t) y$ for $0\le t\le 1$ and $x,y\ge0$.

To estimate the second term, we consider the spectral decomposition of $\tilde{\rho}(t)$ as 
\begin{align}
\label{decomposition_tilde_rho(t)}
\tilde{\rho}(t)= \sum_{q,r} \tilde{p}_{q,r} (t) \ket{q,r}\bra{q,r} ,
\end{align}
where $\ket{q,r}$ is an eigenstate of $\nb_i$ satisfying $\nb_i \ket{q,r} =q\ket{q,r}$.
We then obtain
\begin{align}
\tr \left (  b_i^\dagger  \hat{n}_i^{s-s_1}  b_i \tilde{\rho}(t)  \right)  
&= 
\sum_{q,r | q\ge 1} \tilde{p}_{q,r} (t) q (q-1)^{s-s_1} \ket{q,r}\bra{q,r}  \notag \\
&\le \sum_{q,r} \tilde{p}_{q,r} (t) q^{s-s_1+1} \ket{q,r}\bra{q,r} = M_i^{(s-s_1+1)}(t).
\label{Ineq3_for_lem:commutator_boson_s_power}
\end{align}
By applying Ineqs.~\eqref{Ineq2_for_lem:commutator_boson_s_power} and \eqref{Ineq3_for_lem:commutator_boson_s_power} to Eq.~\eqref{Ineq1_for_lem:commutator_boson_s_power}, we prove Ineq.~\eqref{ineq_lem:expectation_b_j_b_i_poly_n_i}. $\square$

 {~}

\hrulefill{\bf [ End of Proof of Lemma~\ref{lem:expectation_b_j_b_i_poly_n_i}] }

{~}

By applying Lemma~\ref{lem:expectation_b_j_b_i_poly_n_i} with $s_1=1$ to Eq.~\eqref{estimate_s-1_th_moment}, we 
reduce Ineq.~\eqref{derivative_M_S_i_t} to 
\begin{align}
\label{dt_d_F_s_i_t_upp_ineq}
\left| \frac{d}{dt} F^{(s)}_i(t) \right |
&\le  2\bar{J} \sum_{j:\dist_{i, j}=1} \left( \frac{2s-1}{2} M_i^{(s)}(t) + \frac{1}{2}M_j^{(s)}(t)\right) .
\end{align}
So that the above inequality includes only $\{ F^{(s)}_i(t)\}_{i\in \Lambda}$, we use Ineq.~\eqref{second_part_function_s} in Lemma~\ref{lem:function_s}:
\begin{align}
F^{(s)}_i(t) = \tr \left[ f_s(\nb_i) \tilde{\rho}(t)\right] 
\ge  \tr \left[\frac{\nb_i^s-s^s}{4}\tilde{\rho}(t)\right] 
=\frac{M_i^{(s)}(t) -s^s }{4} ,
\end{align}
which yields 
\begin{align}
M_i^{(s)}(t) \le 4 F^{(s)}_i(t)  + s^s . 
\end{align}
By using the above bound, we reduce Ineq.~\eqref{dt_d_F_s_i_t_upp_ineq} to 
\begin{align}
\left| \frac{d}{dt} F^{(s)}_i(t) \right |
&\le  2\bar{J} \sum_{j:\dist_{i, j}=1} \left( \frac{2s-1}{2} [4 F^{(s)}_i(t)  + s^s] + \frac{1}{2}[4 F^{(s)}_k(t)  + s^s] \right) .
\end{align}
By defining $\tilde{F}^{(s)}_i(t)=F_i^{(s)}(t) +s^s/4$, we obtain 
\begin{align}
\left| \frac{d}{dt} \tilde{F}^{(s)}_i(t) \right |
&\le  4\bar{J} \sum_{j:\dist_{i, j}=1} \left[ (2s-1) \tilde{F}^{(s)}_i(t)  + \tilde{F}^{(s)}_k(t) \right] ,
\end{align}
where we use $\frac{d}{dt} \tilde{F}^{(s)}_i(t) = \frac{d}{dt} F^{(s)}_i(t)$.

Then, we can derive an inequality similar to Eq.~\eqref{Moment_1st_Ineq_upp}: 
\begin{align}
|  \vec{\tilde{F}}^{(s)}(t) | \le e^{4(2s-1)\bar{J} d_G t}  e^{4\bar{J} \mathcal{M} t }  \vec{\tilde{F}}^{(s)}(0) ,
\end{align}
where the matrix $\mathcal{M}$ has been defined in \eqref{Moment_1st_Ineq_upp}. 
We also obtain a bound similar to Eq.~\eqref{parameter_ineq_moment_LR}: 
\begin{align} 
\label{parameter_ineq_moment_LR_2}
& [ e^{4\bar{J} \mathcal{M} t}  ]_{i,j} \le C e^{v_0 t - \dist_{i,j}}, \notag \\
&v_0= 4\chi \bar{J} \Delta , \quad \chi\approx 3.59,\quad \Delta = \frac{\|M\|}{2} \le d_G/2 ,\quad  C = \frac{2\chi^2}{\chi-1} < 10.
\end{align}
We introduce the following upper bounds: 
\begin{align}
\label{tilde_F_upper_bound/s}
&\tilde{F}^{(s)}_i(0) \le \tilde{F}^{(s)}_0 \for i\in X^\co, \notag \\ 
&\sum_{i\in X} \tilde{F}^{(s)}_i(0)  \le \tilde{F}^{(s)}_X.
\end{align}
We will calculate $\tilde{F}^{(s)}_0$ and $\tilde{F}^{(s)}_X$ explicitly later.
Using the above inequality, we obtain 
\begin{align}
\label{upp_bound_tilde_F_i_s}
\tilde{F}^{(s)}_i(t) &\le 10 e^{4(2s-1)\bar{J} d_G t}  \sum_{j\in \Lambda} e^{v_0 t - \dist_{i,j}} \tilde{F}^{(s)}_j(0)  \notag \\
&\le  10 e^{4(2s+1)\bar{J} d_G t} \left( \tilde{F}^{(s)}_X e^{-\dist_{i,X}}  + \tilde{F}^{(s)}_0 \sum_{j\in \Lambda}  e^{-\dist_{i,j}} \right) 
\le 10 \tilde{\mathcal{F}}^{(s)}(\dist_{i,X})  e^{4(2s+1)\bar{J} d_G t}  , 
\end{align}
where we use $v_0<8\bar{J} d_G$, and we define $\tilde{\mathcal{F}}^{(s)}(\dist_{i,X}) $ as 
\begin{align}
\label{def_mathcal_F_s_ids_iX}
\tilde{\mathcal{F}}^{(s)}(\dist_{i,X}) := \tilde{F}^{(s)}_X e^{-\dist_{i,X}}  + \lambda_0 \tilde{F}^{(s)}_0 .
\end{align}
Note that $\lambda_0$ has been defined in \eqref{def_gamma_0}.

Because $\tilde{F}^{(s)}_i(t):=F_i^{(s)}(t) +s^s/4$ and $F_i^{(s)}(t)\ge M_i^{(s)}(t)/4 - s^s/4$ from Lemma~\ref{lem:function_s}, 
Ineq.~\eqref{upp_bound_tilde_F_i_s} reduces to 
\begin{align}
\label{upp_bound_M_i_s_t}
M_i^{(s)}(t) \le 40 \tilde{\mathcal{F}}^{(s)}(\dist_{i,X})  e^{4(2s+1)\bar{J} d_G t}   . 
\end{align}
Finally, we obtain the explicit forms of $\tilde{F}^{(s)}_0$ and $\tilde{F}^{(s)}_X$ in Eq.~\eqref{tilde_F_upper_bound/s}. 
First, from $F_i^{(s)}(0)\le M_i^{(s)}(0)$ in Lemma~\ref{lem:function_s}, we have
\begin{align}
\label{tilde_F_s_i_0}
&\tilde{F}^{(s)}_i(0) =F_i^{(s)}(0) +\frac{s^s}{4}  \le M^{(s)}_i(0) +\frac{s^s}{4} = \tilde{F}^{(s)}_0 .
\end{align}
Second, we have 
\begin{align}
\sum_{i\in X} M^{(s)}_i(0) =\sum_{i\in X} \tr (\nb_i^s \tilde{\rho}) 
\le   \tr \left [ \left (  \sum_{i\in X} \nb_i \right)^s \tilde{\rho} \right] =    M^{(s)}_X(0) ,
\end{align}
where we define $M^{(s)}_X(0)$ as 
\begin{align}
M^{(s)}_X(0) :=  \tr \left (  \nb_X^s \tilde{\rho} \right) .
\end{align}
Using $F_i^{(s)}(0)\le M_i^{(s)}(0)$ in Lemma~\ref{lem:function_s}, we obtain 
\begin{align}
\sum_{i\in X} \tilde{F}^{(s)}_i(0)  \le 
\sum_{i\in X} \br{M^{(s)}_i(0) + \frac{s^s}{4}}
\le M^{(s)}_X(0) + \frac{|X| s^s}{4} = \tilde{F}^{(s)}_X(0) .
\end{align}

%\red{Therefore, the quantity $\tilde{\mathcal{F}}^{(s)}(\dist_{i,X})$ in Eq.~\eqref{def_mathcal_F_s_ids_iX} is upper-bounded by using $M^{(s)}_i(0)$ and $M^{(s)}_X(0)$ as follows:
%\begin{align}
%\label{def_mathcal_F_s_ids_iX_upper}
%\tilde{\mathcal{F}}^{(s)}(\dist_{i,X})  \le 
%\br{M^{(s)}_X(0) + |X| s^s/4 } e^{-\dist_{i,X}}  + \lambda_0 \max_{i\in X^\co} \br{M^{(s)}_i(0) +s^s/4}
%\end{align}
%}
In summary, we obtained 
\begin{align}
\label{tilde_F_X_s_0_tilde_F_s_i_0}
\tilde{F}^{(s)}_0 = M^{(s)}_i(0) +\frac{s^s}{4}  ,\quad \tilde{F}^{(s)}_X(0) =M^{(s)}_X(0) + \frac{|X| s^s}{4} .
\end{align}
Therefore, to upper-bound the quantity $\tilde{\mathcal{F}}^{(s)}(\dist_{i,X})$ in Eq.~\eqref{def_mathcal_F_s_ids_iX}, we need to upper-bound $M^{(s)}_i(0)$ and $M^{(s)}_X(0)$. We can prove the following lemma.
\begin{lemma} \label{lem:higher_moments_t=0}
Let $\tilde{\rho}$ be defined as in Eq.~\eqref{tilde_rho_definition/}, that is, $\tilde{\rho}= O_X \rho O_X^\dagger$. 
Then, under the condition given in Eq.~\eqref{condition_for_moment_generating}, $M_i^{(s)}(0)$ ($i\in X^\co$) is upper-bounded by
\begin{align}
\label{moment_upper_bound_single_site}
M_i^{(s)}(0) =\tr(\nb_i^s \tilde{\rho}) \le  \zeta_0^2 e^{c_{0} (\bar{q}+1)} s! c_{0}^{-s-1}.
\end{align}
Next, for $M_X^{(s)}(0)$, we obtain the upper bound of
\begin{align}
\label{moment_upper_bound_subset_site}
M_X^{(s)}(0) \le  4\zeta_0^2(|X|/c_{0})^{s+3}   s!  e^{c_{0} ( \bar{q} +1+ q_0/|X|)} . 
\end{align}
\end{lemma}

\subsubsection{Proof of Lemma~\ref{lem:higher_moments_t=0}}

Let $\Pi_{i,q}$ be a projection onto the eigenspace of $\nb_i$ with eigenvalue $q$:
\begin{align}
\nb_i \Pi_{i,q} = q \Pi_{i,q}.
\end{align}
From Ineq.~\eqref{condition_for_moment_generating}, we obtain 
\begin{align}
&\tr ( \Pi_{i,q} e^{c_{0} (\nb_i-\bar{q}) } \tilde{\rho} \Pi_{i,q}) =\tr (O_X \Pi_{i,q} e^{c_{0} (\nb_i-\bar{q})}  \rho_0  \Pi_{i,q} O_X^\dagger) 
\le \|O_X\|^2 \tr (e^{c_{0} (\nb_i-\bar{q})}\rho_0) \le \zeta_0^2  , \notag \\
&\tr ( \Pi_{i,q} e^{c_{0} (\nb_i-\bar{q})} \tilde{\rho}  \Pi_{i,q})  =\tr (e^{c_{0} (q-\bar{q})}  \Pi_{i,q} \tilde{\rho}  \Pi_{i,q})   ,
\end{align}
where we use $[\Pi_{i,q},O_{X}]=0$ for $i\in X^\co$ and $\| O_X \|=\|U_X^\dagger O_{X_0} U_X\|=\|O_{X_0}\|$ from Eq.~\eqref{O_X_unitary_definition/}.
By combining these two inequalities, we obtain  
\begin{align}
\label{upp_single_site_density_}
\tr ( \Pi_{i,q} \tilde{\rho}   \Pi_{i,q})   \le  \zeta_0^2 e^{-c_{0} (q-\bar{q})}  .
\end{align}
From this inequality, the $s$th order moment is bounded from above by
\begin{align}
\label{der_nb_i_s_rho}
\tr ( \nb_i^s \tilde{\rho}  ) &=   \sum_{q=1}^\infty \tr ( \nb_i^s \Pi_{i,q} \tilde{\rho}  \Pi_{i,q}) 
\le  \zeta_0^2 e^{c_{0} \bar{q}} \sum_{q=1}^\infty  q^s e^{-c_{0} q} 
\le \zeta_0^2 e^{c_{0} \bar{q}} \int_{0}^\infty x^s e^{-c_{0} (x-1)} dx 
= \zeta_0^2 e^{c_{0} (\bar{q}+1)} s! c_{0}^{-s-1}.
\end{align}
We thus prove Ineq.~\eqref{moment_upper_bound_single_site}.

Next, we prove Ineq.~\eqref{moment_upper_bound_subset_site}. 
By using the condition in Eq.~\eqref{cond_u_X_0_spect} and $[U_X,\nb_X]=0$, we obtain 
\begin{align}
\tr (\Pi_{X,q}\tilde{\rho}  \Pi_{X,q}) 
&=  \| \Pi_{X,q} U_X^\dagger O_{X_0} U_X \rho_0 U_X^\dagger u_{X_0}^\dagger U_X \Pi_{X,q}  \|_1 \notag \\
&=  \| \Pi_{X,q} O_{X_0} \Pi_{X, \ge q-q_0}U_X \rho_0 U_X^\dagger  \Pi_{X, \ge q-q_0} O_{X_0}^\dagger \Pi_{X,q}\|_1   ,
\end{align}
where we define $\Pi_{X, \ge q-q_0}:=\sum_{s\ge q-q_0} \Pi_{X,s}$, and $\|\cdot\|_1$ is the trace norm.
Here, we use $\Pi_{X,q} O_{X_0} \Pi_{X, < q-q_0}=0$.
The above equation yields 
\begin{align}
\label{prob_X_q_reduce}
\tr (\Pi_{X,q} \tilde{\rho}  \Pi_{X,q}) \le  \|O_{X_0}\|^2 \tr (\Pi_{X, \ge q-q_0}U_X^\dagger \rho_0 U_X \Pi_{X, \ge q-q_0}) 
=\zeta_0^2 \tr (\Pi_{X, \ge q-q_0} \rho_0  \Pi_{X, \ge q-q_0})    
\end{align}
because $\| \Pi_{X,q} O_{X_0}\| \le \|O_{X_0}\|$, $\|O_1O_2O_3\|_1 \le  \|O_1\|\cdot \|O_3\| \cdot \|O_2\|_1$, and $[U_X,\Pi_{X, \ge q-q_0}]=0$. 

Next, we estimate the following probability for the projection $\Pi_{X,q}$ with respect to $\rho_0$:
\begin{align}
\tr (\Pi_{X,q'} \rho_0  \Pi_{X,q'}) . 
\end{align}
For an arbitrary eigenstate $\ket{q'}$ such that $\nb_X\ket{q'}=q'\ket{q'}$, at least one site has more than $(q'/|X|)$ bosons, and hence 
\begin{align}
\tr (\Pi_{X,q'} \rho_0  \Pi_{X,q'}) &\le \sum_{i\in X} \sum_{s \ge q'/|X|}  \tr (\Pi_{i,s} \rho_0  \Pi_{i,s}) \notag \\
&\le  |X| \sum_{s \ge q'/|X|} e^{-c_{0} (s-\bar{q})} \le  |X| \frac{e^{c_{0} \bar{q}}}{1-e^{-c_{0}}} e^{-c_{0} q'/|X|} ,
\end{align}
where we use Ineq.~\eqref{ineq:condition_for_moment_generating} in the second inequality.
Thus, from Eq.~\eqref{prob_X_q_reduce}, we obtain
\begin{align}
\tr (\Pi_{X,q}\tilde{\rho}  \Pi_{X,q}) &\le \zeta_0^2 |X| \frac{e^{c_{0} \bar{q}}}{1-e^{-c_{0}}}  \sum_{q' \ge q-q_0} e^{-c_{0} q'/|X|}  \le \frac{\zeta_0^2 |X|  e^{c_{0} \bar{q}}}{(1-e^{-c_{0}})(1-e^{-c_{0}/|X|})}  e^{-\frac{c_{0} (q-q_0)}{|X|}} .
\end{align}
Using the above upper bound, we obtain an inequality similar to \eqref{der_nb_i_s_rho}:
\begin{align}
\tr ( \nb_X^s \tilde{\rho}  ) 
&\le  \frac{\zeta_0^2 |X|  e^{c_{0} \bar{q}}}{(1-e^{-c_{0}})(1-e^{-c_{0}/|X|})} \sum_{q=1}^\infty q^s e^{-\frac{c_{0} (q-q_0)}{|X|}}  \notag \\
&\le \frac{\zeta_0^2 |X|  e^{c_{0} ( \bar{q} +1+ q_0/|X|)}}{(1-e^{-c_{0}})(1-e^{-c_{0}/|X|})}  s! (|X|/c_{0})^{s+1}  \le  4\zeta_0^2(|X|/c_{0})^{s+3}   s!  e^{c_{0} ( \bar{q} +1+ q_0/|X|)} ,
\end{align}
where we use $c_{0}\le 1$ and the inequality $1/(1-e^{-x}) \le 2/x$ for $x\le 1$.
This completes the proof. $\square$

 {~}

\hrulefill{\bf [ End of Proof of Lemma~\ref{lem:higher_moments_t=0}] }

{~}

By applying the lemma to~Eq.~\eqref{tilde_F_X_s_0_tilde_F_s_i_0}, we obtain 
\begin{align}
&\tilde{F}^{(s)}_0(0) = \zeta_0^2e^{c_{0} (\bar{q}+1)} s! c_{0}^{-s-1} +s^s/4 , \notag \\
&\tilde{F}^{(s)}_X(0) =4\zeta_0^2 (|X|/c_{0})^{s+3}   s!  e^{c_{0} ( \bar{q} +1+ q_0/|X|)} + |X| s^s/4.
\end{align}
By combining the above inequalities with Eq.~\eqref{def_mathcal_F_s_ids_iX}, we obtain the inequality 
\begin{align}
\tilde{\mathcal{F}}^{(s)}(\dist_{i,X}) 
&\le \left[4\zeta_0^2(|X|/c_{0})^{s+3}   s!  e^{c_{0} ( \bar{q} +1+ q_0/|X|)} + |X| s^s/4 \right] e^{-\dist_{i,X}}  + \lambda_0 (\zeta_0^2 e^{c_{0} (\bar{q}+1)} s! c_{0}^{-s-1} +s^s/4) \notag \\
&\le 8\zeta_0^2 (|X|/c_{0})^{s+3}   s^s  e^{c_{0} ( \bar{q} +1+ q_0/|X|)-\dist_{i,X}}  + 2\lambda_0\zeta_0^2 e^{c_{0} (\bar{q}+1)} s^s  c_{0}^{-s-1} ,
\end{align}
where we use $c_{0}\le 1$, $\zeta_0\ge1$, and $s!\le s^s$.
We thus reduce Ineq.~\eqref{upp_bound_M_i_s_t} to 
\begin{align}
M_i^{(s)}(t) &\le 320\zeta_0^2(|X|/c_{0})^{s+3}   s^s  e^{c_{0} ( \bar{q} +1+ q_0/|X|)-\dist_{i,X}}  e^{4(2s+1)\bar{J} d_G t} +80\lambda_0\zeta_0^2 e^{c_{0} (\bar{q}+1)} s^s  c_{0}^{-s-1}e^{4(2s+1)\bar{J} d_G t}\notag \\
&\le 320\zeta_0^2 c_{0}^{-3}e^{4\bar{J} d_G t+ c_{0} ( \bar{q}+1+q_0/|X|)}|X|^3\left( \frac{ s|X| e^{8\bar{J} d_G t}}{c_{0}} \right)^s e^{-\dist_{i,X}}
+80\lambda_0\zeta_0^2c_{0}^{-1}e^{4 \bar{J} d_G t+ c_{0} (\bar{q}+1)} \left( \frac{s e^{8\bar{J} d_G t}}{c_{0}}\right)^s. \notag 
\end{align}
The RHS of this inequality increases monotonically with $t$.
Hence, when $c_{1}$, $c'_{1}$, and $c''_{1}$ are defined as in Eq.~\eqref{parameters_c_c'_c''_t_0_re}, the above inequality reduces to the main inequality~\eqref{main_ineq_Schuch_boson_extend_re} for $t\le t_0$.
This completes the proof of Proposition~\ref{Schuch_boson_extend}. $\square$

%%%%%%%%%%%%%%%%%%%%%%%%%%%%%%%%%%%%%%%%%%%%%%%%%%%%%%%%%%%%%%%%%%%%%%%%%%%%%%%%%%%%%%%%%%%%%%%%%%%%%%%%%%%%%%%%%%%%%%%%%%%%%%%%%%%%%%%%%%%%%%%%%%%%%%%%%%%%%%%%%%%%%%%%%%%%%%%%%%%%%%%%%%%%%%%%%%%%%%%%%%%%%%%%%%%%%%%%%%%%%%%%%%%%%%%%%%%%%%%%%%%%%%%%%%%%%%%%%%%%%%%%%%%%%%%%%%%%%%%%%%%%%%%%%%%%%%
% end of the proof of Proposition~1 (boson density)
%%%%%%%%%%%%%%%%%%%%%%%%%%%%%%%%%%%%%%%%%%%%%%%%%%%%%%%%%%%%%%%%%%%%%%%%%%%%%%%%%%%%%%%%%%%%%%%%%%%%%%%%%%%%%%%%%%%%%%%%%%%%%%%%%%%%%%%%%%%%%%%%%%%%%%%%%%%%%%%%%%%%%%%%%%%%%%%%%%%%%%%%%%%%%%%%%%%%%%%%%%%%%%%%%%%%%%%%%%%%%%%%%%

\section{Proof of Proposition~\ref{prop:error_time_evolution_effective_Ham}: effective Hamiltonian}
\label{Proof_prop:error_time_evolution_effective_Ham}

\subsection{Restatement}

{\bf Proposition~\ref{prop:error_time_evolution_effective_Ham}}
\textit{
Let us choose $\ell_0$ so that it satisfies 
\begin{align}
\label{condition_ell_0_prop_re}
\ell_0 \ge c_2\log^2 (r) .
\end{align} 
Recall that $r$ has been defined in Eq.~\eqref{tilde_rho_definition/}.
Then, there exists a constant $\eta$ that gives 
 \begin{align}
 \label{ineq:prop:error_time_evolution_effective_Ham_re}
&\left \| \left[ O_X(t) -O_X(\tilde{H}[\tilde{L},\eta\ell_0],t)\right] \rho_0(-t) \right\|_1 \le \frac{1}{2}e^{c_{0} \bar{q}}  \zeta_0 e^{-2\ell_0/\log (r) }
\end{align}
for $t\le t_0$. 
Here, $c_2$ and $\eta$ are $\orderof{1}$ constants which do not depend on $\bar{q}$.}

\subsection{Proof}

First, we choose $\ell_0$ such that the condition~\eqref{condition_dist_i_X_corol} in Corollary~\ref{corol:time_distribution} holds for $\dist_{i,X}=\ell_0$: 
\begin{align}
\label{condition_for_ell_0/3}
\ell_0 \ge 2 \log \left(\frac{\gamma^3 c'_{1}}{c''_{1}  }\right) +  6D \log(r),
\end{align} 
where $r$ has been defined as $X\subseteq i_0[r]$ for an appropriate $i_0\in \Lambda$.
Then, for arbitrary $i\in L_1^\co$ ($L_1=X[\ell_0]$ as in Eq.~\eqref{def_L_1_L_2}), we obtain 
\begin{align}
\label{upper_bound_P_i_t_0_2}
P_{i,\ge z_0}^{(t)}  \le 2c''_{1}e^{c_{0} \bar{q}} \zeta_0^2 \br{\frac{\tilde{c}_{1}\ell_0}{z_0}}^{\tilde{c}'_{1} \ell_0/\log (r)}.
\end{align}
For technical reasons, we adopt the following conditions on $\ell_0$ in addition to Eq.~\eqref{condition_for_ell_0/3}:
\begin{align}
\label{condition_2_for_ell_0/3}
\ell_0\ge \frac{6 \log r}{\tilde{c}'_{1}} ,\quad \ell_0 \ge \log^2 (r) .
\end{align} 
We notice that the first condition is equivalent to $\frac{1}{2}\tilde{c}'_{1} \ell_0/\log (r) \ge 3$. 
The conditions~\eqref{condition_for_ell_0/3} and \eqref{condition_2_for_ell_0/3} are satisfied by Ineq.~\eqref{condition_ell_0_prop_re} if 
we choose the parameter $c_2$ appropriately.

We first notice that if we truncate the boson number using $\bar{\Pi}_{L,q}$ in Eq.~\eqref{def:bar_Pi_X_z}, the quantum state $\tilde{\rho}(-t)$ 
is almost the same, i.e., $\bar{\Pi}_{L,q} \tilde{\rho}(-t)  \approx \tilde{\rho}(-t)$, if $q$ is sufficiently large.  
This point is rigorously justified by the following lemma.
\begin{lemma} \label{lemma:boson_number_concentration_pi_Z_q}
For the time-evolved operator $O_X(t)$ with $t\le t_0$, we obtain the following inequality if $\dist_{X,L} \ge \ell_0$ with Eq.~\eqref{condition_for_ell_0/3}:
 \begin{align}
 \label{lemma:ineq_boson_number_concentration_pi_Z_q}
\|(\bar{\Pi}_{L,q} -1) O_X(t) \rho_0(-t)  \|_1  \le 2c''_{1}e^{c_{0} \bar{q}} \zeta_0 |L| \br{\frac{\tilde{c}_{1}\ell_0}{q+1}}^{\frac{1}{2}\tilde{c}'_{1} \ell_0/\log (r)}
\le 2c''_{1}e^{c_{0} \bar{q}} \zeta_0 |L| \br{\frac{\tilde{c}_{1}\ell_0}{q}}^{\frac{1}{2}\tilde{c}'_{1} \ell_0/\log (r)}
.
\end{align}
\end{lemma}

\subsubsection{Proof of Lemma~\ref{lemma:boson_number_concentration_pi_Z_q}}
We start with the Cauchy--Schwartz inequality as follows: 
 \begin{align}
\| ( \bar{\Pi}_{L,q}   - 1) O_X(t)  \rho_0(-t)  \|_1 
&= \| ( \bar{\Pi}_{L,q}   - 1) O_X(t)  \sqrt{\rho_0(-t) } \sqrt{\rho_0(-t)}  \|_1  \notag \\
&\le \sqrt{\tr \left[ ( \bar{\Pi}_{L,q}   - 1) O_X(t)  \rho_0(-t)   O_X(t)^\dagger ( \bar{\Pi}_{L,q}   - 1)  \right]} \sqrt{ \tr[\rho_0(-t)  ]  }    \notag \\
&=\sqrt{\tr \left[ ( 1- \bar{\Pi}_{L,q}  ) \tilde{\rho}(-t)  \right]} ,
\label{first_inq_lemma_boson_n_concent}
\end{align}
where we  use $\tilde{\rho}(-t)=O_X(t)  \rho_0(-t) O_X(t)^\dagger$ from Eq.~\eqref{def_tilde_rho_t}, and $( 1- \bar{\Pi}_{L,q} )^2=1- \bar{\Pi}_{L,q}$.
For an arbitrary quantum state $\sigma$, we have 
\begin{align}
\tr (\sigma \bar{\Pi}_{L,q}) \ge 1 - \sum_{i\in L} \tr (\sigma \Pi_{i,>q}) ,
\end{align}
and hence we obtain 
 \begin{align}
1-   \tr  ( \bar{\Pi}_{L,q} \tilde{\rho}(-t) ) &\le \sum_{i\in L} \tr [\tilde{\rho}(-t) \Pi_{i,>q}] = \sum_{i\in L} P_{i,> q}^{(t)} \le 2c''_{1}e^{c_{0} \bar{q}}\zeta_0^2  |L| \br{\frac{\tilde{c}_{1}\ell_0}{q+1}}^{\tilde{c}'_{1} \ell_0/\log (r)},
\label{second_inq_lemma_boson_n_concent}
\end{align}
where we use Eq.~\eqref{upper_bound_P_i_t_0_2} in the last inequality, with $z_0=q$. 
By combining Ineqs.~\eqref{first_inq_lemma_boson_n_concent} and \eqref{second_inq_lemma_boson_n_concent}, we prove the main inequality~\eqref{lemma:ineq_boson_number_concentration_pi_Z_q}, where we use $(2c''_{1}  |L| )^{1/2} \le 2c''_{1}  |L|$ from $c''_{1} \ge 1$ [see Eq.~\eqref{parameters_c_c'_c''_t_0}]. 
This completes the proof. $\square$ 

 {~}

\hrulefill{\bf [ End of Proof of Lemma~\ref{lemma:boson_number_concentration_pi_Z_q}] }

{~}

From Lemma~\ref{lemma:boson_number_concentration_pi_Z_q}, if the boson truncation $q$ is sufficiently large, we expect to be able to approximate the time evolution $e^{-iHt}$ using the effective Hamiltonian $\tilde{H}[L,q]$ defined in Eq.~\eqref{def:tilde_Ham_effective}:
 \begin{align}
\tilde{H}[L,q] := \bar{\Pi}_{L,q} H \bar{\Pi}_{L,q}  .
\end{align}
A key ingredient is the following lemma, which specifies the approximation error between $\tilde{\rho}(t)$ and $\tilde{\rho}(\tilde{H}[L,q],t)$.

\begin{lemma} \label{prop:approx_by_hami_tilde}
For $t\le t_0$, the difference between the time evolutions by $H$ and $\tilde{H}[L,q]$ is bounded from above by  
 \begin{align}
 \label{prop:main_ineq_approx_by_hami_tilde}
\| O_X(t) \rho_0(-t) - O_X(\tilde{H}[L,q], t) \rho_0(-t) \|_1 \le  8\sqrt{2} e^{c_{0} \bar{q}}\zeta_0 t_0 c''_{1}  d_G \bar{J} q|L|  (2|L|+q) \br{\frac{\tilde{c}_{1}\ell_0}{q}}^{\frac{1}{2}\tilde{c}'_{1} \ell_0/\log (r)}  .
\end{align}
\end{lemma}

\subsubsection{Proof of Lemma~\ref{prop:approx_by_hami_tilde}} 
We start with the inequality
\begin{align}
\label{first_estimation_rho_0_error_0_2}
&\| O_X(t) \rho_0(-t) - O_X(\tilde{H}[L,q], t)  \rho_0(-t)  \|_1   \notag \\
&\le \left \| e^{iHt} O_X e^{-iHt}  \rho_0(-t) - e^{i\tilde{H}[L,q] t} O_X e^{-iHt}  \rho_0(-t)  \right\|_1  
+ \left \| e^{i\tilde{H}[L,q] t} O_X  ( e^{-iHt} - e^{-i\tilde{H}[L,q] t})  \rho_0(-t) \right \|_1\notag \\
&\le \left \| e^{iHt} O_X \rho_0 - e^{i\tilde{H}[L,q] t} O_X  \rho_0 \right\|_1  
+ \zeta_0 \left \| \rho_0  - e^{-i\tilde{H}[L,q] t} e^{iHt}  \rho_0 \right \|_1 \notag \\
&\le \left \| e^{iHt} O_X \rho_0 - e^{i\tilde{H}[L,q] t} O_X  \rho_0 \right\|_1  
+ \zeta_0 \left \| e^{i\tilde{H}[L,q] t} \rho_0- e^{iHt}  \rho_0 \right \|_1 ,
\end{align}  
where we use $\|O_X\| \le \zeta_0$ in the second inequality. 

%We have 
%\begin{align}
%\label{first_estimation_rho_0_error_0_2}
%e^{-i\tilde{H}[L,q] t} \rho_0(-t) 
%&= \bar{\Pi}_{L,q} e^{-iHt} \bar{\Pi}_{L,q}  \rho_0(-t) \notag \\
%&= \bar{\Pi}_{L,q} e^{-iHt} \rho_0(-t) + \bar{\Pi}_{L,q} e^{-iHt} (\bar{\Pi}_{L,q} -1) \rho_0(-t) \notag \\
%&= e^{-iHt}\rho_0(-t)  + (\bar{\Pi}_{L,q}-1)  \rho_0(-t)  e^{-iHt} + \bar{\Pi}_{L,q} e^{-iHt} (\bar{\Pi}_{L,q} -1) \rho_0(-t),
%\end{align}  
%and hence we obtain
%\begin{align}
%\label{first_estimation_rho_0_error}
%\| e^{-iHt}  \rho_0(-t) - e^{-i\tilde{H}[L,q] t} \rho_0(-t)  \|_1 \le 2\|(\bar{\Pi}_{L,q}-1)  \rho_0(-t)  \|_1 \le 2c''_{1}   |L| \br{\frac{\tilde{c}_{1}\ell_0}{q}}^{\frac{1}{2}\tilde{c}'_{1} \ell_0/\log (r)}.
%\end{align}  
%Note that we can apply Lemma~\ref{lemma:boson_number_concentration_pi_Z_q} to $(\bar{\Pi}_{L,q}-1)  \rho_0(-t)$ 
%because $\rho_0(-t)$ is a special case of $\tilde{\rho}(-t)$ (i.e., $O_X=1$ and $\zeta_0=1$).

%
%We second estimate 
%\begin{align}
%\label{first_estimation_rho_0_error_2}
%\| e^{iHt} O_X e^{-iHt}  \rho_0(-t) - e^{i\tilde{H}[L,q] t} O_X e^{-iHt}  \rho_0(-t)  \|_1 \le
%\| e^{iHt} O_X \rho_0- e^{i\tilde{H}[L,q] t} O_X  \rho_0  \|_1 \cdot \|e^{-iHt}\| ,
%\end{align}  
%where we use $e^{-iHt}  \rho_0(-t)= \rho_0 e^{-iHt}$. 

To estimate the RHS of the above inequality, we need to upper-bound 
\begin{align}
\| e^{iHt} O_X \rho_0- e^{i\tilde{H}[L,q] t} O_X  \rho_0  \|_1 
\end{align}  
for general $O_X$. The second term on the RHS of Eq.~\eqref{first_estimation_rho_0_error_0_2} also reduces to the above form when $O_X=\hat{1}$ is chosen. 
Let us decompose the time to $m_0$ pieces (i.e., $t=m_0 dt$) and take the limit $dt\to 0$.
To estimate the norm $\| e^{iHt} O_X \rho_0 - e^{i\tilde{H}[L,q] t} O_X  \rho_0  \| $, we start with the following identical equation:
\begin{align}
\label{error_term_sum_small_dt}
&e^{iHt} O_X \rho_0 - e^{i\tilde{H}[L,q] t} O_X  \rho_0   \notag \\
&= \sum_{m=0}^{m_0-1} 
\Bigl( e^{i\tilde{H}[L,q] mdt} e^{iH (m_0-m)dt} O_X  - e^{i\tilde{H}[L,q] (m+1)dt} e^{iH (m_0-1-m)dt} O_X   \Bigr) \rho_0  .
\end{align}  
To upper-bound the norm of the above operator, we would like to calculate the norm of 
\begin{align}
\label{upper_bound_t_1_t_2_tildeH_H}
\left(e^{i\tilde{H}[L,q] t_1}  e^{iH t_2} O_X    -e^{i\tilde{H}[L,q] (t_1+ dt) } e^{iH (t_2-dt)} O_X  \right)
 \rho_0  
\end{align}
for arbitrary $t_1$ and $t_2$ such that $t_1+t_2=t$.

From $e^{i\tilde{H}[L,q] t_1} = e^{i\tilde{H}[L,q] t_1} \bar{\Pi}_{L,q} $, we first obtain 
\begin{align}
e^{i\tilde{H}[L,q] t_1}  e^{iH t_2}
=&e^{i\tilde{H}[L,q] t_1} \bar{\Pi}_{L,q} e^{iH dt} e^{iH (t_2-dt)}  \notag \\
=&e^{i\tilde{H}[L,q] t_1} \left[\bar{\Pi}_{L,q} (1+i H_{0,L^\co}+iV) \bar{\Pi}_{L,q} dt + \bar{\Pi}_{L,q}  i ( H_{0,L} + H_{0,\partial L}) dt \right]  e^{iH (t_2-dt)}   
+ \orderof{dt^2} \notag \\
=&e^{i\tilde{H}[L,q] t_1} \left[\bar{\Pi}_{L,q} (1+i H) \bar{\Pi}_{L,q} dt + \bar{\Pi}_{L,q}  i ( H_{0,L} + H_{0,\partial L}) (1-\bar{\Pi}_{L,q}) dt \right]  e^{iH (t_2-dt)}  
+ \orderof{dt^2} \notag \\
=&e^{i\tilde{H}[L,q] (t_1+dt)} e^{iH (t_2-dt)}  +e^{i\tilde{H}[L,q] t_1}  \left[\bar{\Pi}_{L,q}  i ( H_{0,L} + H_{0,\partial L}) (1-\bar{\Pi}_{L,q}) dt \right]  e^{iH (t_2-dt)
}   + \orderof{dt^2},
\end{align}
where we use $[H_{0,L^\co}, \bar{\Pi}_{L,q}]=0$, $[V, \bar{\Pi}_{L,q}]=0$, and $\bar{\Pi}_{L,q}^2 =\bar{\Pi}_{L,q}$.
Then, the upper bound of the norm of Eq.~\eqref{upper_bound_t_1_t_2_tildeH_H} is given by  
\begin{align}
\label{upper_bound__error_t_1_t_2}
&\left \| e^{i\tilde{H}[L,q] t_1}  \left[\bar{\Pi}_{L,q}  i ( H_{0,L} + H_{0,\partial L}) (1-\bar{\Pi}_{L,q})  \right]  e^{iH (t_2-dt)}  O_X   \rho_0   \right \|_1dt + \orderof{dt^2} \notag \\
\le & \left \|\left[ \bar{\Pi}_{L,q} (H_{0,L} + H_{0,\partial L}) (1-\bar{\Pi}_{L,q})\right]  O_X (t_2-dt)  \rho_0(-t_2+dt)\right\|_1 dt   + \orderof{dt^2} \notag \\
\le &\| \bar{\Pi}_{L,q}  H_{0,L} \| \cdot   \| (1-\bar{\Pi}_{L,q}) O_X (t_2-dt)  \rho_0(-t_2+dt) \|_1dt  \notag \\
&+ \left \| \bar{\Pi}_{L,q} H_{0,\partial L} (1-\bar{\Pi}_{L,q}) O_X (t_2-dt)  \rho_0(-t_2+dt) \right\|_1dt + \orderof{dt^2} ,
\end{align}
where we use $e^{iH (t_2-dt)}  O_X   \rho_0  =  O_X (t_2-dt)  \rho_0(-t_2+dt) e^{-iH (t_2-dt)}$ in the first inequality.
By using the definition~\eqref{def:bar_Pi_X_z}, we have 
\begin{align}
\label{upper_bound_b_i_b_j_dagger}
\| \bar{\Pi}_{L,q}  b_i b_j^\dagger \| \le \sqrt{q(q+1)} \le \sqrt{2}q
\end{align}
for arbitrary $i,j \in L$, where we use $\| b_i \Pi_{i,z} \| \le \sqrt{z}$ and $\| b_i^\dagger \Pi_{i,z} \| \le \sqrt{z+1}$. 
This inequality yields 
\begin{align}
\| \bar{\Pi}_{L,q}  H_{0,L} \| \le  \sqrt{2}q \sum_{\langle i,j \rangle :i,j \in L}2 |J_{i,j}|  \le  2\sqrt{2} d_G \bar{J} q  |L|.
\end{align}
Recall that $\bar{J}$ is the upper bound of $J_{i,j}$, and $d_G$ is the maximum degree of the lattice.
Hence, the first term in Eq.~\eqref{upper_bound__error_t_1_t_2} is upper-bounded by
 \begin{align}
\label{upper_bound__error_t_1_t_2_first_term}
\| \bar{\Pi}_{L,q}  H_{0,L} \| \cdot   \| (1-\bar{\Pi}_{L,q}) O_X (t_2-dt)  \rho_0(-t_2+dt) \|_1dt \le 4 \sqrt{2} e^{c_{0} \bar{q}}c''_{1} \zeta_0 d_G \bar{J} q    |L|^2 \br{\frac{\tilde{c}_{1}\ell_0}{q}}^{\frac{1}{2}\tilde{c}'_{1} \ell_0/\log (r)},
\end{align} 
 where we use Lemma~\ref{lemma:boson_number_concentration_pi_Z_q} to upper-bound $\| (1-\bar{\Pi}_{L,q}) O_X (t_2-dt)  \rho_0(-t_2+dt) \|_1$ ($t_2 \le  t_0$).

We next consider the upper bound of the second term in Eq.~\eqref{upper_bound__error_t_1_t_2}. 
Here, the Hamiltonian $H_{0,\partial L}$ includes boundary interactions such as $J_{i,j} b_i b_j^\dagger$ with $i\in L$ and $j\notin L$.
Because the norm of $\bar{\Pi}_{L,q} b_i b_j^\dagger$ ($i\in L$, $j\notin L$) is unbounded, we need to consider
\begin{align}
\label{upper_bound__error_t_1_t_2_second_term_0}
&\left \| \bar{\Pi}_{L,q} b_i b_j^\dagger (1-\bar{\Pi}_{L,q}) O_X (t_2-dt)  \rho_0(-t_2+dt)  \right\|_1 \notag \\
&= \left \| \bar{\Pi}_{L,q}  b_i b_j^\dagger (1-\bar{\Pi}_{L,q})  \sum_{x=0}^\infty \Pi_{j,x} O_X (t_2-dt)  \rho_0(-t_2+dt)  \right\|_1  \notag \\
&\le 
 \| \bar{\Pi}_{L,q} b_i b_j^\dagger  \Pi_{j,\le q} \| \cdot \|  (1-\bar{\Pi}_{L,q})O_X (t_2-dt)  \rho_0(-t_2+dt)  \|_1 \notag \\
& \quad  + \sum_{x=q+1}^\infty \| \bar{\Pi}_{L,q}b_i b_j^\dagger\Pi_{j,x} \| \cdot \|  \Pi_{j,x} O_X (t_2-dt)  \rho_0(-t_2+dt)  \|_1  ,
\end{align}
where we use $[\Pi_{j,x}, \bar{\Pi}_{L,q}]=0$, which is obtained directly from the definition~\eqref{def:bar_Pi_X_z}.
For the first term, we can apply the same analysis as in Eq.~\eqref{upper_bound__error_t_1_t_2_first_term}:
\begin{align}
\| \bar{\Pi}_{L,q} b_i b_j^\dagger  \Pi_{j,\le q} \| \cdot \|  (1-\bar{\Pi}_{L,q})  O_X (t_2-dt)  \rho_0(-t_2+dt)  \|_1 \le  2\sqrt{2}e^{c_{0} \bar{q}} c''_{1}\zeta_0 q |L| \br{\frac{\tilde{c}_{1}\ell_0}{q}}^{\frac{1}{2}\tilde{c}'_{1} \ell_0/\log (r)}.
\end{align}

For the second term in RHS of Ineq.~\eqref{upper_bound__error_t_1_t_2_second_term_0}, because we obtain the inequality
\begin{align}
 \| \bar{\Pi}_{L,q}  b_i b_j^\dagger \Pi_{j,x} \| \le \sqrt{q(x+1)} \le \sqrt{2xq}
 \end{align}
 for $x\ge q+1$, we have 
 \begin{align}
\sum_{x=q+1}^\infty \| \bar{\Pi}_{L,q}  b_i b_j^\dagger \Pi_{j,x} \| \cdot \|   \Pi_{j,x} O_X (t_2-dt)  \rho_0(-t_2+dt)  \|_1  
\le   2\sqrt{2}e^{c_{0} \bar{q}} c''_{1} \zeta_0 \sum_{x=q+1}^\infty \sqrt{xq}\br{\frac{\tilde{c}_{1}\ell_0}{x} }^{\frac{1}{2}\tilde{c}'_{1} \ell_0/\log (r)} .
\end{align}
Here, we use the fact that for $L=\{j\}$, we have $\bar{\Pi}_{\{j\},q}= \Pi_{j,\le q}$, and hence 
 \begin{align}
\|   \Pi_{j,x} O_X (t_2-dt)  \rho_0(-t_2+dt)  \|_1  
\le \| \Pi_{j,\ge x} O_X (t_2-dt)  \rho_0(-t_2+dt)  \|_1 
&=\| (1-\bar{\Pi}_{\{j\},x-1}) O_X (t_2-dt)  \rho_0(-t_2+dt)  \|_1 \notag \\
& \le 2 e^{c_{0} \bar{q}} c''_{1} \zeta_0 \br{\frac{\tilde{c}_{1}\ell_0}{x} }^{\frac{1}{2}\tilde{c}'_{1} \ell_0/\log (r)} ,
\end{align}
where we use Ineq.~\eqref{lemma:ineq_boson_number_concentration_pi_Z_q} in the last inequality.
For an arbitrary positive $s$ ($s\ge 3$), we have 
\begin{align}
\sum_{x=q+1} x^{-s+1/2} \le \int_q^\infty x^{-s+1/2} dx = \frac{1}{s-3/2} q^{-s+3/2}   \le q^{-s+3/2},
\end{align} 
where we use the condition $s\ge 3$ in the last inequality.
The condition~\eqref{condition_2_for_ell_0/3} implies $\frac{1}{2}\tilde{c}'_{1} \ell_0/\log (r) \ge 3$; hence, we have
 \begin{align}
2\sqrt{2}e^{c_{0} \bar{q}} c''_{1} \zeta_0  \sum_{x=q+1}^\infty  \sqrt{xq}\br{\frac{\tilde{c}_{1}\ell_0}{x} }^{\frac{1}{2}\tilde{c}'_{1} \ell_0/\log (r)} 
\le 2\sqrt{2}e^{c_{0} \bar{q}}c''_{1} \zeta_0 q^{2} \br{\frac{\tilde{c}_{1}\ell_0}{q}}^{\frac{1}{2}\tilde{c}'_{1} \ell_0/\log (r)} .
\end{align}
We thus reduce Ineq.~\eqref{upper_bound__error_t_1_t_2_second_term_0} to 
\begin{align}
\label{upper_bound__error_t_1_t_2_second_term_1}
&\left \| \bar{\Pi}_{L,q} b_i b_j^\dagger (1-\bar{\Pi}_{L,q}) O_X (t_2-dt)  \rho_0(-t_2+dt) \right\|_1 
\le 2\sqrt{2}e^{c_{0} \bar{q}}c''_{1} \zeta_0q  (|L|+q)\br{\frac{\tilde{c}_{1}\ell_0}{q}}^{\frac{1}{2}\tilde{c}'_{1} \ell_0/\log (r)} .
\end{align}
When the above inequality is used,  the second term in \eqref{upper_bound__error_t_1_t_2} is upper-bounded by
\begin{align}
\label{upper_bound__error_t_1_t_2_second_term_2}
 \left \| \bar{\Pi}_{L,q} H_{0,\partial L} (1-\bar{\Pi}_{L,q}) O_X (t_2-dt)  \rho_0(-t_2+dt) \right\|_1
&\le  2\sqrt{2}e^{c_{0} \bar{q}}c''_{1} \zeta_0q  (|L|+q) \br{\frac{\tilde{c}_{1}\ell_0}{q}}^{\frac{1}{2}\tilde{c}'_{1} \ell_0/\log (r)} 
\sum_{\substack{\langle i, j \rangle \\ i\in L, j\in L^\co }}2 |J_{i,j}| \notag \\
&\le  4\sqrt{2}e^{c_{0} \bar{q}}c''_{1} \zeta_0 d_G \bar{J} q|L|  (|L|+q)  \br{\frac{\tilde{c}_{1}\ell_0}{q}}^{\frac{1}{2}\tilde{c}'_{1} \ell_0/\log (r)}  .
\end{align}

By combining Ineqs.~\eqref{upper_bound__error_t_1_t_2_first_term} and \eqref{upper_bound__error_t_1_t_2_second_term_2}, 
we reduce Ineq.~\eqref{upper_bound__error_t_1_t_2} to 
\begin{align}
&\left \| \left(e^{i\tilde{H}[L,q] t_1}  e^{iH t_2} O_X    -e^{i\tilde{H}[L,q] (t_1+ dt) } e^{iH (t_2-dt)} O_X  \right)\rho_0  \right \|_1 \notag \\
&\le 4\sqrt{2}e^{c_{0} \bar{q}}c''_{1} \zeta_0 d_G \bar{J} q|L|  (2|L|+q)   \br{\frac{\tilde{c}_{1}\ell_0}{q}}^{\frac{1}{2}\tilde{c}'_{1} \ell_0/\log (r)}   dt + \orderof{dt^2}.
\end{align}
By applying the above inequality to Eq.~\eqref{error_term_sum_small_dt}, we finally obtain 
\begin{align}
\label{Pre_prop:main_ineq_approx_by_hami_tilde}
\| e^{iHt} O_X \rho_0 - e^{i\tilde{H}[L,q] t} O_X  \rho_0  \|_1  
&\le 4\sqrt{2}e^{c_{0} \bar{q}}  t c''_{1} \zeta_0 d_G \bar{J} q|L|  (2|L|+q) \br{\frac{\tilde{c}_{1}\ell_0}{q}}^{\frac{1}{2}\tilde{c}'_{1} \ell_0/\log (r)}    + \orderof{dt} \notag \\
&\le4\sqrt{2}e^{c_{0} \bar{q}} t_0 c''_{1} \zeta_0 d_G \bar{J} q|L|  (2|L|+q) \br{\frac{\tilde{c}_{1}\ell_0}{q}}^{\frac{1}{2}\tilde{c}'_{1} \ell_0/\log (r)} ,
\end{align}  
where we take the limit $dt\to 0$ (i.e., $N\to \infty$) and use $t\le t_0$.
When we consider the case $O_X=\hat{1}$, we change the above inequality only by taking $\zeta_0=1$.  
By combining the above inequality with Eq.~\eqref{first_estimation_rho_0_error_0_2}, 
we obtain the main inequality~\eqref{prop:main_ineq_approx_by_hami_tilde}: 
\begin{align}
&\| O_X(t) \rho_0(-t) - O_X(\tilde{H}[L,q], t)  \rho_0(-t)  \|_1  
\le 8\sqrt{2}e^{c_{0} \bar{q}} t_0 c''_{1}  \zeta_0d_G \bar{J} q|L|  (2|L|+q) \br{\frac{\tilde{c}_{1}\ell_0}{q}}^{\frac{1}{2}\tilde{c}'_{1} \ell_0/\log (r)}.
\end{align}  
This completes the proof. $\square$

 {~}

\hrulefill{\bf [ End of Proof of Lemma~\ref{prop:approx_by_hami_tilde}] }

{~}

In Lemma~\ref{prop:approx_by_hami_tilde}, we choose $L=\tilde{L}$, and we choose $q$ such that 
 \begin{align}
 \label{condition_pro_prop:error_time_evolution_effective_Ham}
& 8\sqrt{2}e^{c_{0} \bar{q}}\zeta_0 t_0 c''_{1}  d_G \bar{J} q|\tilde{L}|  (2|\tilde{L}|+q) \br{\frac{\tilde{c}_{1}\ell_0}{q}}^{\frac{1}{2}\tilde{c}'_{1} \ell_0/\log (r)} 
\le \frac{1}{2}e^{c_{0} \bar{q}}\zeta_0 e^{-2\ell_0/\log (r)}  \notag \\
&\longrightarrow 
8\sqrt{2}  t_0 c''_{1}  d_G \bar{J} q|\tilde{L}|  (2|\tilde{L}|+q) \br{\frac{\tilde{c}_{1}\ell_0}{q}}^{\frac{1}{2}\tilde{c}'_{1} \ell_0/\log (r)} 
\le \frac{1}{2} e^{-2\ell_0/\log (r)} .
\end{align}
Because $|\tilde{L}| \le \gamma (r+\ell_0)^D$, if $\ell_0 \gtrsim \log^2(r)$, we can find $q\propto \ell_0$, which satisfies the above inequality.  
Now, the length $\ell_0$ is chosen such that the conditions~\eqref{condition_for_ell_0/3} and \eqref{condition_2_for_ell_0/3} hold, and hence 
there exist constants $c_2$ and $\eta$ satisfying  
\begin{align} 
\label{error_effective_time_evolu_simplified}
\| O_X(t) \rho_0(-t)  -   O_X(\tilde{H}[\tilde{L},\eta \ell_0], t) \rho_0(-t) \|_1  
\le \frac{1}{2}e^{c_{0} \bar{q}}\zeta_0 e^{-2\ell_0/\log (r)} ,
\end{align}  
for $\ell_0 \ge c_2 \log^2(r)$.  
Because the inequality~\eqref{condition_pro_prop:error_time_evolution_effective_Ham} does not contain $\bar{q}$, the constants 
$c_2$ and $\eta$ are constants of  $\orderof{1}$ which do not depend on $\bar{q}$.
%By using the inequality~\eqref{error_effective_time_evolu_simplified}, we obtain 
% \begin{align}
%&\left \|e^{-iHt}\tilde{\rho}e^{iHt} -O_X(\tilde{H}[\tilde{L},\eta\ell_0],t)^\dagger  \rho_0 O_X(\tilde{H}[\tilde{L},\eta\ell_0],t) \right\| \notag \\
%&\le \| \tilde{\rho}(t) - \tilde{\rho}(\tilde{H}[\tilde{L},\eta \ell_0],t) \|_1  + 
%\| \tilde{\rho}(\tilde{H}[\tilde{L},\eta \ell_0],t) -O_X(\tilde{H}[\tilde{L},\eta\ell_0],t)^\dagger  \rho_0 O_X(\tilde{H}[\tilde{L},\eta\ell_0],t)  \|_1 \notag \\
%&\le \frac{1}{4}e^{-2\ell_0/\log (r)}+ \| \rho_0(\tilde{H}[\tilde{L},\eta \ell_0],t) -  \rho_0 \|_1 \le \frac{1}{2}e^{-2\ell_0/\log (r)}
%\end{align}
%where in the second inequality we use 
% \begin{align} 
%&\| \tilde{\rho}(\tilde{H}[\tilde{L},\eta \ell_0],t) -O_X(\tilde{H}[\tilde{L},\eta\ell_0],t)^\dagger  \rho_0 O_X(\tilde{H}[\tilde{L},\eta\ell_0],t)  \|_1\notag \\
%&=\left \| O_X(\tilde{H}[\tilde{L},\eta\ell_0],t)^\dagger  \br{ \rho_0(\tilde{H}[\tilde{L},\eta \ell_0],t)  - \rho_0 } O_X(\tilde{H}[\tilde{L},\eta\ell_0],t)   \right \|_1 \notag\\
%&= \| \rho_0(t)-\rho_0(\tilde{H}[\tilde{L},\eta \ell_0],t)  \|_1 \le \frac{1}{4}e^{-2\ell_0/\log (r)}.
%\end{align}
%Note that $\rho_0(t_0)=\rho_0$ and $\rho_0$ is a special case of $\tilde{\rho}$, namely $u_{X_0}=U_X=1$ in Eq.~\eqref{O_X_unitary_definition/}, and hence we can apply 
%the inequality~\eqref{error_effective_time_evolu_simplified}.
We thus prove the main inequality~\eqref{ineq:prop:error_time_evolution_effective_Ham_re}. 
This completes the proof of Proposition~\ref{prop:error_time_evolution_effective_Ham}. $\square$

%%%%%%%%%%%%%%%%%%%%%%%%%%%%%%%%%%%%%%%%%%%%%%%%%%%%%%%%%%%%%%%%%%%%%%%%%%%%%%%%%%%%%%%%%%%%%%%%%%%%%%%%%%%%%%%%%%%%%%%%%%%%%%%%%%%%%%%%%%%%%%%%%%%%%%%%%%%%%%%%%%%%%%%%%%%%%%%%%%%%%%%%%%%%%%%%%%%%%%%%%%%%%%%%%%%%%%%%%%%%%%%%%%%%%%%%%%%%%%%%%%%%%%%%%%%%%%%%%%%%%%%%%%%%%%%%%%%%%%%%%%%%%%%%%%%%%%%%%%%%%%%%%%%%%%%%%%%%%%%%%%%%%%%%%%%%%%%%%%%%%%%%%%%%%%%%%%%%%%%%%%%%%%%%%%%%%%%%%%%%%%%%%%%%%%%%%%%%%%%%%%%%%%%%%%%%%%%%%%%%%%%%%%%%%%%%%%%%%%%%%%%%%%%%%%%%%%%%%%%%%%%%%%%%%%%%%%%%%%%%%%%%%%%%%%%%%%%%%%%%%%%%

\section{Proof of Proposition~\ref{prop:short_time_Lieb--Robinson}: Lieb--Robinson bound for the effective Hamiltonian}
\label{Proof_prop:short_time_Lieb--Robinson}

\subsection{Restatement}

{\bf Proposition~\ref{prop:short_time_Lieb--Robinson}.}
\textit{
When $L=L_2' :=X[2\ell_0-2k]$ ($\subset L_2$) is chosen, the approximation error in Eq.~\eqref{approximate_L_1_H_L_tau} is bounded from above by
 \begin{align}
 \label{ineq:prop:short_time_Lieb--Robinson_re}
\left\| \tilde{O}_{L_1} (\tilde{H}_\tau, 0\to t) - \tilde{O}_{L_1} (\tilde{H}_{L'_2,\tau} , 0\to t) \right\| \le 
2e^3 \zeta_0 c_3 t  |\partial L_2'| \ell_0 e^{-\ell_0/(2k)}
\end{align}
under the condition
  \begin{align}
t\le \frac{1}{e c_3'} , 
\end{align} 
where $c_3:= 4\bar{J}  \eta \gamma (2k)^D d_G$, and $c_3' := 16ekc_3 \gamma (2k)^D$.
}

\subsection{Proof}

In the Hamiltonian $\tilde{H}_\tau $, as in Eq.~\eqref{def:tilde_H_tau}, namely, 
 \begin{align}
 \label{def:tilde_H_tau_2}
\tilde{H}_\tau =\sum_{\langle i, j\rangle } J_{i,j} e^{i \tilde{V}[\tilde{L},\eta\ell_0] \tau }  \bar{\Pi}_{\tilde{L},\eta\ell_0}(b_i^\dagger b_j + {\rm h.c.})\bar{\Pi}_{\tilde{L},\eta\ell_0} e^{-i \tilde{V}[\tilde{L},\eta\ell_0] \tau} = \sum_{Z\subset\Lambda: \diam (Z) \le 2k} \tilde{h}_{Z,\tau},
\end{align} 
we have 
\begin{align}
\label{extensivity_tilde_H_tau}
\|\tilde{h}_{Z,\tau} \| \le \bar{J} \max_{i,j\in \tilde{L}} \left( \left\| \bar{\Pi}_{\tilde{L},\eta\ell_0}   (b_i^\dagger b_j + {\rm h.c.})  \bar{\Pi}_{\tilde{L},\eta\ell_0} \right\|  \right) 
\le 4\bar{J}  \eta\ell_0 \quad (Z \subset \tilde{L}) ,
\end{align} 
where we use the inequality $\bar{\Pi}_{\tilde{L},\eta\ell_0} b_i^\dagger b_j \bar{\Pi}_{\tilde{L},\eta\ell_0}  \le 2\eta\ell_0 $ for $i,j \in \tilde{L}$ [see also Ineq.~\eqref{upper_bound_b_i_b_j_dagger}].
For an arbitrary site $i_0$ such that $i_0[2k] \subset \tilde{L}$, 
an arbitrary subset $Z$ such that $Z\ni i_0$ satisfies $Z\subset \tilde{L}$ since $\diam(Z)\le 2k$. 
Therefore, by using the inequality~\eqref{extensivity_tilde_H_tau}, we have   
\begin{align}
\label{g_extensive_ness_finite} 
\sum_{Z: Z\ni i_0}\| \tilde{h}_{Z,\tau} \| 
&\le  \sum_{\langle i, j\rangle :  i,j \in i_0[2k]} \left\| J_{i,j} e^{i \tilde{V}[\tilde{L},\eta\ell_0] \tau }    (b_i^\dagger b_j + {\rm h.c.}) e^{-i \tilde{V}[\tilde{L},\eta\ell_0] \tau} \right \| \notag \\
&\le \sum_{\langle i, j\rangle :  i,j \in i_0[2k]} 4\bar{J}  \eta\ell_0 
\le \gamma (2k)^D d_G  \cdot 4\bar{J}  \eta\ell_0  =: c_3 \ell_0 ,
\end{align} 
where we define $c_3:= 4\bar{J}  \eta \gamma (2k)^D d_G$, which is an $\orderof{1}$ constant. 

In the following, we first estimate the norm 
 \begin{align}
 \label{norm_error_tilde_H_tau_L2}
&\| \tilde{O}_{L_1} (\tilde{H}_\tau, 0\to t) - \tilde{O}_{L_1} (\tilde{H}_{L,\tau} , 0\to t) \|  
\end{align} 
for general $L$ such that $L\supseteq X$, and we consider the case of $L=\tilde{L}$ later, where
$\tilde{H}_{L,\tau}$ is the subset Hamiltonian as in Eq.~\eqref{subset_Hamiltonian_L_tau}.  
To this end, we first prove the following lemma.

\begin{lemma} \label{lem:Lieb--Robinson_start}
Let $H_\tau$ be an arbitrary time-dependent Hamiltonian in the form
\begin{align}
\label{lemma_Ham_tauH}
H_\tau = \sum_{Z\subset \Lambda} h_{Z,\tau} . 
\end{align}
We also write a subset Hamiltonian on $L$ as follows:
\begin{align}
\label{subset_Hamiltonian_L_tau}
H_{L,\tau} = \sum_{Z:Z\subset L} h_{Z,\tau}.
\end{align}
Then, for an arbitrary subset $L$ such that $L \supseteq X$, we obtain 
\begin{align}
\label{lem:ineq:Lieb--Robinson_start}
\| O_X(H_\tau, 0\to t) - O_X(H_{L,\tau}, 0\to t) \|  \le \sum_{Z: Z \in \mathcal{S}_{L}} \int_0^t \left\| [h_{Z,x}(H_{L,\tau}, x\to 0), O_X] \right\| dx ,
\end{align}
where $\mathcal{S}_{L}$ is defined as a set of subsets $\{Z\}_{Z\subseteq \Lambda}$, which overlap the surface region of $L$:
\begin{align}
\mathcal{S}_{L}:= \{ Z\subseteq \Lambda | Z \cap L \neq \emptyset, Z \cap L^\co \neq \emptyset\} .
\end{align}
\end{lemma}

\subsubsection{Proof of Lemma~\ref{lem:Lieb--Robinson_start}}

By using the notation in Eq.~\eqref{notation_time_dependent_operator}, we first decompose the unitary operator $U_{H_\tau, 0\to t}$ as follows:
\begin{align}
U_{H_\tau, 0\to t}&=  \mathcal{T} e^{-i \int_0^t [H_{\partial L,x} +H_{L,x} + H_{L^\co,x}] dx} \notag \\
 &=\mathcal{T} \exp\left[-i \int_0^t U_{H_{L,\tau}+H_{L^\co,\tau}, 0\to x} H_{\partial L,x} U_{H_{L,\tau}+H_{L^\co,\tau}, 0\to x}^\dagger dx\right] \mathcal{T} e^{-i \int_0^t [H_{L,x} + H_{L^\co,x}] dx}   \notag \\
 &=\mathcal{T} \exp\left[-i \int_0^t H_{\partial L,x}(H_{L,\tau}+H_{L^\co,\tau}, x\to 0)dx\right] U_{H_{L^\co,\tau}, 0\to t} U_{H_{L,\tau}, 0\to t} \notag \\
 &=: \tilde{U}_{0\to t} U_{H_{L^\co,\tau}, 0\to t} U_{H_{L,\tau}, 0\to t}  ,
\end{align}  
where we use $[H_{L,\tau}, H_{L^\co,\tau}]=0$ for arbitrary $\tau$ in the third equation.
By using the above notation, we obtain 
\begin{align}
O_X(H_\tau , 0\to t) =&U_{H_\tau, 0\to t}^\dagger  O_X U_{H_\tau, 0\to t}\notag \\
 =&U_{H_{L,\tau}, 0\to t}^\dagger  U_{H_{L^\co,\tau}, 0\to t} ^\dagger  \tilde{U}_{0\to t}^\dagger  O_X \tilde{U}_{0\to t} U_{H_{L^\co,\tau}, 0\to t} U_{H_{L,\tau}, 0\to t} \notag \\
 =&U_{H_{L,\tau}, 0\to t}^\dagger  U_{H_{L^\co,\tau}, 0\to t} ^\dagger  \tilde{U}_{0\to t}^\dagger  [O_X, \tilde{U}_{0\to t}] U_{H_{L^\co,\tau}, 0\to t} U_{H_{L,\tau}, 0\to t}  \notag \\
 &+ U_{H_{L,\tau}, 0\to t}^\dagger  U_{H_{L^\co,\tau}, 0\to t} ^\dagger O_X U_{H_{L^\co,\tau}, 0\to t} U_{H_{L,\tau}, 0\to t}\notag \\
 =&U_{H_{L,\tau}, 0\to t}^\dagger  U_{H_{L^\co,\tau}, 0\to t} ^\dagger  \tilde{U}_{0\to t}^\dagger  [O_X, \tilde{U}_{0\to t}] U_{H_{L^\co,\tau}, 0\to t} U_{H_{L,\tau}, 0\to t} \notag \\
 &+U_{H_{L,\tau}, 0\to t}^\dagger  O_X  U_{H_{L,\tau}, 0\to t},
\end{align}
where, in the third equation, we use $\tilde{U}_{0\to t}^\dagger \tilde{U}_{0\to t} =1$, and in the fourth equation, we use $[O_X, U_{H_{L^\co,\tau}, 0\to t}]=0$, because $X\subseteq L$ (i.e., $X\cap L^\co =\emptyset$).
Therefore, we obtain the inequality 
\begin{align}
&\| O_X(H_\tau, 0\to t) - O_X(H_{L,\tau}, 0\to t) \| \le \|[O_X, \tilde{U}_{0\to t}] \| .
\end{align}
By expanding the commutator $\|[O_X, \tilde{U}_{0\to t}] \|$, we obtain
\begin{align}
\|[O_X, \tilde{U}_{0\to t}] \|   
&\le \int_0^t \| [O_X,  H_{\partial L,x}(H_{L,\tau}+H_{L^\co,\tau}, x\to 0) ] \| dx \notag \\ 
&\le \sum_{Z: Z \in \mathcal{S}_{L}} \int_0^t \left\| [h_{Z,x}(H_{L,\tau}+H_{L^\co,\tau}, x\to 0) , O_X] \right\| dx \notag \\
&=\sum_{Z: Z \in \mathcal{S}_{L}} \int_0^t \left\| [h_{Z,x}(H_{L,\tau}, x\to 0) , O_X] \right\| dx ,
\end{align}
where in the last equation we use 
\begin{align}
\left\| [h_{Z,x}(H_{L,\tau}+H_{L^\co,\tau}, x\to 0), O_X] \right\|  
= &\left\| [ U_{H_{L^\co,\tau}, x\to 0}^\dagger U_{H_{L,\tau}, x\to 0}^\dagger  h_{Z,x}U_{H_{L,\tau}, x\to 0} U_{H_{L^\co,\tau}, x\to 0} , O_X] \right\|   \notag \\
=&\left\| [ U_{H_{L,\tau}, x\to 0}^\dagger  h_{Z,x}U_{H_{L,\tau}, x\to 0} , U_{H_{L^\co,\tau}, x\to 0} O_XU_{H_{L^\co,\tau}, x\to 0}^\dagger ] \right\|  \notag \\
=&\left\| [ h_{Z,x}(H_{L,\tau}, x\to 0) , O_X] \right\| .
\end{align}
We thus obtain the main inequality~\eqref{lem:ineq:Lieb--Robinson_start}. This completes the proof. $\square$

 {~}

\hrulefill{\bf [ End of Proof of Lemma~\ref{lem:Lieb--Robinson_start}] }

{~}

In Lemma~\ref{lem:Lieb--Robinson_start}, we choose $O_X=\tilde{O}_{L_1}$ [see Eq.~\eqref{def_tilde_O_L_1}] and the subset $L$ as $L_2'$:
 \begin{align}
L_1 \subset L_2' := X[2\ell_0 - 2k] \subset L_2 , 
\end{align} 
where $L_1=X[\ell_0] \subset L_2'$ because of the condition~\eqref{new_condition_ell_0_LR}. 
We then upper-bound the norm~\eqref{norm_error_tilde_H_tau_L2} as follows:
 \begin{align}
 \label{norm_calculation_LR_h_Z_tilde}
\| \tilde{O}_{L_1} (\tilde{H}_\tau, 0\to t) - \tilde{O}_{L_1} (\tilde{H}_{L'_2,\tau} , 0\to t) \|  
\le \sum_{Z: Z \in \mathcal{S}_{L'_2}} \int_0^t \left\| [ \tilde{h}_{Z,x}(\tilde{H}_{L'_2,\tau}, x\to 0), \tilde{O}_{L_1}] \right\| dx .
\end{align} 
We note that if $Z \in \mathcal{S}_{L'_2}$, we can ensure $Z\in \tilde{L}$ because of $\diam(Z)\le 2k$.
For the estimation of the RHS of Ineq.~\eqref{norm_calculation_LR_h_Z_tilde}, we generally consider the norm of the commutator, as follows:
 \begin{align}
\left\| [ O_Z (\tilde{H}_{L'_2,\tau}, x \to 0), \tilde{O}_{L_1}] \right\|  .
 \label{norm_calculation_LR}
\end{align} 
We prove the following lemma.

\begin{lemma} \label{lem:Lieb--Robinson_calculation}
Let us choose the subset $Z$ such that $Z\in \mathcal{S}_{L_2'}$, and $\diam(Z)\le 2k$.
Then, for an arbitrary operator $O_Z$, we have an upper bound of 
 \begin{align}
 \label{ineq_lem:Lieb--Robinson_calculation}
\left\| [ O_Z (\tilde{H}_{L'_2,\tau}, x \to 0), \tilde{O}_{L_1}] \right\| \le 2e^3 \zeta_0 \|O_Z\| e^{-\ell_0/(2k)}
\end{align} 
under the condition 
 \begin{align}
 \label{cond_for_time_t_LR_short_time}
x\le \frac{1}{e c_3'} , \quad c_3' := 16ekc_3 \gamma (2k)^D,
\end{align} 
where $c_3$ has been defined in Eq.~\eqref{g_extensive_ness_finite}. 
\end{lemma}

 \begin{figure}[tt]
\centering
\includegraphics[clip, scale=0.4]{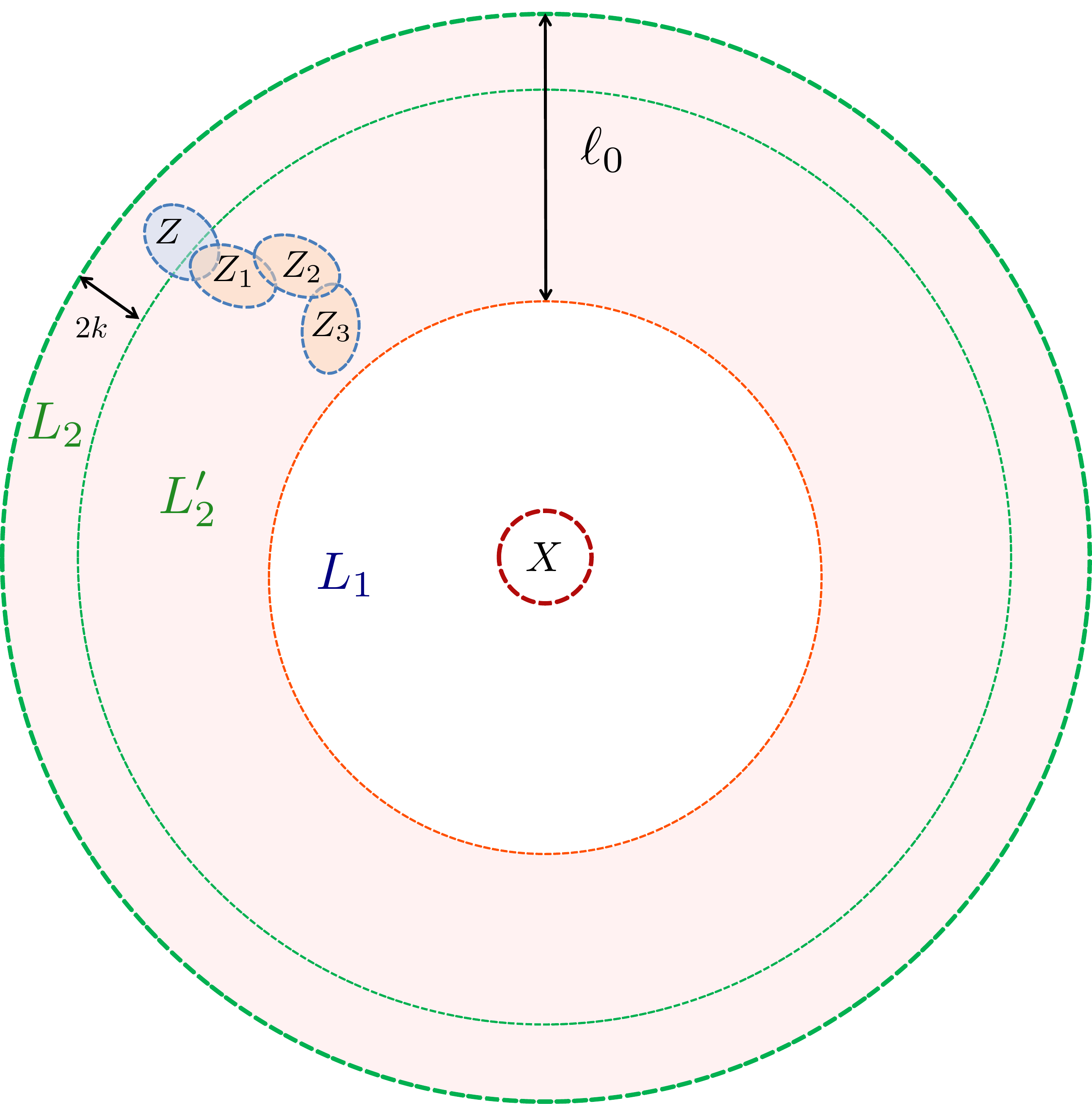}
\caption{Subsets $Z$ and $\{Z_1,Z_2,\ldots,Z_{m^\ast}\}$ in Eq.~\eqref{upper_mathcal_L_m_ast} for $m^\ast=3$.
The subset $Z$ is defined on the boundary region of $L_2'=X[2\ell_0-2k]$, and hence $Z\subset L_2$ because of $\diam(Z)\le 2k$. 
The other subsets $\{ Z_s \}_{s=1}^{m^\ast}$ satisfy $Z_1 \cap Z \neq \emptyset$, $Z_2 \cap Z_1 \neq \emptyset$ , $\ldots$, and 
$Z_{m^\ast} \cap Z_{m^\ast-1} \neq \emptyset$.
The value of $m^\ast$ is taken so that $Z_{m^\ast}$ is included in the region $\tilde{L} = L_2\setminus L_1$, where the boson number is truncated.
}
\label{fig_Lieb--Robinson_setup}
\end{figure}

\subsubsection{Proof of Lemma~\ref{lem:Lieb--Robinson_calculation}}

To estimate the upper bound of the norm~\eqref{norm_calculation_LR}, 
we use the standard recursive approach to prove the Lieb--Robinson bound~\cite{ref:LR-bound72,ref:Hastings2006-ExpDec,ref:Nachtergaele2006-LR}. 
We start with the following inequality [see Ineqs.~(S.256 and S.284) in the supplementary material of Ref.~\cite{PhysRevX.10.031010}]:
 \begin{align}
 \label{start_Lieb_Robinson_PRX}
\left\| [ O_Z (\tilde{H}_{L'_2,\tau}, x \to 0), \tilde{O}_{L_1}] \right\| \le \mathcal{L}_{m^\ast},
\end{align} 
where we choose $m^\ast$ such that $m^\ast \le \dist_{Z,L_1}/(2k)$, and $\mathcal{L}_{m^\ast}$ is defined as 
\begin{align} 
\mathcal{L}_{m^\ast}=2^{m^\ast+1} \|O_Z\|  \cdot \|\tilde{O}_{L_1}\|&\sum_{\substack{Z_1 \subset L'_2 \\ Z_1 \cap Z \neq \emptyset}}\int_0^x \|\tilde{h}_{Z_1,\tau_1}\|  d\tau_1 
\sum_{\substack{Z_2 \subset L'_2 \\ Z_{2} \cap Z_{1} \neq \emptyset}} \int_0^{\tau_1}  \|\tilde{h}_{Z_{2},\tau_2} \| d\tau_2 \notag \\
& \cdots   \sum_{\substack{Z_{m^\ast} \subset L'_2 \\ Z_{m^\ast} \cap Z_{m^\ast-1} \neq \emptyset}}   \int_0^{\tau_{m^\ast-1}}  \| \tilde{h}_{Z_{m^\ast},\tau_{m^\ast}} \| d\tau_{m^\ast} .
\label{mathdal_L_m_def_finite_________1}
\end{align} 
Note that each of the interaction terms $\tilde{h}_{Z,\tau}$ ($Z\subset L'_2$) is given in Eq.~\eqref{def:tilde_H_tau_2}.
Because of the condition $m^\ast \le \dist_{Z,L_1}/(2k)$ and $\diam(Z)\le 2k$, all the subsets $\{Z_1,Z_2,\ldots,Z_{m^\ast}\}$ in the above summations are included in the subsets $\tilde{L}$ (see also Fig.~\ref{fig_Lieb--Robinson_setup}).
Hence, we have 
 \begin{align}
\sum_{\substack{Z_s \subset L'_2 \\ Z_s \cap Z_{s-1} \neq \emptyset}}  \|\tilde{h}_{Z_s} \| 
\le \sum_{i\in Z_{s-1}} \sum_{\substack{Z_s \subset L'_2 \\ Z_s: Z_s\ni i }} 
 \|\tilde{h}_{Z_s} \|  \le  \sum_{i\in Z_{s-1}} c_3 \ell_0 \le  c_3 \gamma (2k)^D\ell_0
\end{align} 
for $s=1,2,\ldots, m^\ast$,
where we use Ineq.~\eqref{g_extensive_ness_finite} and the inequality $|Z_{s-1}| \le \gamma (2k)^D$ because of $\diam(Z_{s-1})\le 2k$.
We then obtain  
\begin{align} 
\label{upper_mathcal_L_m_ast}
\mathcal{L}_{m^\ast}\le \frac{2[2x c_3 \gamma (2k)^D\ell_0]^{m^\ast}}{m^\ast!}  \|O_Z\|  \cdot \|\tilde{O}_{L_1}\|
&\le 2 \zeta_0 \left[\frac{2ex c_3 \gamma (2k)^D\ell_0}{m^\ast}\right]^{m^\ast} \|O_Z\|   \notag \\
&=: 2 \zeta_0 \left(\frac{x c_3'\ell_0}{8k m^\ast}\right)^{m^\ast}\|O_Z\|   ,
\end{align} 
where we use $\|\tilde{O}_{L_1}\| =\zeta_0$ [see Eq.~\eqref{def_tilde_O_L_1}], and $m^\ast!\ge (m^\ast/e)^{m^\ast}$.
Note that we have defined $c_3'= 16ekc_3 \gamma (2k)^D$ in the statement of Proposition~\ref{prop:short_time_Lieb--Robinson}.  

Here, the distance $\dist_{Z,L_1}$ is larger than $\ell_0 -4k$ because $Z\in \mathcal{S}_{L_2'}$, and hence we obtain
 \begin{align}
m^\ast =  \left \lfloor \frac{\dist_{Z,L_1}}{2k} \right \rfloor  \ge  \left \lfloor \frac{\ell_0 -4k}{2k} \right \rfloor \ge \frac{\ell_0-6k}{2k}  \ge \frac{\ell_0}{8k},
\end{align} 
where we use the condition~\eqref{new_condition_ell_0_LR} in the last inequality. 
Using the above inequality for $m^\ast$ and the condition~\eqref{cond_for_time_t_LR_short_time} for $x$, we have 
\begin{align} 
\frac{x c_3'\ell_0}{8k m^\ast} \le x c_3' \le e^{-1} ,
\end{align} 
which reduces Ineq.~\eqref{upper_mathcal_L_m_ast} to 
 \begin{align} 
\mathcal{L}_{m^\ast}\le 2 \zeta_0 \|O_Z\|  e^{-m^\ast} \le 2 \zeta_0 \|O_Z\| e^{-\frac{\ell_0}{2k} +3} .
\end{align} 
By combining the above inequality with Eq.~\eqref{start_Lieb_Robinson_PRX}, we obtain the main inequality~\eqref{ineq_lem:Lieb--Robinson_calculation}.
This completes the proof. $\square$

 {~}

\hrulefill{\bf [ End of Proof of Lemma~\ref{lem:Lieb--Robinson_calculation}] }

{~}

Because $x\le t\le 1/(e c_3')$, we use Lemma~\ref{lem:Lieb--Robinson_calculation} to reduce the upper bound~\eqref{norm_calculation_LR_h_Z_tilde} to 
  \begin{align}
 \label{norm_error_tilde_H_tau_L2_2_2}
\| \tilde{O}_{L_1} (\tilde{H}_\tau, 0\to t) - \tilde{O}_{L_1} (\tilde{H}_{L'_2,\tau} , 0\to t) \|  
&\le \sum_{Z: Z \in \mathcal{S}_{L'_2}} \int_0^t 2e^3\zeta_0 \| \tilde{h}_{Z,x}\|  e^{-\ell_0/(2k)}  dx  .
\end{align} 
For an arbitrary site $i'$ such that $i' \in \partial L_2'$, we have $i'[2k] \subset \tilde{L}$, and hence we can use Ineq.~\eqref{g_extensive_ness_finite} and obtain
\begin{align}
\sum_{Z: Z \in \mathcal{S}_{L'_2}} \| \tilde{h}_{Z,x}\| \le \sum_{i' \in \partial L_2'} \sum_{Z\ni i'} \| \tilde{h}_{Z,x}\| \le 
\sum_{i' \in \partial L_2'}  c_3 \ell_0 \le  c_3 |\partial L_2'|\ell_0  . 
\end{align} 
We therefore reduce the upper bound~\eqref{norm_error_tilde_H_tau_L2_2_2} to 
  \begin{align}
 \label{norm_error_tilde_H_tau_L2_lemma_apply}
\| \tilde{O}_{L_1} (\tilde{H}_\tau, 0\to t) - \tilde{O}_{L_1} (\tilde{H}_{L'_2,\tau} , 0\to t) \|  
\le 2e^3\zeta_0 c_3 t  |\partial L_2'| \ell_0 e^{-\ell_0/(2k)} .
\end{align} 
This completes the proof of Proposition~\ref{prop:short_time_Lieb--Robinson}. $\square$

%
%
%
%
%%%%%%%%%%%%%%%%%%%%%%%%%%%%%%%%%%%%%%%%%%%%%%%%%%%%%%%%%%%%%%%%%%%%%%%%%%%%%%%%%%%%%%%%%%%%%%%%%%%%%%%%%%%%%%%%%%%%%%%%%%%%%%%%%%%%%%%%%%%%%%%%%%%%%%%%%%%%%%%%%%%%%%%%%%%%%%%%%%%%%%%%%%%%%%%%%%%%%%%%%%%%%%%%%%%%%%%%%%%%%%%%%%%%%%%%%%%%%%%%%%%%%%%%%%%%%%%%%%%%%%%%%%%%%%%%%%%%%%%%%%%%%%%%%%%%%%%%%%%%%%%%%%%%%%%%%%%%%%%%%%%%%%%%%%%%%%%%%%%%%%%%%%%%%%%%%%%%%%%%%%%%%%%%%%%%%%%%%%%%%%%%%%%%%%%%%%%%%%%%%%%%%%%%%%%%%%%%%%%%%%%%%%%%%%%%%%%%%%%%%%%%%%%%%%%%%%%%%%%%%%%%%%%%%%%%%%%%%%%%%%%%%%%%%%%%%%%%%%%%%%%%%%%%%%%%%%%%%%%%%%%%%%%%%%%%%%%%%%%%%%%%%%%%%%%%%%%%%%%%%%%%%%%%%%%%%%%%%%%%%%%%%%%%%%%%
%

\section{Quench} \label{Sec:LR_Quench}

\subsection{Setup of the quench dynamics}

We consider the quench of the Hamiltonian from $H \to H'$ with $H'$ given by
\begin{align}
\label{quench_h_X_0_def}
H'= H + h_{X_0} = H_0' + V', \quad X_{0}\subset i_0[r_0],
\end{align}
where $h_{X_0}$ is assumed to have the form of Eq.~\eqref{def:Ham}. 
In addition, $H_0'$ and $V'$ are the quenched Hamiltonians, which include free boson hopping and boson--boson interactions, respectively.
We define the function $Q(q)$ as the upper bound of the norm of $h_{X_0}\bar{\Pi}_{X_0,q}$:
\begin{align}
\label{quench_func_def}
\|h_{X_0}\bar{\Pi}_{X_0,q} \| \le   Q(q) , 
\end{align}
where $\bar{\Pi}_{X_0,q}$ has been defined in Eq.~\eqref{def:bar_Pi_X_z}.
The function $Q(q)$ characterizes the norm of the quench Hamiltonian when the boson number is truncated up to $q$.  
In Eq.~\eqref{def:Ham}, the boson--boson interactions can take arbitrary forms such as $e^{n_i n_j}$, 
but we assume here that $h_{X_0}$ includes only a finite-degree polynomial of the boson number operators, which also ensures that $Q(q)$ is given by a finite-degree polynomial 
[i.e., $Q(q)={\rm poly}(q)$]. 

\subsection{Main theorem}

We assume that the initial state $\rho_0$ is a steady state under the Hamiltonian $H$. 
After the quench of the Hamiltonian, the state $\rho_0$ no longer satisfies $[\rho_0,H']\neq 0$, and it evolves with time.
Our purpose is to find the approximation error as 
\begin{align}
\label{approx_error_quench}
\norm{\rho_0(H',t)  - U_{i_0[R]} \rho_0   U^\dagger_{i_0[R]} }_1 ,
\end{align}
where $U_{i_0[R]}$ is appropriately defined on the subset $i_0[R]$.
 
Intuitively, from Theorem~\ref{main_theorem_long_time_LR}, the quantity~\eqref{approx_error_quench} is expected to obey the same upper bound as Eq.~\eqref{main_theorem_short_time_LR_main_ineq}. 
However, the situation is not that simple. 
In considering $\rho_0(H',t)$, we may consider 
\begin{align}
e^{-iH't}  \rho_0= \mathcal{T} e^{-\int_0^t h_{X_0}(H, -x) dx }   e^{-iHt} \rho_0 = 
\mathcal{T} e^{-\int_0^t h_{X_0}(H, -x) dx } \rho_0  e^{-iHt}  ,
\end{align}
where we use $[\rho_0,H]=0$ in the last equation. 
From Theorem~\ref{main_theorem_long_time_LR}, we can upper-bound 
$$
\norm{ \br{h_{X_0}(H, -x) - U_{i_0[R]} h_{X_0}  U^\dagger_{i_0[R]} } \rho_0}_1.
$$
To approximate $\mathcal{T} e^{-\int_0^t h_X(H, -x) dx }$, we need to consider  
\begin{align}
h_{X_0}(H,-t_1) h_{X_0}(H,-t_2)  \cdots  h_{X_0}(H,-t_m) \rho_0 
\end{align}
with $t_1 \le t_2 \le \cdots \le t_m$. 
The approximation of $h_{X_0}(H,-t_1) h_{X_0}(H,-t_2)  \cdots  h_{X_0}(H,-t_m) $ onto the region $i_0[R]$ is 
nontrivial only from Theorem~\ref{main_theorem_long_time_LR}.
In addition, we need to consider that the norm of $h_{X_0}$ is not finitely bounded in general.
In fact, we can address these problems and prove the following theorem.

\begin{theorem} \label{thm:quench_boson_LR}
Let $h_{X_0}$ be an arbitrary operator in the form of~\eqref{def:Ham} that satisfies the condition \eqref{quench_func_def}.
Then, for an arbitrary quantum state $\rho_0$ satisfying $[\rho_0,H]=0$ and assuming that~\eqref{only_the_assumption_initial} holds, 
the time evolution $\rho_0(t)$ is approximated using the local unitary operator $U_{i_0[R]} $ supported on $i_0[R]$ with the following approximation error:
\begin{align}
\label{main_theorem_short_time_LR_main_ineqquench}
&\norm{\rho_0(H',t)  - U_{i_0[R]} \rho_0   U^\dagger_{i_0[R]} }_1 \le \exp\br{c_{0} \bar{q}- C'_1 \frac{(R-r_0)}{t\log (R)} + C'_2 \log(R)} \quad (t\ge 1) ,
\end{align}
where $C'_1$ and $C'_2$ are constants of $\orderof{1}$ which are independent of $\bar{q}$ and depend only on the details of the system. 
Moreover, the computation cost to construct the unitary operator $U_{i_0[R]}$ is at most
 \begin{align}
 \label{boson_space_prop_comp_cost_quench_thm}
 \exp\left [ \mathcal{O}\br {R^D  \log(R)} \right]  .
\end{align}
\end{theorem}

{\bf Remark.} 
For $r_0=\orderof{1}$, $\bar{q}=\orderof{1}$, and $D=1$, in order to obtain a fixed error $\epsilon$, we need to choose $R$ as 
\begin{align}
R\approx t \log^2(t) +  t\log(1/\epsilon) \log\log(1/\epsilon),
\end{align}
and hence the time complexity is given by
 \begin{align}
\exp \left[ t \log^3(t) +t \log(1/\epsilon) \log\log^2(1/\epsilon) \right].
\end{align}

\subsection{Proof of Theorem~\ref{thm:quench_boson_LR}}

The proof is obtained using an approach similar to that used for Theorem~\ref{main_theorem_long_time_LR}. 
The approximation for the short-time evolution is crucial. 
We can prove the following proposition (see Sec.~\ref{proof_prop_main_theorem_short_time_LR_quench} for the proof).
\begin{prop} \label{main_theorem_short_time_LR_quench}
 Let $u_X$ be an arbitrary unitary operator such that $[u_{X},\nb_X]=0$, $X\subseteq i[r]$, and $\Pi_{X,\ge q_0}u_X=0$ for fixed $q_0$. 
Then, the time evolution $\tilde{\rho}(H',t)$ with
\begin{align}
\tilde{\rho}:= u_X \rho_0 u_X^\dagger 
\end{align}
is approximated using a unitary operator $U_{X[\ell]}$ ($[U_{X[\ell]},\nb_{X[\ell]}]=0$) as follows:
\begin{align}
\label{ineq:main_theorem_short_time_LR_quench}
&\left \| \tilde{\rho}(t)-  U_{X[\ell]} \rho_0 U_{X[\ell]}^\dagger \right\|_1
\le  e^{c_{0} \bar{q}} e^{-\ell/\log (r)} 
\end{align}
for $t\le \Delta t_0$, 
where $\Delta t_0=\orderof{1}$, and the length $\ell$ is chosen such that it satisfies
 \begin{align}
 \label{condition_for_lenfth_R_quench}
\ell \ge  C_0'\log^2(r),
\end{align}
with $C_0'=\orderof{1}$, which does not depend on $\bar{q}$.
Moreover, the unitary operator $U_{X[\ell]}$ satisfies 
 \begin{align}
 \label{boson_space_prop}
U_{X[\ell]} \Pi_{X[\ell],\ge \max(q_0, \eta' \ell |X|/2 )} =0
\end{align}
for $\eta'=\orderof{1}$.
In addition, the computational cost of preparing $U_{X[\ell]}$ is  
 \begin{align}
  \label{boson_space_prop_comp_cost}
 \exp\left [ \mathcal{O}\br { (r+\ell)^D  \log(r+\ell+q_0)} \right]  .
\end{align}
\end{prop}

{~}\\

\noindent 
From the above proposition, we can easily prove Theorem~\ref{thm:quench_boson_LR} by connecting the short-time evolution as described in Sec.~\ref{sec:proof:Lieb--Robinson bound for short-time evolution}. 
We adopt the same decomposition of the time and the length as in Sec.~\ref{sec:proof:Lieb--Robinson bound for short-time evolution}.
First, we can derive an inequality similar to Eq.~\eqref{unitary_connect_upper_bound} as follows:
 \begin{align}
 \label{unitary_connect_upper_bound_quenc} 
&\norm{ \rho_0(H',t)  - U_{X_{m_t}}^{(m_t)} \rho_0    U_{X_{m_t}}^{(m_t)\dagger} }_1
\le 
\sum_{m=0}^{m_t}   \norm {  \br{U_{X_{m-1}}^{(m-1)} \rho_0    U_{X_{m-1}}^{(m-1)\dagger}}(H',\Delta t)  - U_{X_{m}}^{(m)} \rho_0    U_{X_{m}}^{(m)\dagger} }_1,
\end{align}
where each of the unitary operators $\{ U_{X_{m}}^{(m)} \}_{m=1}^{m_t}$ gives the approximation of 
 \begin{align}
\br{ U_{X_{m-1}}^{(m-1)} \rho_0    U_{X_{m-1}}^{(m-1)\dagger}} (H',\Delta t)  \approx U_{X_{m}}^{(m)} \rho_0    U_{X_{m}}^{(m)\dagger} . 
\end{align}
By using Proposition~\ref{main_theorem_short_time_LR_quench} iteratively, 
we prove an inequality similar to Ineq.~\eqref{proof_of_theorem_main_last}:
 \begin{align}
&\norm{ \rho_0(H',t)  - U_{X_{m_t}}^{(m_t)} \rho_0    U_{X_{m_t}}^{(m_t)\dagger} }_1
\le 
 \exp\left(c_{0} \bar{q} -\frac{\Delta t (R-r_0)}{t\log(R)}+\frac{1}{\log(R)} +(C_0'+1) \log(R) \right) ,
\end{align}
which reduces to the main inequality~\eqref{main_theorem_short_time_LR_main_ineqquench} by appropriately choosing $C_1'$ and $C_2'$.
In addition, from Eq.~\eqref{boson_space_prop_comp_cost}, the construction of each operator $\{ U_{X_{m}}^{(m)} \}_{m=1}^{m_t}$ 
has a maximum computational cost of 
 \begin{align}
  \label{boson_space_prop_comp_cost_quench_total_time}
 \exp\left [ \mathcal{O}\br {R^D  \log(R)} \right]  .
\end{align}
We thus prove Theorem~\ref{thm:quench_boson_LR}. $\square$

\subsection{Proof of Proposition~\ref{main_theorem_short_time_LR_quench}}  
\label{proof_prop_main_theorem_short_time_LR_quench}

 \begin{figure}[tt]
\centering
\includegraphics[clip, scale=0.4]{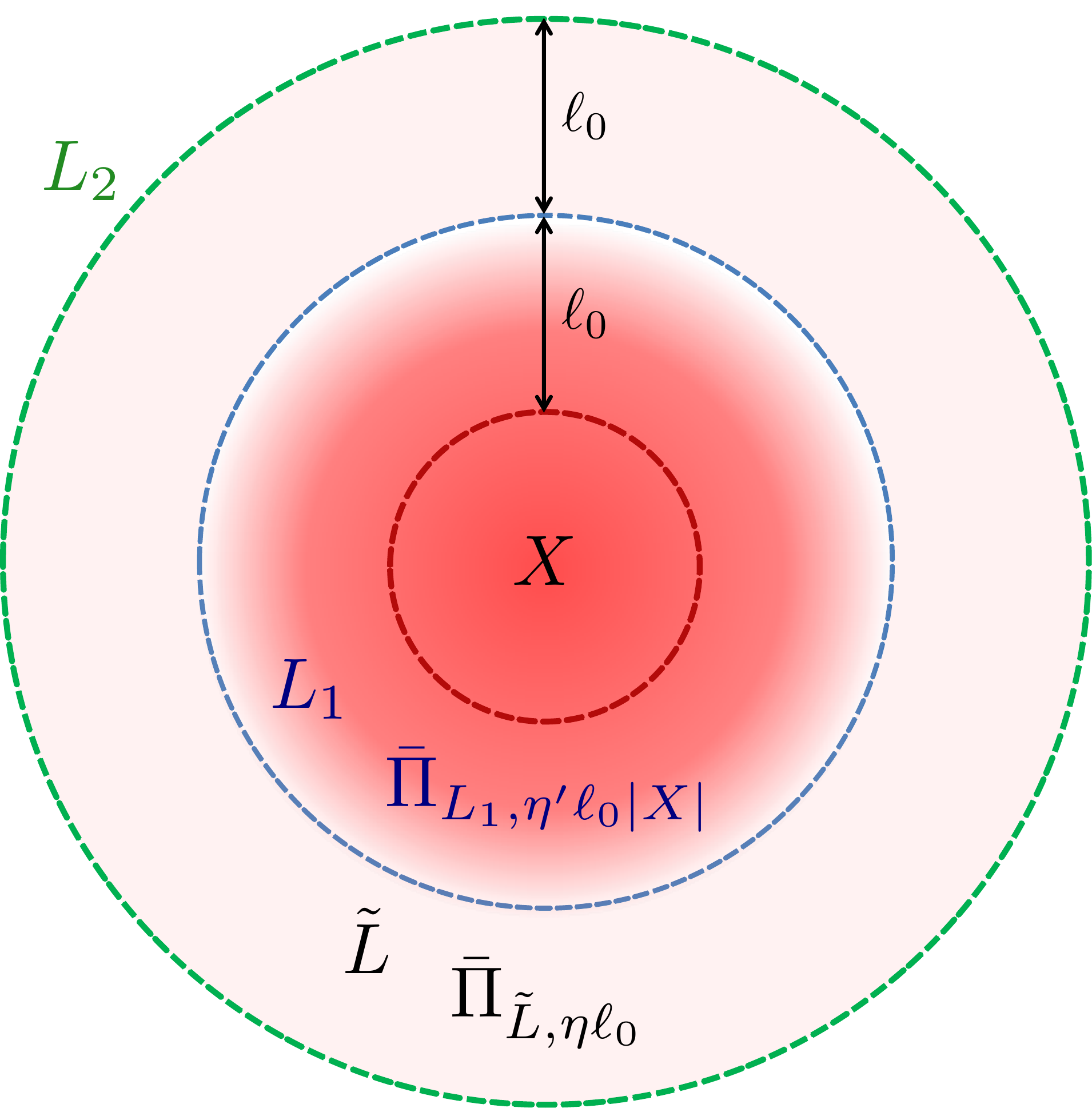}
\caption{Boson number truncation. In the proof of Proposition~\ref{main_theorem_short_time_LR_quench}, 
we truncate the boson number in the region $L_1$ in addition to that in $\tilde{L}$ [see Eq.~\eqref{def:bar_Pi_X_z_new}].  
In the region $L_1$, the boson number distribution decays exponentially beyond $\orderof{|X|}$ [see Ineq.~\eqref{boson_dist_region_L1}]. 
}
\label{fig_boson_truncation_quench}
\end{figure}

We adopt the same definition of $\tilde{\rho}(H',t)$ as in Eq.~\eqref{def_tilde_rho_t}: 
\begin{align}
\tilde{\rho}(H',t)= e^{-iH't} u_X \rho_0 u_X^\dagger e^{iH't}.
\end{align}
Then, Proposition~\ref{Schuch_boson_extend} and Corollary~\ref{corol:time_distribution} hold for $\tilde{\rho}(H',t)$ for $t\le t_0$ because the quenched Hamiltonian has the form of Eq.~\eqref{def:Ham}.

In the second step, we also define the effective Hamiltonian by truncating the boson number in a particular region.
Following Sec.~\ref{sec:def_Effective Hamiltonian}, we define the regions $L_1$, $L_2$, and $\tilde{L}$ in the same ways as in Eqs.~\eqref{def_L_1_L_2}
and \eqref{def:tilde_L}.
The main difference from the case of Subtheorem~\ref{main_theorem_short_time_LR} is 
that we truncate the boson number on the subset $L_1$ in addition to the region $\tilde{L}$ (see Fig.~\ref{fig_boson_truncation_quench}).
There are two main reasons for this additional truncation: 
i) we need to upper-bound the norm of $h_{X_0}$ ($X_0 \in L_1$) in Eq.~\eqref{quench_h_X_0_def}, and
ii) to estimate the computational cost of constructing the unitary operator, we need to restrict the maximum boson number in the region $L_1$.

Here, we perform boson number truncation in the region $L_2$. 
In the region $L_1$, the bosons can become concentrated on one site, and hence we need to choose a sufficiently large truncation number.
We note that the moment function $M^{(s)}_i(t)$ in the region $L_1$ can also be estimated by Proposition~\ref{Schuch_boson_extend}, which gives 
\begin{align}
\label{main_ineq_Schuch_boson_extend_i_in_X_0}
M_i^{(s)}(t) &\le  c'_{1}e^{c_0\bar{q}}\zeta_0^2 |X|^3 (c_{1}s |X| )^{s} +c''_{1}e^{c_0\bar{q}}\zeta_0^2 (c_{1}s)^{s} \quad (i\in L_1), 
\end{align}
where we take $\dist_{i,X}=0$ in Ineq.~\eqref{main_ineq_Schuch_boson_extend}. 
The above inequality yields 
\begin{align}
\label{boson_dist_region_L1}
P_{i,\ge z_0}^{(t)} = e^{c_0\bar{q}-\orderof{z_0/|X|}} \quad (i\in L_1)
\end{align}
if we use the Markov inequality.

We define the boson truncation operator $\bar{\Pi}_{L_2,q,q'}$ as follows:
 \begin{align} \label{def:bar_Pi_X_z_new}
\bar{\Pi}_{L_2,q,q'} :=\bar{\Pi}_{\tilde{L},q} \bar{\Pi}_{L_1,q'} ,
\end{align} 
where $\bar{\Pi}_{L,q}$ has been defined in Eq.~\eqref{def:bar_Pi_X_z}. 
Recall that $L_2=L_1\sqcup \tilde{L}$. 
By using this notation, we also define the effective Hamiltonian $\tilde{H}'[L_2,q,q']$ for the Hamiltonian $H'$ in the same way as in Eq.~\eqref{def:tilde_Ham_effective}.

To estimate the difference between the time evolutions by $H'$ and $\tilde{H}'[L_2,q,q']$, 
we can use the proof technique we used for Proposition~\ref{prop:error_time_evolution_effective_Ham},
which yields the following inequality:
 \begin{align}
 \label{ineq:lemma:error_time_evolution_effective_Ham}
&\left \| \left[ u_X(H',-t) - u_X(\tilde{H}'[L_2,\eta \ell_0,\eta' \ell_0|X|],-t) \right]\rho_0(H',t) 
% \rho_0[L_2,\eta\ell_0, \eta'\ell_0|X|],t)  
\right\|_1 \le \frac{1}{8} e^{c_0\bar{q}} e^{-2\ell_0/\log (r) }
\end{align}
for $t\le t_0$, where $\ell_0$ is chosen as in Eq.~\eqref{condition_ell_0_prop}, 
\begin{align}
\ell_0 \ge c_2\log^2 (r) ,
\end{align} 
and $\eta$ and $\eta'$ are constants of $\orderof{1}$, which are chosen appropriately.
In the following, for simplicity, we describe 
 \begin{align}
 \label{notation_H'_tilde}
\tilde{A}[L_2,\eta\ell_0, \eta'\ell_0|X|] \to \tilde{A}   \quad ({\rm i.e.,} \ \tilde{H}'[L_2,\eta\ell_0, \eta'\ell_0|X|] \to \tilde{H}' )
\end{align}
for an arbitrary operator $A$ by omitting the information on the boson number truncation $[L_2,\eta\ell_0, \eta'\ell_0|X|]$. 

We focus on the time evolution of $\tilde{\rho}$ by $\tilde{H}'$:
\begin{align}
\tilde{\rho}(\tilde{H}',t)= u_X(\tilde{H}',-t) \rho_0(\tilde{H}',t) [u_X(\tilde{H}',-t)]^\dagger .
\end{align}
We first consider the local approximations of $u_X(\tilde{H}',-t)$. 
For the unitary operator, we can prove the statement in Proposition~\ref{prop:short_time_Lieb--Robinson}, which gives 
 \begin{align}
 \label{u_X_approx_U_1_L_2}
&\norm{u_X(\tilde{H}',-t) - U^\dagger_{1,L_2}u_X U_{1,L_2}} 
\le 2e^3  c_3 t  |\partial L_2'| \ell_0 e^{-\ell_0/(2k)}, \notag \\
&c_3:= 4\bar{J}  \eta \gamma (2k)^D d_G, \quad c_3' := 16ekc_3 \gamma (2k)^D   
\end{align}  
for $t\le 1/(ec_3')$
with $[U_{1,L_2},\nb_{L_2}]=0$, 
where we set $\zeta_0=1$ in Ineq.~\eqref{ineq:prop:short_time_Lieb--Robinson} because $\|u_X\|=1$. 
Note that the statement in Proposition~\ref{prop:short_time_Lieb--Robinson} is not affected by the boson truncation on $L_1$, and hence Ineq.~\eqref{u_X_approx_U_1_L_2} does not depend on $\eta'$ (see also Sec.~\ref{Proof_prop:short_time_Lieb--Robinson}).
The unitary operator $U_{1,L_2}$ was given explicitly in Eq.~\eqref{unitary_explicit_form_L_2} in Sec.~\ref{sec:Lieb--Robinson bound for the effective Hamiltonian}.
Then, the unitary operator $U_{1,L_2}$ has the following form:
 \begin{align}
  \label{u_X_approx_U_1_L_2_choice1}
e^{-i \tilde{V}'_{X[k]} t} \mathcal{T}e^{-i\int_0^t e^{i \tilde{V}' \tau} \tilde{H}'_{0,L_2'} e^{-i \tilde{V}' \tau} d\tau}  ,
\end{align}  
where  $L_2':=X[2\ell_0-2k]$. 
When the above unitary operator acts on $u_X$ in Eq.~\eqref{u_X_approx_U_1_L_2}, it is equivalent to   
\begin{align}
  \label{u_X_approx_U_1_L_2_choice2}
e^{-i \tilde{V}'_{L_2} t} \mathcal{T}e^{-i\int_0^t e^{i \tilde{V}'_{L_2} \tau} \tilde{H}'_{0,L_2'} e^{-i \tilde{V}'_{L_2} \tau} d\tau} 
=e^{-i(\tilde{H}'_{0,L_2'}+\tilde{V}'_{L_2})t} =:  U_{1,L_2} . 
\end{align}  

By combining Ineqs.~\eqref{ineq:lemma:error_time_evolution_effective_Ham} and  \eqref{u_X_approx_U_1_L_2}, we obtain
 \begin{align}
\left \| \left[ u_X(H',-t) - U^\dagger_{1,L_2}u_X U_{1,L_2}  \right]\rho_0(H',t) \right\|_1
 \le \frac{1}{8}e^{c_0\bar{q}} e^{-2\ell_0/\log (r) }+2e^3  c_3 t  |\partial L_2'| \ell_0 e^{-\ell_0/(2k)} ,
\end{align}
which yields 
 \begin{align}
 \label{approx_who_H'_U_L_2'}
\left \|\tilde{\rho}(H',t)   - \br{ U^\dagger_{1,L_2}u_X U_{1,L_2}} \rho_0(H',t) \br{ U^\dagger_{1,L_2}u_X U_{1,L_2} }^\dagger \right\|_1
 \le \frac{1}{4}e^{c_0\bar{q}}e^{-2\ell_0/\log (r) }+4e^3  c_3 t   |\partial L_2'| \ell_0 e^{-\ell_0/(2k)} ,
\end{align}
where we use the equation $\tilde{\rho}(H',t)=u_X(H',-t)  \rho_0(H',t) u_X(H',-t)^\dagger $.

Here, the initial state $\rho_0$ is \textit{not invariant} under the time evolution of $e^{-iH't}$.
Therefore, the remaining task is to estimate the approximate error of 
 \begin{align}
\rho_0(H',t)   \approx U_{2,L_2} \rho_0 U_{2,L_2}^\dagger,  
\end{align}
where $U_{2,L_2}$ is appropriately chosen. 
\begin{lemma} \label{lemma:error_time_evolution_rho_0_effective}
We can find a unitary operator $U_{2,L_2}$ that approximates $\rho_0(H',t)$ with an error of 
 \begin{align}
 \label{ineq_lemma:error_time_evolution_rho_0_effective}
\norm{ \rho_0(H',t) - U_{2,L_2} \rho_0 U_{2, L_2}^\dagger }_1 
\le  \frac{1}{2}e^{c_0\bar{q}} e^{-2\ell_0/\log (r) }  +  4e^3 Q(\eta'\ell_0|X| ) c_3 t^2  |\partial L_2'| \ell_0 e^{-\ell_0/(2k)} ,
\end{align}
where the function $Q(q)$ is given by Eq.~\eqref{quench_func_def}.
\end{lemma}

\subsubsection{Proof of Lemma~\ref{lemma:error_time_evolution_rho_0_effective}}

We start with the equations 
\begin{align}
&e^{-iH' t}\rho_0 = e^{-i\tilde{H}' t}\rho_0 + (e^{-iH' t} - e^{-i\tilde{H}' t} )\rho_0 ,  \\
&e^{-i\tilde{H}' t}\rho_0  = \mathcal{T} e^{-i \int_0^\tau \tilde{h}_{X_0}(\tilde{H},-\tau)} e^{-i\tilde{H}t } \rho_0  \notag \\ 
&\quad\quad \quad \ \ =\mathcal{T} e^{-i \int_0^\tau \tilde{h}_{X_0}(\tilde{H},-\tau) d\tau} e^{-i Ht } \rho_0
 -\mathcal{T} e^{-i \int_0^\tau \tilde{h}_{X_0}(\tilde{H},-\tau) d\tau} ( e^{-i\tilde{H}t } - e^{-iHt})  \rho_0,   \\
&\mathcal{T} e^{-i \int_0^\tau \tilde{h}_{X_0}(\tilde{H},-\tau) d\tau} e^{-i Ht } \rho_0 \notag \\
&\quad\quad \quad \ \  = \mathcal{T} e^{-i \int_0^\tau u_{L_2,\tau}^\dagger \tilde{h}_{X_0} u_{L_2,\tau} d\tau } e^{-i Ht } \rho_0 
+ \br{\mathcal{T} e^{-i \int_0^\tau \tilde{h}_{X_0}(\tilde{H},-\tau) d\tau} - \mathcal{T} e^{-i \int_0^\tau u_{L_2,\tau}^\dagger \tilde{h}_{X_0} u_{L_2,\tau} d\tau} } e^{-i Ht } \rho_0 , \label{third_equation_/lemma_rho_0_eff}
 \end{align}  
 where the unitary operators $u_{L_2,\tau}$ in Eq.~\eqref{third_equation_/lemma_rho_0_eff} are appropriately chosen.
 By combining the above equations, we can derive the following inequality:
 \begin{align}
 \label{start_ineq_quench_rho_0_evolution}
&\norm{ e^{-iH' t}\rho_0 -  \mathcal{T} e^{-i \int_0^\tau u_{L_2,\tau}^\dagger \tilde{h}_{X_0} u_{L_2,\tau} d\tau}  e^{-i Ht } \rho_0  }_1 \notag \\
\le&\norm{ (e^{-iH' t} - e^{-i\tilde{H}' t} )\rho_0}_1 + \norm{( e^{-i\tilde{H}t } - e^{-iHt})  \rho_0 }_1 +\norm{\br{\mathcal{T} e^{-i \int_0^\tau \tilde{h}_{X_0}(\tilde{H},-\tau) d\tau} - \mathcal{T} e^{-i \int_0^\tau u_{L_2,\tau}^\dagger \tilde{h}_{X_0} u_{L_2,\tau} d\tau } } }.
\end{align}  
 The norms $\| (e^{-iH' t} - e^{-i\tilde{H}' t} )\rho_0\|_1$ and $ \|( e^{-i\tilde{H}t } - e^{-iHt})  \rho_0 \|_1$ can be derived by using the analyses that were used to obtain~\eqref{Pre_prop:main_ineq_approx_by_hami_tilde}. 
For the $\eta$ and $\eta'$ chosen in Ineq.~\eqref{ineq:lemma:error_time_evolution_effective_Ham}, we obtain the upper bound as  
\begin{align}
 \label{2_ineq_quench_rho_0_evolution}
\norm{ (e^{-iH' t} - e^{-i\tilde{H}' t} )\rho_0}_1 \le \frac{1}{8}e^{c_0\bar{q}} e^{-2\ell_0/\log (r) } ,\quad 
 \norm { ( e^{-i\tilde{H}t } - e^{-iHt})  \rho_0 }_1 \le \frac{1}{8}e^{c_0\bar{q}} e^{-2\ell_0/\log (r) } .
\end{align}
Therefore, our task is to estimate the third term on the RHS of Ineq.~\eqref{start_ineq_quench_rho_0_evolution}.

To this end, we use the following lemma.
\begin{claim} \label{lemma:error_time_evolution_A_B}
Let $A_\tau$ and $B_\tau$ be arbitrary time-dependent operators with continuous time dependence.
Then, the difference between unitary operations by $A_\tau$ and $B_\tau$ is upper-bounded by 
\begin{align}
\label{ineq_lemma:error_time_evolution_A_B}
\| U_{A_\tau, 0\to t} -  U_{B_\tau, 0\to t} \| \le \int_0^t \| A_\tau - B_\tau\| d\tau 
\end{align}
for arbitrary $t$, 
where we use the notation in Eq.~\eqref{notation_time_dependent_operator}. 
\end{claim}

{~}\\

\noindent 
\textit{Proof of Claim~\ref{lemma:error_time_evolution_A_B}.}
For the proof, we first consider
 \begin{align}
U_{A_\tau, 0\to t+dt} 
&= e^{-i A_{t+dt} dt} U_{A_\tau, 0\to t} +\orderof{dt^2} \notag \\
&= e^{-i A_{t+dt} dt} U_{B_\tau, 0\to t} + e^{-i A_{t+dt} dt} \br{ U_{A_\tau, 0\to t} - U_{B_\tau, 0\to t} }+\orderof{dt^2}  \notag \\
&= U_{B_\tau, 0\to t+dt} + \br{ e^{-i A_{t+dt} dt} -e^{-i B_{t+dt} dt}} U_{B_\tau, 0\to t} + e^{-i A_{t+dt} dt} \br{ U_{A_\tau, 0\to t} - U_{B_\tau, 0\to t} } +\orderof{dt^2} .\end{align}
Hence, if we define $\mathcal{G}(t):=\norm{U_{A_\tau, 0\to t} - U_{B_\tau, 0\to t} }$, we have 
 \begin{align}
\frac{d\mathcal{G}(t)}{dt} \le  \| A_{t} -B_{t}\| .
\end{align}
By integrating the above inequality, we obtain the main inequality~\eqref{ineq_lemma:error_time_evolution_A_B}. This completes the proof. $\square$

{~}\\

\noindent 
Using the above lemma, we obtain 
 \begin{align}
\norm{\br{\mathcal{T} e^{-i \int_0^t \tilde{h}_{X_0}(\tilde{H},-\tau) d\tau } - \mathcal{T} e^{-i \int_0^t u_{L_2,\tau}^\dagger \tilde{h}_{X_0} u_{L_2,\tau} d\tau} } } 
\le \int_0^t  \norm{ \tilde{h}_{X_0}(\tilde{H},-\tau)  -u_{L_2,\tau}^\dagger \tilde{h}_{X_0} u_{L_2,\tau}  } d\tau .
\end{align}  
To approximate $\tilde{h}_{X_0}(\tilde{H},-\tau)$ by $u_{L_2,\tau}^\dagger \tilde{h}_{X_0} u_{L_2,\tau}$, we can use Proposition~\ref{prop:short_time_Lieb--Robinson}, which 
yields 
 \begin{align}
\norm{ \tilde{h}_{X_0}(\tilde{H},-\tau)  -u_{L_2,\tau}^\dagger \tilde{h}_{X_0} u_{L_2,\tau} } 
\le 2e^3 \|\tilde{h}_{X_0}\| c_3 \tau  |\partial L_2'| \ell_0 e^{-\ell_0/(2k)}.
\end{align}  
Here, the unitary operator $u_{L_2,\tau}$ is given by 
\begin{align}
\label{U_2_approx_U_2_L_2_choice0}
u_{L_2,\tau}= e^{i(\tilde{H}'_{0,L_2'}+\tilde{V}'_{L_2})\tau} ,
\end{align} 
where we follow the steps as in Eqs.~\eqref{u_X_approx_U_1_L_2_choice1} and \eqref{u_X_approx_U_1_L_2_choice1}. 
We therefore obtain 
 \begin{align}
  \label{3_ineq_quench_rho_0_evolution}
\norm{\br{\mathcal{T} e^{-i \int_0^t \tilde{h}_{X_0}(\tilde{H},-\tau) d\tau} - \mathcal{T} e^{-i \int_0^t u_{L_2,\tau}^\dagger \tilde{h}_{X_0} u_{L_2,\tau} d\tau } } } 
\le 2e^3 \|\tilde{h}_{X_0}\| c_3 t^2  |\partial L_2'| \ell_0 e^{-\ell_0/(2k)}.
\end{align}  

We choose $U_{2,L_2}$ as 
 \begin{align}
\label{U_2_approx_U_2_L_2_choice___2}
U_{2,L_2} = \mathcal{T} e^{-i \int_0^t u_{L_2,\tau}^\dagger \tilde{h}_{X_0} u_{L_2,\tau} d\tau } . 
\end{align}  
By combining Ineqs.~\eqref{2_ineq_quench_rho_0_evolution} and \eqref{3_ineq_quench_rho_0_evolution} with Ineq.~\eqref{start_ineq_quench_rho_0_evolution}, we obtain  
 \begin{align}
 \label{last_ineq_lemma_pfoof_rho_0_H'}
\norm{ e^{-iH' t}\rho_0 -  U_{2,L_2}  e^{-i Ht } \rho_0  }_1
\le \frac{1}{4}e^{c_0\bar{q}}e^{-2\ell_0/\log (r) }  +  2e^3 \|\tilde{h}_{X_0}\| c_3 t^2  |\partial L_2'| \ell_0 e^{-\ell_0/(2k)}  .
\end{align}  
By explicitly writing $\|\tilde{h}_{X_0}\|$ without the simplified notation of Eq.~\eqref{notation_H'_tilde}, we obtain
 \begin{align}
\|\tilde{h}_{X_0}\| \to \|\tilde{h}_{X_0}[L_2,\eta\ell_0, \eta'\ell_0|X|]\| = \|\tilde{h}_{X_0} \bar{\Pi}_{X_0, \eta'\ell_0|X|} \| \le Q(\eta'\ell_0|X|) ,
\end{align}  
where we use Ineq.~\eqref{quench_func_def}. 
Because of $e^{-i Ht } \rho_0=\rho_0 e^{-i Ht }$, the above inequality reduces Ineq.~\eqref{last_ineq_lemma_pfoof_rho_0_H'} to the main inequality~\eqref{ineq_lemma:error_time_evolution_rho_0_effective}. This completes the proof. $\square$

 {~}

\hrulefill{\bf [ End of Proof of Lemma~\ref{lemma:error_time_evolution_rho_0_effective}] }

{~}

By applying Ineq.~\eqref{ineq_lemma:error_time_evolution_rho_0_effective} to Ineq.~\eqref{approx_who_H'_U_L_2'}, we obtain 
 \begin{align}
 \label{ineq:main_theorem_short_time_LR_quench_0}
&\left \|\tilde{\rho}(H',t)   - \br{ U^\dagger_{1,L_2}u_X U_{1,L_2}} U_{2,L_2} \rho_0 U_{2, L_2}^\dagger  \br{ U^\dagger_{1,L_2}u_X U_{1,L_2} }^\dagger \right\|_1 \notag \\
& \le \frac{3}{4}e^{c_0\bar{q}} e^{-2\ell_0/\log (r) }+4e^3  c_3 t (1+t Q(\eta'\ell_0|X|) )  |\partial L_2'| \ell_0 e^{-\ell_0/(2k)} 
\end{align}
for $t\le 1/(ec_3')$.
Finally, we need to choose $C_0'$ in the condition~\eqref{condition_for_lenfth_R_quench}. 
Because $Q(q)$ is given by a finite-degree polynomial, the second term in the above inequality is roughly given by
 \begin{align}
{\rm poly}(\ell_0) \cdot {\rm poly}(r)  e^{-\ell_0/(2k)} ,
\end{align}
where we use $X\subseteq i[r]$. Hence, we can find an $\orderof{1}$ constant $C_0'$ such that for $\ell_0 \ge (C_0'/2) \log^2(r)$,
 \begin{align}
  \label{ineq:main_theorem_short_time_LR_quench_0_cond}
4e^3  c_3 t (1+t Q(\eta'\ell_0|X|) )  |\partial L_2'| \ell_0 e^{-\ell_0/(2k)} \le \frac{1}{4} e^{-2\ell_0/\log (r) }
\le \frac{1}{4}e^{c_0\bar{q}}  e^{-2\ell_0/\log (r) }
\end{align}
holds. 
If we write 
 \begin{align}
\label{U_X_2ell_0_form}
U_{X[2\ell_0]} := U^\dagger_{1,L_2}u_X U_{1,L_2}  U_{2,L_2} ,
\end{align}
Ineq.~\eqref{ineq:main_theorem_short_time_LR_quench_0} with Ineq.~\eqref{ineq:main_theorem_short_time_LR_quench_0_cond} gives the main inequality~\eqref{ineq:main_theorem_short_time_LR_quench} with $\ell=2\ell_0$ and $\Delta t_0=1/(ec_3')$. 
Note that we have defined $X[2\ell_0]=L_2$.

Finally, we consider the time complexity of preparing the unitary operator $U_{X[2\ell_0]}$.
From Eqs.~\eqref{u_X_approx_U_1_L_2_choice2}, \eqref{U_2_approx_U_2_L_2_choice0}, and \eqref{U_2_approx_U_2_L_2_choice___2}, 
the unitary operator~\eqref{U_X_2ell_0_form} is given by
 \begin{align}
\label{U_X_2ell_0_form_final}
U_{X[2\ell_0]} := e^{i(\tilde{H}'_{0,L_2'}+\tilde{V}'_{L_2})t}u_X e^{-i(\tilde{H}'_{0,L_2'}+\tilde{V}'_{L_2})t}  
\mathcal{T} \exp\left[-i \int_0^t e^{i\br{\tilde{H}'_{0,L_2'}+\tilde{V}'_{L_2}}\tau} \tilde{h}_{X_0} e^{-i\br{\tilde{H}'_{0,L_2'}+\tilde{V}'_{L_2}}\tau} d\tau \right].
\end{align}
From this form and the initial condition $u_X \Pi_{X,\ge q_0}=0$ (see the statement in Proposition~\ref{main_theorem_short_time_LR_quench}), we can immediately obtain the equation~\eqref{boson_space_prop}. 

For any operator $O_{L_2}$ supported on $L_2$, after the boson number truncation $\bar{\Pi}_{L_2,q,q'}$, the number of parameters needed to describe
$\bar{\Pi}_{L_2,q,q'} O_{L_2} \bar{\Pi}_{L_2,q,q'}$ is at most $[\max(q,q')]^{|L_2|}$.
To describe the initial unitary operator $u_X$,  
the number of parameters is less than $q_0^{|X|}$ because of the condition~$\Pi_{X,\ge q_0}u_X=0$ in the statement.
%Other possibility is that the boson truncation $q_0$ for $u_{X}$ may be larger than $\max(q,q')$. 
%In this case, the number of parameters is $q_0^{|L_2|}$.  
Now, we have $q=\eta \ell_0$ and $q'=\eta' \ell_0 |X|$ as in Ineq.~\eqref{ineq:lemma:error_time_evolution_effective_Ham}, and hence the time complexity of preparing 
$U_{X[2\ell_0]}$ is at most 
 \begin{align}
\left[\max (q_0, \eta' \ell_0 |X| )\right]^{\orderof{|L_2|}} 
&= \exp\left [ \mathcal{O}\br { (r+\ell_0)^D  \log(r+\ell_0+q_0)} \right]  = 
      \exp\left [ \mathcal{O}\br { (r+\ell)^D  \log(r+\ell+q_0)} \right] 
\end{align}
for $\ell=2\ell_0$.
This completes the proof of Proposition~\ref{main_theorem_short_time_LR_quench}. $\square$

\end{widetext}

\end{document}